\documentclass[a4paper,11pt]{article}
\usepackage{jheppubMod}
\usepackage{amsmath,amsfonts,amssymb,graphicx,subfigure}
\usepackage{datetime}
\usepackage{slashed}
\usepackage{multirow}
\usepackage{numprint}
\usepackage{lineno}
\usepackage[svgnames]{xcolor}
\usepackage{hyperref} 
\hypersetup{colorlinks=true, linkcolor=blue, citecolor=ForestGreen,
filecolor=magenta, urlcolor=ForestGreen,
  pdftitle={A Zee-Babu case study for the LHC, R. Ruiz --
  JHEP 10 (2022) 200, 
  [arXiv:2206.14833]},pdfauthor={R. Ruiz},
  pdfsubject={Subject},pdfcreator={Creator},pdfproducer={Producer},
  pdfkeywords={Zee-Babu}{Neutrino Masses}{Large Hadron Collider}{Beyond the Standard Model Physics}
}

\graphicspath{{Plots/PDF/}}

\newcommand{\orcid}[1]{\,\href{https://orcid.org/#1}{\includegraphics[width=9pt]{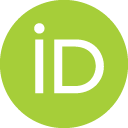}}}
\newcommand{\orcidRR}{0000-0002-3316-2175} 

\newcommand{\eV}{{\rm ~eV}}
\newcommand{\keV}{{\rm ~keV}}
\newcommand{\MeV}{{\rm ~MeV}}
\newcommand{\GeV}{{\rm ~GeV}}
\newcommand{\TeV}{{\rm ~TeV}}
\newcommand{\pb}{{\rm ~pb}}
\newcommand{\fb}{{\rm ~fb}}
\newcommand{\ab}{{\rm ~ab}}
\newcommand{\zb}{{\rm ~zb}}
\newcommand{\invfb}{{\rm ~fb^{-1}}}
\newcommand{\invab}{{\rm ~ab^{-1}}}

\newcommand{\libName}{{\texttt{SM\_ZeeBabu}\ }}

\definecolor{darkgreen}{rgb}{0.0, 0.2, 0.13}
\definecolor{darkmagenta}{rgb}{0.55, 0.0, 0.55}
\definecolor{amber}{rgb}{1.0, 0.6, 0.0}
\newcommand{\confirm}[1]{{\color{black}#1}}

\title{Doubly Charged Higgs Boson Production at Hadron Colliders II:
A Zee-Babu Case Study}

\author{Richard Ruiz\ \orcid{\orcidRR}}

\affiliation{Institute of Nuclear Physics -- Polish Academy of Sciences {\rm (IFJ PAN)},\\ ul. Radzikowskiego, 31-342, Krak{\'o}w, Poland}

\emailAdd{rruiz@ifj.edu.pl}

\abstract{
Motivated by searches for so-called leptonic scalars at the LHC and the recent measurement of the $W$ boson's mass at the Tevatron, we revisit the phenomenology of the Zee-Babu model for neutrino masses and the ability to differentiate it from the Type II Seesaw model at the LHC.
We conclude that this task is much more difficult than previously believed.
All inputs equal in the two scenarios, we find that total and differential rates for producing pairs of doubly and singly charged scalars are identical in shape and only differ in normalization.
The normalization is given by the ratio of hadronic cross sections and  can be unity. 
Differences in cross sections are small and can be hidden by unknown branching rates.
This holds for Drell-Yan, $\gamma\gamma$ fusion, and $gg$ fusion, as well as observables at LO and NLO in QCD. 
This likeness allows us to reinterpret Run II limits on the Type II Seesaw and estimate projections for the HL-LHC.
Using updated neutrino oscillation data, we also find that some collider observables, e.g., lepton flavor-violating branching ratios, are now sufficiently precise to provide a path forward. Other means of discrimination are also discussed.
As a byproduct of this work, we report the availability of new Universal \texttt{FeynRules}  Object libraries, the \texttt{SM\_ZeeBabu} UFO, that enable fully differential simulations up to NLO+LL(PS) with tool chains employing \texttt{MadGraph5\_aMC@NLO}.
}

\preprint{IFJPAN-IV-2022-10 \quad \today}
\arxivnumber{2206.14833}

\begin{document}
\setcounter{page}{1}
\maketitle
\setcounter{page}{2}
\flushbottom

\section{Introduction}\label{sec:intro}

It is a fact of life that neutrino oscillation data support large mixing angles and at least two neutrinos having tiny masses~\cite{Ahmad:2002jz, Ashie:2005ik}, whereas the Standard Model of particle physics (SM) postulates that all three neutrinos are massless Weyl fermions. 
As the SM is an otherwise successful description of data across many scales, one is inclined to extend the model in order to reconcile this discrepancy and reproduce the well-established~\cite{Esteban:2020cvm,Gonzalez-Garcia:2021dve} Pontecorvo-Maki-Nakagawa-Sakata (PMNS) paradigm~\cite{Pontecorvo:1957qd,Pontecorvo:1957cp,Maki:1962mu} from more fundamental principles.
What is less clear, is which, if any, of the many neutrino mass models throughout the literature are at least partially correct.

Among the most studied neutrino mass models are those that conjecture the existence of new fermions, such as right-handed (RH) neutrinos $(\nu_R)$ and vector-like leptons.
While motivated, the most minimal~\cite{Ma:1998dn} incarnations of these, e.g., the Types I~\cite{Minkowski:1977sc, Yanagida:1979as, GellMann:1980vs, Glashow:1979nm, Mohapatra:1979ia, Shrock:1980ct, Schechter:1980gr} and III~\cite{Foot:1988aq} Seesaws models, introduce   coupling and/or mass hierarchies. Subsequently, additional states and interactions are typically needed to soften these hierarchies or render them experimentally testable.
In light of this and of the limited guidance provided by data and theory, it is important to take a broad, complementary approach to neutrino mass models, including exploring scenarios without $\nu_R$.

Along these lines, the Type II Seesaw model~\cite{Konetschny:1977bn, Cheng:1980qt,Lazarides:1980nt,Schechter:1980gr, Mohapatra:1980yp} and Zee-Babu model~\cite{Zee:1985rj,Zee:1985id,Babu:1988ki} are typical scenarios that can reproduce oscillation data without invoking $\nu_R$.
Instead, these models postulate the existence of so-called leptonic scalars, i.e., scalars that carry nonzero lepton number (LN), that also couple directly to electroweak (EW) gauge bosons. 
In the Type II case, the existence of a scalar SU$(2)_L$ triplet $\hat{\Delta}$ is hypothesized and left-handed (LH) Majorana neutrino masses are sourced at tree level from the vacuum expectation value (vev) of $\hat{\Delta}$.
In the Zee-Babu case, the existence of two scalar SU$(2)_L$ singlets $k,h$, which carry hypercharge, are hypothesized and LH Majorana neutrino masses are generated radiatively  at two loops.
In both cases, neutrino masses are proportional to a parameter $\mu_{\not L}$ that signals the scale at which LN is broken.

It is notable that these two models share similar phenomenology. Both predict, for example, the existence of singly and doubly charged scalars, an absence of sterile neutrino mixing, charged lepton flavor violation (cLFV),
as well as lepton number violation (LNV).
Despite these similarities, the number of theoretical studies and experimental searches dedicated to the Type II Seesaw far exceed those for the Zee-Babu model; for reviews, see Refs.~\cite{Deppisch:2015qwa,Cai:2017jrq,Cai:2017mow,Coy:2021hyr}.
This asymmetry is despite radiative neutrino mass models generically inducing neutrino non-standard interactions with matter~\cite{Ohlsson:2009vk,Babu:2019mfe},
despite models of leptonic scalars generically predicting new phenomena at low- and high-energy experiments~\cite{deGouvea:2019qaz,Dev:2021axj,Mandal:2022zmy},
and despite the Zee-Babu model specifically being testable at current and future lepton-flavor experiments~\cite{Babu:2002uu,AristizabalSierra:2006gb,Nebot:2007bc,Ohlsson:2009vk,Araki:2010kq,Schmidt:2014zoa,Herrero-Garcia:2014usa,Okada:2021aoi}
as well as at the Large Hadron Collider (LHC) and its high luminosity upgrade (HL-LHC)~\cite{Babu:2002uu,Nebot:2007bc,Herrero-Garcia:2014usa,Alcaide:2017dcx,Cai:2017jrq,Cai:2017mow}.

In this work, we revisit the phenomenology of the Zee-Babu model and the extent to which it can be distinguished from Type II Seesaw at the hadron colliders.
This study is motivated, in part, by recent the measurement of the $W$ boson's mass at the CDF experiment~\cite{CDF:2022hxs}.
With high significance, the collaboration reports a mass larger than predicted by precision EW data, but also one that fits naturally in the Type II Seesaw~\cite{Almeida:2022lcs,Fan:2022yly,Senjanovic:2022zwy,Bagnaschi:2022whn,Heeck:2022fvl,Cai:2022cti}.
Subsequently, the measurement is reigniting interest in searches for doubly and singly charged scalars at the LHC~\cite{CMS:2012dun,ATLAS:2018ceg,ATLAS:2021jol,ATLAS:2022yzd}.
If a discovery of exotically charged scalars follows soon at the LHC, it will be paramount to exclude one or both models~\cite{delAguila:2013mia,Geib:2015tvt,King:2014uha}.
The aim of the work is to provide some guidance in this direction.

Comparative studies of doubly charged scalars in the Zee-Babu model and other scalar extensions of the SM have been conducted in the past~\cite{Gunion:1996pq,Nebot:2007bc,delAguila:2013yaa,delAguila:2013mia,Schmidt:2014zoa,Herrero-Garcia:2014usa,RodriguezChala:2014oqq,Alcaide:2017dcx,Gluza:2020qrt,Dziewit:2021pak}. 
These, however, are largely restricted to comparing inclusive cross sections via Drell-Yan (DY) at leading order (LO) in quantum chromodynamics (QCD), 
or brute-force recasting using simulated events at LO with parton shower-matching at leading logarithmic  accuracy (LO+LL(PS)).
In this work, we report an interesting observation.
Namely, for the DY, gluon fusion, and photon fusion mechanisms, and for all inputs equal, 
hadronic cross sections (Sec.~\ref{sec:typeii_normalization}) and differential (Sec.~\ref{sec:typeii_kinematics}) distributions of scalars in the Zee-Babu and Type II models differ at most by a uniform scaling factor.
That is to say, for fixed masses, production channel, etc., the shapes of kinematical distributions in the two scenarios are the same. Therefore, fiducial cross sections predicted by one theory can be obtained for the other by a na\"ive re-scaling.
This holds for observables at LO+LL(PS) and next-to-leading order in QCD with PS-matching (NLO+LL(PS)).
For some processes, the scale factor, which is given by the ratio of hadronic cross sections, is precisely unity at LO in the EW theory.
For others, it is nearly independent of scalar mass but inherits a weak sensitivity via the dynamics of parton density functions (PDFs). 
In all cases, differences in cross sections between the two models are sufficiently small that they can be hidden by unknown branching rates.

This observation means that experimentally distinguishing the Zee-Babu and Type II Seesaw models will be more difficult than previously believed. 
However, our findings also show that LHC searches for 
charged scalars decaying directly to leptons in the Type II Seesaw can automatically be reinterpreted in the context of the Zee-Babu model.   
Using Ref.~\cite{ATLAS:2022yzd}, we estimate (Sec.~\ref{sec:typeii_limits}) that $k^{\mp\mp}$ masses as large as {$m_k = 890\GeV$} and decay rates as small as {BR$(k^{\mp\mp}\to\ell^\mp\ell'^\mp)=16\%$} for $\ell\in\{e,\mu\}$ are excluded by the ATLAS experiment with $\mathcal{L}\approx 139\invfb$ of data at $\sqrt{s}=13\TeV$.
Assuming constant analysis and detector performance, we project 
this can reach {$m_k = 1110\GeV$} and {BR$(k^{\mp\mp}\to\ell^\mp\ell'^\mp)=8\%$} with $\mathcal{L}=3\invab$ at $\sqrt{s}=13\TeV$.
Furthermore, with updated neutrino oscillation data, we find (Sec.~\ref{sec:typeii_decay}) that predictions for flavor-violating $h^{\mp}\to \ell^\mp \nu_{\ell'}$ decays in the Zee-Babu are now sufficiently precise to have discriminating power. As a byproduct of this work, we report the  availability of new Universal \texttt{FeynRules}  Object libraries~\cite{Degrande:2011ua}, the \libName UFO, that enable simulations up to NLO+LL(PS) with Monte Carlo tool chains involving \texttt{MadGraph5\_aMC@NLO}~\cite{Stelzer:1994ta,Alwall:2014hca}.
We highlight that previously published UFOs capable of simulating Zee-Babu scalars~\cite{delAguila:2013mia} can only achieve simulations up to LO+LL(PS).

This report continues according to the following:
In Sec.~\ref{sec:theory_model}, we describe the theoretical framework in which we work.
In Sec.~\ref{sec:setup}, we summarize our computational setup and the tuning of our Monte Carlo (MC) tool chain. In Sec.~\ref{sec:pheno}, we broadly revisit and update the phenomenology of Zee-Babu model, including the first NLO in QCD predictions for Zee-Babu scalars at the LHC.
We present our main results in Sec.~\ref{sec:typeii}, where we discuss similarities and differences of Zee-Babu and Type II scalars at hadron colliders.
We summarize and conclude in Sec.~\ref{sec:conclude}.


\section{Theoretical framework: the Zee-Babu model}\label{sec:theory_model}

The Zee-Babu model~\cite{Zee:1985rj,Zee:1985id,Babu:1988ki} 
extends the SM by two complex scalars, $k$ and $h$, with the quantum number assignments $(\mathbf{1},\mathbf{1},-2)$ and $(\mathbf{1},\mathbf{1},-1)$  under the SM gauge group $\mathcal{G}_{\rm SM}=$SU$(3)_c$ $\otimes$SU$(2)_L$ $\otimes$U$(1)_Y$.
Neither carries color or weak isospin but both are charged under weak hypercharge. $k$ and $h$ are assigned lepton number $L=+2$, which is normalized such that SM leptons carry $L=+1$. In terms of the SM Lagrangian $(\mathcal{L}_{\rm SM})$,
the Lagrangian of the Zee-Babu model  $(\mathcal{L}_{\rm ZB})$ is
\begin{align}
\label{eq:lag_full}
 \mathcal{L}_{\rm ZB} = \mathcal{L}_{\rm SM} + \mathcal{L}_{\rm Kin.} + \mathcal{L}_{\rm Yuk.} + \mathcal{L}_{\rm ZB\ scalar}
 + \delta\mathcal{L}_{\nu}
 \ .
\end{align}

The kinetic part of the Lagrangian for $k$ and $h$ is given by the following covariant derivatives    
\begin{align}
\label{eq:lag_kin}
  \mathcal{L}_{\rm Kin.} &= (D_\mu k)^\dagger (D^\mu k) + (D_\mu h)^\dagger (D^\mu h), 
  \quad\text{with}\quad
  D_\mu = \partial_\mu +i g_Y \hat{Y} B_\mu\ .
\end{align}
Here, the weak hypercharge operator is normalized such that the electromagnetic charge operator is $\hat{Q}=\hat{T}_L^3+\hat{Y}$, and $Y_k = -2\ (Y_h = -1)$. 
The weak hypercharge coupling is denoted by $g_Y\approx 0.36$.
As neither $k$ nor $h$ mix with SM states, the mass eigenstates, denoted by $k^{--}$ and $h^-$, are aligned with their gauge states and carry the electric charges $Q_k=-2$ and $Q_h=-1$, respectively.
(In the following, we often omit the charge symbols $\mp$ when referring to mass eigenstates.)

After EW symmetry breaking (EWSB), the hypercharge field $B_\mu$ mixes with the weak isospin field $W_\mu^3$ and can be decomposed in terms of the usual mass eigenstates $B_\mu = \cos\theta_W A_\mu - \sin\theta_W Z_\mu$, where $\theta_W$ is the weak mixing angle. 
In the mass basis, Eq.~\eqref{eq:lag_kin} becomes
\begin{align}
& \mathcal{L}_{\rm Kin.} \overset{\rm EWSB}{=} 
 (\partial_\mu S^\dagger) (\partial^\mu S) 
 + e^2 Q_s^2 (S^\dagger S) A_\mu A^\mu
 \nonumber
 + e^2 Q_s^2\ \tan^2\theta_W\  (S^\dagger S) Z_\mu Z^\mu 
  \\
 \label{eq:lag_kin_ewsb}
 &
  + 2e^2 Q_s^2\ \tan\theta_W\   (S^\dagger S) A_\mu Z^\mu 
 -i e Q_S \left(A_\mu - \tan\theta_W Z_\mu\right)\
 \left((\partial^\mu S^\dagger) S - S^\dagger (\partial^\mu S)\right)
 \ .
\end{align}
As usual, $e=g_Y \cos\theta$ is the electromagnetic coupling.
The list $(S,S^\dagger,Q_s)$ is shorthand for the mass eigenstates and charge $(k^{--},k^{++},Q_k)$ and $(h^-,h^+,Q_h)$.
In anticipation of Sec.~\ref{sec:typeii_normalization}, we highlight that 
$h$ and $k$ couple to the $Z$ boson only through $B-W^3$ mixing since they are SU$(2)_L$ singlets.
And since the weak mixing angle (modulo running) is only about $\theta_W\approx29^\circ$, $h$ and $k$ inherently couple  more to the photon than to the $Z$.
For example: in comparison to the four-point $S-S-A-A$ vertex,
the $S-S-A-Z$ vertex is suppressed by a factor of $\tan\theta_W\approx 1/\sqrt{3}\approx0.55$, and
the $S-S-Z-Z$ vertex is down $\tan^2\theta_W\approx 1/3$.
Similarly, the $S-S-Z$ vertex is suppressed by $\tan\theta_W\approx 1/\sqrt{3}$ in comparison to the $S-S-A$ vertex.

The Yukawa part of $\mathcal{L}_{\rm ZB}$ 
describes the coupling of SM leptons to $k$ and $h$. It is given by
\begin{align}
 \mathcal{L}_{\rm Yuk.} & \ ~  = 
 f_{ij}\ \overline{\tilde{L}^i} L^j h^\dagger
 +
 g_{ij}\ \overline{(e_R^c)^i} e_R^j k^\dagger + \text{H.c.}
  \label{eq:lag_yukawa}
 \\
 &
 \overset{\rm EWSB}{=}
 f_{\ell\ell'}\ \left(\overline{(\ell_L^c)}\nu_{\ell'} - \overline{(\nu_\ell^c)}\ell'_L\right) h^\dagger 
  +
 g_{\ell\ell'}\ \overline{(\ell_R^c)} \ell'_R k^\dagger + \text{H.c.}
\end{align}
Here, $(L^i)^T = (\nu_L^i, e_L^i)$ is the SM LH lepton doublet with generation index $i=1,\dots,3$;
$\tilde{L}^i \equiv i\sigma^2 (L^c)^i$ is the usual rotation in SU$(2)_L$ space but of $L$'s charge conjugate; and
$e_R^i$ is the SM RH charged lepton.
The Yukawa couplings to $h$ are given by $f_{ij}$,
a $3\times3$, complex matrix that is anti-symmetric, i.e., $f_{ij}=-f_{ji}$.
The Yukawa couplings to $k$ are given by $g_{ij}$, a $3\times3$, complex matrix symmetric, i.e., $g_{ij} = g_{ji}$.
After EWSB, the chiral states $e^i_{L/R}$ can be rotated trivially into their flavor/mass eigenstates $\ell=e,\mu,\tau$.
At this point neutrinos are still massless, meaning that their gauge and flavor states are also aligned.
Formally, this leads to the redefinition of $g_{\ell\ell'}=R^*_{\ell i}g_{ij}R_{j\ell'}$ and $f_{\ell\ell'}=R^*_{\ell i}f_{ij}R'_{j\ell'}$, where $R$ and $R'$ are the identity matrix. The Yukawa couplings $g_{\ell\ell'}$ and $f_{\ell\ell'}$ induce LFV in decays and transition of $\tau$ and $\mu$. While this is constrained by experimental searches for LFV, $k$ and $h$ with masses below 1 TeV are still allowed~\cite{Nebot:2007bc,Ohlsson:2009vk}.

The scalar potential of $k$ and $h$, including couplings to the SM Higgs doublet $\Phi$, is given by
\begin{align}
- \mathcal{L}_{\rm ZB\ scalar} &=\ 
\tilde{m}_k^2 k^\dagger k  +\ \tilde{m}_h^2 h^\dagger h\
+\ \lambda_k (k^\dagger k)^2\ +\ \lambda_{h} (h^\dagger h)^2\
+\ \lambda_{hk} (k^\dagger k)(h^\dagger h)\
\nonumber\\
&
+\ \left(\mu_{\not L}\ h h k^\dagger + \text{H.c.}\right)\
+\ \lambda_{kH} (k^\dagger k) \Phi^\dagger \Phi\ 
+\ \lambda_{hH} (h^\dagger h) \Phi^\dagger \Phi 
\ .
\end{align}
The states $H^0$, $G^\pm$, and $G^0$ are the usual SM Higgs and Goldstone bosons, with 
$\sqrt{2} \Phi^T = (-i\sqrt{2} G^+, v + H^0 + iG^0)$ and
$v = \sqrt{2}\langle\Phi\rangle\approx 246\GeV$.
After EWSB, one has in the mass basis
\begin{align}
\label{eq:lag_scalar}
  -  \mathcal{L}_{\rm ZB\ scalar}  &\overset{\rm EWSB}{=}\
 m_k^2 k^{++} k^{--}\  +\ m_h^2 h^+ h^-\
+\ \lambda_k (k^{++} k^{--})^2\
+\ \lambda_{h} (h^+ h^-)^2\
\nonumber\\
&
+\ \lambda_{hk}\ k^{++} k^{--}\ h^+ h^-\
+\ \left(\mu_{\not L}\ h^- h^- k^{++} + \text{H.c.}\right)\
\nonumber\\
&
+\ \frac{\lambda_{kH}}{2}\ k^{++} k^{--} \left(2vH^0 + H^0H^0\right)\
+\ \frac{\lambda_{hH}}{2}\ h^+ h^- \left(2vH^0 + H^0H^0\right)\
\nonumber\\
&
+\ \frac{\lambda_{kH}}{2}\ k^{++} k^{--} \left(2G^-G^+ + G^0G^0\right)\
+\ \frac{\lambda_{hH}}{2}\ h^+ h^-\ \left(2G^-G^+ + G^0G^0\right) 
\ .
\end{align}
The physical masses of $k$ and $h$ are, respectively,
\begin{align}
m_k^2 = \tilde{m}^2_k + \frac{\lambda_{kH}}{2}v^2
\quad\text{and}\quad 
m_h^2 = \tilde{m}^2_h + \frac{\lambda_{hH}}{2}v^2
\ .
\end{align}
Demands for a first-order EW phase transition favors lighter masses, with $m_k, m_h<300\GeV$~\cite{Phong:2015vlk}.

A few comments:
(i)
In this work, we adopt the conventional assignments of LN wherein leptons carry $L=+1$, $h$ and $k$ carry $L=+2$, and antiparticle states carry $L<0$. This implies that the three-point vertex $h-h-k$, which is proportional to dimensionful parameter $\mu_{\not L}$,  violates LN explicitly by $\Delta L = \pm2$ units.
We identify $\mu_{\not L}$ as the scale of LNV.
In the $\mu_{\not L}\to0$ limit, LN is conserved in the Zee-Babu model; conversely, in the limit where $\mu_{\not L}$ is fixed but either $k$ or $h$ is infinitely heavy, i.e., the decoupling limit~\cite{Appelquist:1974tg}, the $h-h-k$ vertex vanishes and leads to LN conservation.
(ii) The assignment also implies that the Yukawa interactions in Eq.~\eqref{eq:lag_yukawa} conserve LN.
However, in the absence of $\mu_{\not L}$, $g$, $f$, or the Yukawa couplings $y_\ell$ between the SM Higgs and charged leptons, LN can be redefined such that it is conserved~\cite{Nebot:2007bc}.
(iii) In $\mathcal{L}_{\rm ZB\ scalar}$, the coupling normalizations follow Refs.~\cite{Nebot:2007bc}.
In this convention, the $h-h-k$ Feynman rule, $\Gamma_{h-h-k}=-i(2!)\mu_{\not L}$, carries a factor of $(2!)$ that would otherwise cancel in other normalizations.
(iv) In this model, neutrinos are massless  at tree level after EWSB, i.e., $\delta\mathcal{L}_{\nu}=0$. Unlike the Types I-III Seesaws, they are generated radiatively. Discussion of this is postponed to Sec.~\ref{sec:pheno_nuMasses}.


\section{Computational setup and Monte Carlo tuning}\label{sec:setup}

In order to assist reproducing our results, we now document our Monte Carlo tool chain (Sec.~\ref{sec:setup_mc}), our SM inputs  (Sec.~\ref{sec:setup_sm}), and our benchmark Zee-Babu inputs (Sec.~\ref{sec:setup_zb}).


\subsection{Monte Carlo tool chain}\label{sec:setup_mc}

To study the Zee-Babu model numerically,  we transcribe the tree-level Lagrangian with Goldstone boson couplings in Eq.~\eqref{eq:lag_full} into \texttt{FeynRules} v2.3.36~\cite{Christensen:2008py,Alloul:2013bka}.
For the SM Lagrangian, we use the implementation available in \texttt{FeynRules}, the file \texttt{sm.fr} v1.4.7.
We phenomenologically parameterize the Lagrangian and set $\delta\mathcal{L}_\nu=0$. QCD ultraviolet and $R_2$ counter terms up to $\mathcal{O}(\alpha_s)$  are computed using \texttt{NLOCT} v1.02~\cite{Degrande:2014vpa} and \texttt{FeynArts} 3.11~\cite{Hahn:2000kx}.
Feynman rules up to one loop in $\alpha_s$ are then generated and packaged into a series of Universal \texttt{FeynRules} Output (UFO) libraries~\cite{Degrande:2011ua} that we collectively label the \libName UFO libraries\footnote{The UFO libraries \texttt{SM\_ZeeBabu\_NLO}, \texttt{SM\_ZeeBabu\_XLO}, etc., and the associated \texttt{FeynRules} generation files are publicly available on the \texttt{FeynRules} model database at the URL \href{https://feynrules.irmp.ucl.ac.be/wiki/ZeeBabu}{https://feynrules.irmp.ucl.ac.be/wiki/ZeeBabu}.}. In this work, we use the \texttt{SM\_ZeeBabu\_NLO} UFO, which enables the computation of tree-induced processes up to NLO+LL(PS) in QCD and QCD loop-induced processes up to LO+LL(PS).
We have checked that our implementation of the model agrees with partonic expressions for $k$ and $h$ pair production at LO~\cite{Gunion:1996pq}, 
as well as with hadronic-level rates for $k$ pair production at LO~\cite{delAguila:2013mia,Geib:2015tvt,King:2014uha}.

For computing matrix element and generating events,
we use \texttt{MadGraph5\_aMC@NLO} (\texttt{mg5amc}) v3.4.0~\cite{Stelzer:1994ta,Alwall:2014hca}, 
which employs \texttt{MadLoop}~\cite{Hirschi:2011pa,Hirschi:2015iia} and  
the MC@NLO formalism~\cite{Frixione:2002ik} as implemented in \texttt{MadFKS}~\cite{Frixione:1995ms,Frixione:1997np,Frederix:2009yq}.
The interface between the UFO and \texttt{mg5amc} is handled by \texttt{ALOHA}~\cite{deAquino:2011ub}.
Events are parton showered using \texttt{Pythia} v8.306~\cite{Bierlich:2022pfr}, with underlying event / multi-particle interactions and QED showering enabled.
Hadrons are clustered using the anti-$k_T$ sequential clustering algorithm~\cite{Cacciari:2008gp} as implemented in \texttt{FastJet}~\cite{Cacciari:2005hq,Cacciari:2011ma}.
A customized analysis\footnote{Scripts and analyses libraries written for this study are available publicly from the URL:\\ \href{https://gitlab.cern.ch/riruiz/public-projects/-/tree/master/ZeeBabu_LHC_Update}{https://gitlab.cern.ch/riruiz/public-projects/-/tree/master/ZeeBabu\_LHC\_Update}.} is used to analyze hadron-level events with the \texttt{Histogram with Uncertainties} (HwU)~\cite{Alwall:2014hca} platform.


\subsection{Standard Model  inputs}\label{sec:setup_sm}

For numerical results, we assume $n_f=5$  massless quarks and the following SM inputs~\cite{ParticleDataGroup:2020ssz}:
\begin{subequations}
\label{eq:input_sm}
\begin{align}
\sin^2\theta_W &= 0.23126, \ \alpha^{-1}_{\rm QED}(M_Z) = 127.94, \
M_Z = 91.1876\GeV, \ \Gamma_Z=  2.4952\GeV,
\\
m_t(m_t) &= 173.3\GeV, \ \Gamma_t=1.350\GeV, \
m_H=125.7\GeV, \ \Gamma_W= 2.085\GeV\ . 
\end{align}
\end{subequations}
At tree level, this corresponds to 
$M_W\approx 79.95\GeV$, $v\approx 245\GeV$, and $G_F \approx 1.17456\times10^{-5}\GeV^{-2}$. For select results, we use the following charged lepton masses:
\begin{align}
 \label{eq:input_lepton}
 m_e = 511\keV,\quad m_\mu \approx 106\MeV, \quad m_\tau \approx 1.78\GeV\ .
\end{align}
We otherwise assume charged lepton are massless\footnote{By default, we use the \texttt{SM\_ZeeBabu\_NLO} UFO, which assumes massless leptons.
Wherever charged lepton masses are relevant, we use the
\texttt{SM\_ZeeBabu\_MassiveLeptons\_NLO} UFO, which is an otherwise identical UFO.}.
We approximate the  Cabbibo-Kobayashi-Maskawa matrix by the identity matrix.
For hadronic cross sections we use the MMHT 2015 QED NLO (\texttt{lhaid=26000}) and next-to-next-to-leading order (NNLO) (\texttt{lhaid=26300}) PDF sets \cite{Harland-Lang:2019pla}.
Both PDF sets employ to the LUXqed formalism to determine the photon PDF~\cite{Manohar:2016nzj,Manohar:2017eqh} and use $\alpha_s(M_Z)\approx0.1180$.
PDFs and $\alpha_s(\mu_r)$ are evolved using \texttt{LHAPDF} v6.3.0~\cite{Buckley:2014ana}. PDF uncertainties are extracted using eigenvector sets~\cite{Martin:2009iq} as implemented in \texttt{LHAPDF}.
For all DY processes we use the NLO PDF set;
for all non-DY processes we use the NNLO PDF set.

For DY calculations at LO and NLO, we set the central collinear factorization $(\mu_f)$ and renormalization $(\mu_r)$ scales to be half the sum of transverse energies of final state particles:
\begin{subequations}
\label{eq:scale_choice}
\begin{align}
 \mu_f, \mu_r &= \zeta \times \mu_0, \quad\text{where}\quad \zeta=1\quad \text{and}
\quad
 \mu_0 = \frac{1}{2} \sum_{f\in\{\text{final state}\}} \sqrt{m_f^2 + p_{Tf}^2}\ .
\end{align}
For all other calculations, we set the two scales equal to the scale of hard scattering process:
\begin{align}
 \mu_0 &= Q \equiv \sqrt{ q^2}, \quad\text{where}\quad
 q^\mu =\sum_{f\in\{\text{final state}\}} p_f^\mu \ .
\end{align}
\end{subequations}
The 9-point scale uncertainty is obtained by varying $\zeta$ over the discrete range $\zeta\in\{0.5,1.0,2.0\}$.  

Finally, the shower scale $\mu_s$ is set according to its default prescription~\cite{Alwall:2014hca}.
To steer the shower, we use the MSTW 2008  LO PDF set (\texttt{lhaid=21000})~\cite{Martin:2009iq} and the
ATLAS A14 central tune (\texttt{Tune:pp = 20})~\cite{TheATLAScollaboration:2014rfk}. We do not estimate the uncertainty associated with PS modeling.


\subsection{Zee-Babu inputs}\label{sec:setup_zb}

As neutrinos are effectively massless on momentum scales observed at the LHC and to minimize potential theoretical biases, we take a phenomenological approach and neglect neutrino masses for collider computations.
In practice, this means that the relationships in Eqs.~\eqref{eq:nuMasses_full}, \eqref{eq:yukawa_ratio_no}, and \eqref{eq:yukawa_ratio_io} are not imposed. This allows us to vary nonzero $f_{\ell\ell'}$ and $g_{\ell\ell'}$ freely and independently.

Unless specified, we assume the following model benchmark inputs 
\begin{align}
 m_k =  500\GeV,\  m_h = 300\GeV,\ \mu_{\not L}=1\TeV,\
 \{\lambda\} =1,\ g_{\ell\ell'}=1,\ f_{\ell\ell'}=(1-\delta_{\ell\ell'}), 
\label{eq:benchmark_inputs}
 \end{align}
where $\{\lambda\}$ represents all the scalar couplings in the Lagrangian $\mathcal{L}_{\rm ZB~scalar}$ of Eq.~\eqref{eq:lag_scalar}.
In Table~\ref{tab:ufoInputs}, we summarize the external inputs of the {\libName} UFO and their default values.

\begin{table*}[t!]
\begin{center}
\resizebox{\textwidth}{!}{
\begin{tabular}{| c  | c c c | c  c c | c  c c |}
\hline\hline
\multirow{2}{*}{Particle information}
&
$k^{--}\ (k^{++})$ & \texttt{k--}\ (\texttt{k++}) & PID: 61\ (-61) &
$m_k$ & \texttt{mkZB} & 500\GeV &
$\Gamma_k$ & \texttt{wkZB} & 1 GeV 
\\
 &
$h^{-}\ (h^{+})$ & \texttt{h-}\ (\texttt{h+})   & PID: 38\ (-38) &
$m_h$ & \texttt{mhZB} & 300\GeV &
$\Gamma_h$ & \texttt{whZB} & 1 GeV 
\\
\hline
\multirow{2}{*}{Scalar potential couplings} &
$\lambda_{h}$ & \texttt{lamhZB} & 1 &
$\lambda_{k}$ & \texttt{lamkZB} & 1 &
$\mu_{\not L}$ & \texttt{muZB} & 1\TeV 
\\
 &
$\lambda_{hk}$ & \texttt{lamhZBkZB} & 1 & 
$\lambda_{hH}$ & \texttt{lamhZBH} & 1 &
$\lambda_{kH}$ & \texttt{lamkZBH} & 1 
\\
\hline
Antisymmetric Yukawa couplings &
$f_{e\mu}$ & \texttt{femu} & 1 &
$f_{e\tau}$ & \texttt{fetau} & 1 &
$f_{\mu\tau}$ & \texttt{fmutau} & 1 
\\
\hline
\multirow{2}{*}{Symmetric Yukawa couplings} &
$g_{ee}$ & \texttt{gee} & 1 &
$g_{e\mu}$ & \texttt{gemu} & 1 &
$g_{e\tau}$ & \texttt{getau} & 1 
\\
&
$g_{\mu\mu}$ & \texttt{gmumu} & 1 &
$g_{\mu\tau}$ & \texttt{gmutau} & 1 &
$g_{\tau\tau}$ & \texttt{gtautau} & 1 
\\
\hline\hline
\end{tabular}
} 
\caption{Inputs of the Zee-Babu Lagrangian (left symbol) as given in Sec.~\ref{sec:theory_model}, their corresponding identifier (center in \texttt{typewriter} font) in the {\libName} UFO, and their default value in the UFO.
}
\label{tab:ufoInputs}
\end{center}
\end{table*}


\section{Phenomenology of the canonical Zee-Babu model}\label{sec:pheno}

In this section, we revisit the non-collider and collider phenomenology of the Zee-Babu model.
We start in Sec.~\ref{sec:pheno_nuMasses} with a discussion of neutrino masses and then summarize the decay properties of $k$ and $h$  in Sec.~\ref{sec:pheno_decay}.
Theoretical constraints from partial wave unitarity are obtained in Sec.~\ref{sec:pheno_unitarity}.
Finally, we present updated cross section predictions for the LHC in Sec.~\ref{sec:pheno_lhc}.
We stress that several findings here have not previously been  reported in the literature.


\subsection{Neutrino masses}\label{sec:pheno_nuMasses}

In the Zee-Babu model, there are no $\nu_R$ and therefore no Dirac neutrino masses. Likewise, $k$ and $h$ cannot contract with $L$ and $\Phi$ so as to generate LH Majorana masses at tree level.
Instead, the $\mu_{\not L}$ term in $\mathcal{L}_{\rm ZB~scalar}$ induces LH 
Majorana masses at two loops. In the flavor basis and with flavor indices $i,j,a,b\in\{e,\mu,\tau\}$, neutrino masses are described by Lagrangian 
~\cite{Zee:1985id,Babu:1988ki}
\begin{subequations}
\begin{align}
\delta\mathcal{L}_{\nu}^{\rm 2-loop} &= -\frac{1}{2} \overline{(\nu_L^c)^i}\ \left(\mathcal{M}_\nu^{\rm flavor}\right)_{ij}\ \nu_L^j + \text{H.c.}\ ,
\\
\left(\mathcal{M}_\nu^{\rm flavor}\right)_{ij} &= 16\mu_{\not L}\ f_{ia}\ m_a\ g_{ab}^*\ \mathcal{I}_{ab}(r)\ m_b\ f_{jb}.
\end{align}
\end{subequations}
The integral factor $\mathcal{I}_{ab}(r)$ is approximately given by the expression~\cite{McDonald:2003zj,Nebot:2007bc}
\begin{subequations}
\begin{align}
 \mathcal{I}_{ab}(r) &\approx \frac{\pi^2}{3(16\pi^2)^2}\frac{\delta_{ab}}{M^2_{\max}} \tilde{I}(r), \quad \text{where}
 \quad
\tilde{I}(r) \approx \left\{\begin{matrix}
1+\frac{3}{\pi^2}(\log^2 r - 1), & r\gg 1\\ 
1, & r\to0
\end{matrix}\right. \ .
\end{align}
\end{subequations}
Here, $r = (m_k^2 / m_h^2)$ and $M_{\max}=\max(m_k,m_h)$.
Rotating neutrinos from their flavor states $(\nu_\ell)$ into their mass states $(\nu_m)$ via the PMNS matrix~\cite{Pontecorvo:1957qd,Pontecorvo:1957cp,Maki:1962mu}, i.e.,
\begin{align}
 \nu_L^i = U^{\rm PMNS}_{im}\ \nu_m, \quad m=1,\dots,3,
\end{align}
allows one to diagonalize the mass matrix.
The result is
\begin{align}
\label{eq:nuMasses_massBasis}
 \left(\mathcal{M}_\nu^{\rm mass}\right) &= \text{diag}(m_1,m_2,m_3)
 \\
 &= 
 U^{\rm PMNS*}_{mi}\
 \left(\mathcal{M}_\nu^{\rm flavor}\right)_{ij}\
 U^{\rm PMNS}_{jm}
 \\
 &= 
 \label{eq:nuMasses_full}
 (16\mu_{\not L})\ \times  
U^{\rm PMNS*}_{mi}\
 f_{ia}\ m_a\ g_{ab}^*\ \mathcal{I}_{ab}\ m_b\ f_{jb}\
 U^{\rm PMNS}_{jm}\ 
 \\ 
  \sim\  &  \mathcal{O}\left(10^{-3}\right)\times \mathcal{O}\left(f^2 g\right)\times  \mu_{\not L }\times \left(\frac{m_a m_b}{M^2_{\max}}\right) .
\end{align}

For Yukawa couplings $f\sim g\sim\mathcal{O}(0.1)$, and charged lepton masses $(m_a m_b)\sim \mathcal{O}(0.1)\GeV^2$, neutrino masses that are naturally $\mathcal{O}(1)\eV$  can be obtained from Zee-Babu mass scales $\mu_{\not L}, M_{\rm max}$ that are $\mathcal{O}(100)\GeV$.
Given presently available oscillation data~\cite{Gonzalez-Garcia:2021dve}, the above expression impose meaningful constraints on $f_{ij}$ and $g_{ij}$.
However, exploring the rich complementarity of oscillation data, low-energy flavor data, and high-energy collider data for the Zee-Babu model is outside the scope of this work. Such studies have been conducted in Refs.~\cite{Nebot:2007bc,Ohlsson:2009vk,Schmidt:2014zoa,Okada:2021aoi}.
Nevertheless, we comment on a nontrivial correlation between oscillation data and the Yukawa couplings.

The antisymmetric nature of $f_{\ell\ell'}$ implies a zero determinant:
\begin{align}
 \det(f_{\ell\ell'})=\det(f_{\ell'\ell})=\det(-f_{\ell\ell'})=(-1)^3\det(f_{\ell\ell'})=0\ ,
\end{align}
and subsequently that 
$\det(\mathcal{M}_\nu^{\rm mass})=0$.
This forces at least one neutrino to be massless. Thus, the mass spectrum is fixed by the measured atmospheric and solar 
mass splittings, up to the mass-ordering ambiguity.
Moreover, one can build eigenvector equations that relate elements of $f_{\ell\ell'}$ and $U^{\rm PMNS}$. For the normal ordering (NO) of neutrino masses, one has~\cite{Nebot:2007bc}
\begin{subequations}
\label{eq:yukawa_ratio_no}
\begin{align}
\frac{f_{e\tau}}{f_{\mu\tau}} &= \tan\theta_{12}\frac{\cos\theta_{23}}{\cos\theta_{13}} + \tan\theta_{13}\sin\theta_{23} e^{-i\delta}\ , 
\\
\frac{f_{e\mu}}{f_{\mu\tau}} &= \tan\theta_{12}\frac{\cos\theta_{23}}{\cos\theta_{13}} - \tan\theta_{13}\sin\theta_{23} e^{-i\delta}\ ,
\end{align}
\end{subequations}
while for the inverse order (IO), one has 
\begin{subequations}
\label{eq:yukawa_ratio_io}
\begin{align}
 \frac{f_{e\tau}}{f_{\mu\tau}} &= -\frac{\sin\theta_{23}}{\tan\theta_{13}}e^{-i\delta},
 \\
  \frac{f_{e\mu}}{f_{\mu\tau}} &= \frac{\cos\theta_{23}}{\tan\theta_{13}}e^{-i\delta},
  \\
  \frac{f_{e\tau}}{f_{e\mu}} &= -\tan\theta_{23}.
\end{align}
\end{subequations}
Relationships for $g_{\ell\ell'}$ also exist but are more complicated. For both mass orderings, consistency between oscillation data and the hierarchy of charged lepton masses leads to scaling~\cite{Nebot:2007bc}:
\begin{align}
\label{eq:yukawa_ratio_g}
 g_{\tau\tau}\ :\ g_{\mu\tau}\ :\ g_{\mu\mu} \sim 
 \frac{m_\mu^2}{m_\tau^2}\ : 
 \frac{m_\mu}{m_\tau}\ : 1\ .
\end{align}


\subsection[{Decay channels of k and h}]{Decay channels of $k^{\mp\mp}$ and $h^\mp$}\label{sec:pheno_decay}

We now comment on the leading and sub-leading decays of $k^{\mp\mp}$ and $h^\mp$. While formula for two-body partial widths $(\Gamma)$ have been documented before, not all of the following properties have been reported. For an $n_f$-body state $f$, the $i\to f$ partial width is given by the formula
\begin{subequations}
\begin{align}
 \Gamma(i\to f) &= \int dPS_{n_f}\ \frac{d\Gamma}{dPS_{n_f}},
 \quad
\frac{d\Gamma}{dPS_{n_f}} = \frac{1}{2m_i}\frac{1}{\mathcal{S}_i} \sum_{\rm dof} \vert \mathcal{M}(i\to f)\vert^2,
\\
dPS_{n_f} &= (2\pi)^4\  \delta^{4}\left(p_i-\sum_{k=1}^{n_f} p_k\right)\ \prod_{k=1}^{n_f}\frac{d^3 p_k}{(2\pi)^3\ 2E_k}\ .
\label{eq:def_phase_space}
\end{align}
\end{subequations}
Here, $\mathcal{M}$ is the $i\to f$ matrix element; the summation in the first line is over discrete degrees of freedom (dof); e.g., helicity, $\mathcal{S}_i$ is the spin-averaging multiplicity; and $dPS_{n_f}$ is the phase space integration measure.
The total width $(\Gamma_i^{\rm Tot.})$ of $i$  and the $i\to f$ branching rate (BR) are then:
\begin{align}
 \Gamma_i^{\rm Tot.} &= \sum_f\ \Gamma(i\to f),
 \quad\text{and}\quad
  {\rm BR}(i\to f) &= \frac{\Gamma(i\to f)}{\Gamma_i^{\rm Tot.}} \ .
\end{align}

Starting with the state $k^{\mp\mp}$, the leading two- and three-body decay channels include
\begin{subequations}
 \begin{align}
  k^{\mp\mp} &\to \ell^\mp \ell'^\mp\ ,
  \\
  k^{\mp\mp} &\to h^\mp h^\mp\ ,
  \\
  k^{\mp\mp} &\to h^\mp \ell^\mp\ \overset{(-)}{\nu_{\ell'}}\ ,
  \\
  k^{\mp\mp} &\to h^\mp h^\mp Z/H^0/\gamma\ ,
  \\
  k^{\mp\mp} &\to \ell^\mp\ \overset{(-)}{\nu_{\ell'}}\ W^\mp\ . \end{align}
\end{subequations}
The first proceeds by the symmetric Yukawa coupling $g_{\ell\ell'}$ and conserves LN since $k^{\mp\mp}$ carries $L=\pm2$.
The second proceeds through $\mu_{\not L}$ and violates LN since $h^\mp$ 
also carries $L=\pm2$. 
The last three are radiative corrections to the first two 
and are  coupling or phase-spaced suppressed.

The $k^{\mp\mp} \to \ell^\pm \ell'^\pm$ partial width with full lepton-mass dependence is given by
\begin{subequations}
\begin{align}
 \Gamma(k^{\mp\mp}\to \ell^\pm \ell'^\pm) &=  \frac{\vert g_{\ell\ell'}\vert^2}{4\pi(1+\delta_{\ell\ell'})}\ m_k\ 
 \times   (1-r_\ell - r_{\ell'})\ \lambda^{1/2}(1,r_\ell,r_{\ell'})\ ,
 \\
 \text{where}\quad \lambda(x,y,z) &= (x-y-z)^2 - 4yz\ ,  \quad\text{and}\quad 
 r_i = \frac{m_i^2}{m_k^2}\ .
\end{align}
\end{subequations}
The Kronecker $\delta_{\ell\ell'}$ accounts for $1/(2!)$ symmetry factor 
for identical particles in the final state. 

Assuming $m_k > 2m_h$, the $k^{\mp\mp}\to h^\mp h^\mp$ partial width is given by
\begin{align}
 \Gamma(k^{\mp\mp}\to h^\mp h^\mp) &= \frac{1}{8\pi}\ \left(\frac{\mu^2_{\not L}}{m_k}\right)\sqrt{1-4r_h}\ ,
 \quad\text{where}\quad r_h = \frac{m_h^2}{m_k^2}\ .
\end{align}
Unusually, this decay is inversely proportional to the mass of $k$; normally, partial widths grow as a positive power of a parent particle's mass. 
Subsequently, the $k^{\mp\mp}\to h^\mp h^\mp$ partial width can be suppressed if the scale of LNV is much smaller than $m_k$.
At the same time, the branching rate can be competitive, or even dominant, if the couplings $g_{\ell\ell'}$ and $f_{\ell\ell'}$ 
are sufficiently small.

For the benchmark inputs in Eqs.~\eqref{eq:input_lepton} and \eqref{eq:benchmark_inputs}, one finds the following branching rates:
\begin{subequations}
\label{eq:decay_rate_k}
\begin{align}
 {\rm BR}(k^{--}\to\ell^- \ell'^-)\quad & \sim 22\%, \quad \text{for}\ \ell\neq\ell'
 \\
  {\rm BR}(k^{--}\to\ell^- \ell'^-)\quad & \sim 11\%, \quad \text{for}\ \ell=\ell'
 \\
 {\rm BR}(k^{--}\to\ell^- \nu_{\ell'}h^-) & \sim 0.14\%,
 \\
 {\rm BR}(k^{--}\to\ell^- \nu_{\tau}W^-) & \sim 1\cdot10^{-5}\ \%,
 \\
  {\rm BR}(k^{--}\to\ell^- \nu_{\mu}W^-) & \sim 4\cdot10^{-8}\ \%,
 \\
  {\rm BR}(k^{--}\to\ell^- \nu_{e}W^-) & \sim 9\cdot10^{-13}\ \% .
\end{align}
\end{subequations}
For our values of $m_k$ and $m_h$, the two-body $k^{--}\to h^- h^-$ decay is kinematically forbidden. However, the largeness of $\mu_{\not L}$ enhances the three-body, LN-violating $k^{--}\to\ell^- \nu_{\ell'}h^-$ decay to the per mil level. The hierarchy displayed by three-body decays $k^{--}\to\ell^- \nu_{\ell'}W^-$ reflects the fact the rates are proportional to charged lepton masses.
More specifically, the $k^{--}\to\ell^- \nu_{\ell'}W^-$ decay proceeds through the intermediate step $k^{--}\to \ell^- \ell'^{-*} \to \ell^- \nu_{\ell'}W^-$,
which is mediated by the coupling of $k^{--}$ to a RH lepton $\ell'^-_R$ and the coupling of $W^-$ to LH leptons. This implies that $\ell'^{-*}$ must propagate in its RH helicity state, and hence that the amplitude scales with its mass.
We caution that these rates are only illustrative.
They assume that all nonzero $g_{\ell\ell'}$ and $f_{\ell\ell'}$ are 
unity; realistic values must be more varied to satisfy oscillation and flavor data~\cite{Ohlsson:2009vk}.

\begin{table*}[t!]
\begin{center}
\resizebox{\textwidth}{!}{
\begin{tabular}{r c | c || c | c || c | c | c }
\hline\hline
\multirow{2}{*}{$m_k\quad$} & 
\multirow{2}{*}{$m_{h}$} & 
\multirow{2}{*}{$\mu_{\not L}$} & 
\multirow{2}{*}{$\Gamma^{\rm Tot.}_k$} &
\multirow{2}{*}{$\Gamma^{\rm Tot.}_h$} &
$\Gamma(k^{\mp\mp}\to \ell^\pm \ell'^\pm)$ & 
$\Gamma(k^{\mp\mp}\to h^\mp h^\mp)$  &
\multirow{2}{*}{$\Gamma(h^{\pm}\to \ell^\pm \nu_{\ell'})$} 
\\
        &   &   &   &   &
$\times(\delta_{\ell\ell'}+1)/\vert g_{\ell\ell'}\vert^2$ & 
$/ \vert f_{\ell\ell'}\vert^2$  &
\\ \hline
500\GeV     & 100\GeV & 1\TeV   & 252\GeV   &47.7\GeV   & 39.8\GeV\ (16\%) & 72.9\GeV\ (29\%) & 7.96\GeV\ (17\%)\\
1\TeV       & 100\GeV & 100\GeV & 358\GeV   &47.7\GeV   & 79.6\GeV\ (22\%) & 390\MeV\ (0.11\%)  & 7.96\GeV\ (17\%)\\
1.25\TeV    & 500\GeV & 100\GeV & 448\GeV   &239\GeV    & 99.5\GeV\ (22\%) & 191\MeV\ (0.04\%)  & 39.8\GeV\ (17\%)\\
3\TeV       & 1\TeV   & 100\GeV & 1.07\TeV  &477\GeV    & 239\GeV\ (22\%)  & 98.9\MeV\ (0.01\%) & 79.6\GeV\ (17\%) \\
\hline\hline
\end{tabular}
} 
\caption{
For masses $m_k$, $m_h$ and coupling $\mu_{\not L}$ (columns 1-3),
the total widths $\Gamma_k^{\rm Tot.}, \Gamma_h^{\rm Tot.}$ (column 4-5), assuming $g_{\ell\ell'}, f_{\ell\ell'} =1$, as well as the
normalized partial widths for 
$k^{\mp\mp}\to \ell^\pm\ell'^\pm$ (column 6),
$k^{\mp\mp}\to h^\mp h^\mp$ (column 7), and
$h^\mp \to \ell^\pm \nu_{\ell'}$ (column 8).
In parentheses are the branching rates.
}
\label{tab:widths}
\end{center}
\end{table*}

Assuming that $m_k< m_h$, the leading two- and three-body decay channels are
\begin{subequations}
\begin{align}
 h^\mp &\to \ell^\pm \nu_{\ell'}\ ,
 \\
 h^\mp &\to \ell^\pm \nu_{\ell'} Z/H^0/\gamma\ ,
 \\
 h^\mp &\to \nu_\ell \nu_{\ell'} W^\pm\ ,
 \\
 h^\mp &\to k^{\mp\mp}\ell^\mp \nu_{\ell'}\ .
\end{align}
\end{subequations}
In analogy to $k^{\mp\mp}$, the first channel proceeds through the antisymmetric Yukawa coupling $f_{\ell\ell'}$. The last three can be classified as being radiative corrections to the first channel and therefore are suppressed. The final proceeds through the LN-violating $k-h-h$ vertex.

The $h^{\pm}\to \ell^\pm \nu_{\ell'}$ partial width 
with charged lepton mass dependence is given by
\begin{align}
 \Gamma(h^{\pm}\to \ell^\pm \nu_{\ell'}) &= \frac{\vert f_{\ell\ell'}\vert^2}{4\pi}\ m_h\ (1-\tilde{r}_\ell)^2\ ,
 \quad\text{where}\quad
 \tilde{r}_\ell = \frac{m_\ell^2}{m_h^2}\ .
\end{align}
For the benchmark inputs listed in Eqs.~\eqref{eq:input_lepton} and \eqref{eq:benchmark_inputs}, one finds the following branching rates
\begin{align}
\label{eq:decay_rate_h}
{\rm BR}(h^- \to \ell^- \nu_{\ell'})\sim 16\%
\quad\text{and}\quad
{\rm BR}(h^- \to \nu_\ell \nu_{\ell'} W^-)\sim 0.08\% \ .
\end{align}
These channels proceed through LH chiral states and so little dependence on $m_\ell$ is observed.

For representative masses $m_k$, $m_h$ and coupling $\mu_{\not L}$ (columns 1-3), we summarized in Table~\ref{tab:widths} the total widths $\Gamma_k^{\rm Tot.}, \Gamma_h^{\rm Tot.}$ (column 4-5), assuming $g_{\ell\ell'}, f_{\ell\ell'} =1$.
We also summarize the (normalized) partial widths for the processes
$k^{\mp\mp}\to \ell^\pm\ell'^\pm$ (column 6),
$k^{\mp\mp}\to h^\mp h^\mp$ (column 7), and
$h^\mp \to \ell^\pm \nu_{\ell'}$ (column 8).
In parentheses are the corresponding branching rates assuming the benchmarks listed in Eq.~\eqref{eq:benchmark_inputs}.
For these inputs, we find that the total widths of $k$ and $h$ span approximately
$\Gamma_k^{\rm Tot.} \sim 250\GeV-1\TeV$
and 
$\Gamma_h^{\rm Tot.} \sim 47\GeV-475\TeV$.
These translate to characteristic lifetimes of $d = \tau c = \hbar c/\Gamma^{\rm Tot.} \sim 10^{-4}-10^{-3}$ fm.
Even for Yukawa couplings as small as $g_{\ell\ell'},f_{\ell\ell'}\sim 10^{-4} ~ (10^{-6})$, 
lifetimes would still be below 1 nm (on the order of microns).

We postpone further exploitation of correlations among $k$ and $h$ decays to Sec.~\ref{sec:typeii_decay}.


\subsection{Constraints from partial wave unitarity}\label{sec:pheno_unitarity}

\begin{figure}[!t]
\includegraphics[width=\textwidth]{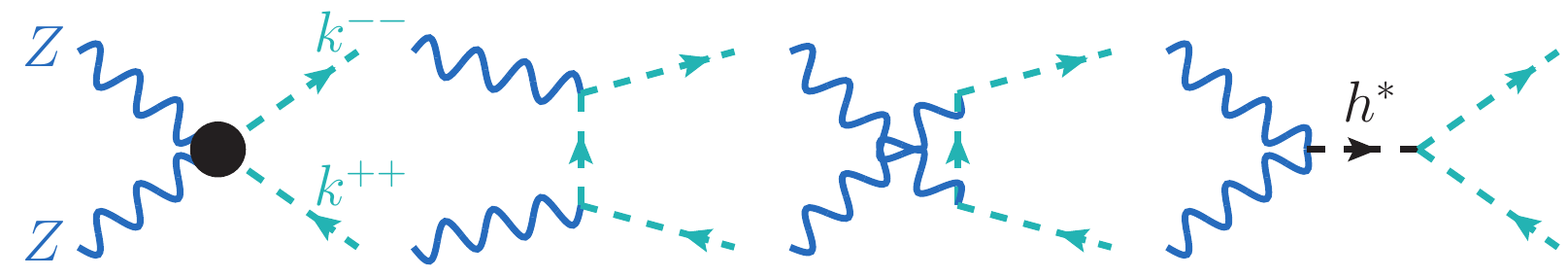} 
\caption{Born diagram for $ZZ\to k^{--}k^{++}$ in the Zee-Babu model. Graphs drawn with \texttt{JaxoDraw}~\cite{Binosi:2008ig}. }
\label{fig:diagram_ZeeBabu_zz_kk_scatt}
 \end{figure}

Before presenting predictions for the LHC, we consider constraints from  partial-wave unitarity. For some Seesaw scenarios, the $J=0$ partial wave in  vector boson scattering (VBS) is known to be interesting at high energies~\cite{Dicus:2004rg,Dicus:2005ku,Fuks:2020att,Fuks:2020zbm}. For the Zee-Babu model, we focus on the channels
\begin{subequations}
\begin{align}
 Z Z\ &\to\ k^{--}k^{++},\ h^- h^+,
 \\
 W^+ W^-\ &\to\ k^{--}k^{++},\ h^- h^+.
\end{align}
\end{subequations}
As depicted in Fig.~\ref{fig:diagram_ZeeBabu_zz_kk_scatt}, 
the Born-level matrix element for $ZZ\to k^{++}k^{--}$ is facilitated by 
a 4-point term $(\mathcal{M}_{4})$,
as well as 
$t$- $(\mathcal{M}_{t})$,
$u$- $(\mathcal{M}_{u})$,
and $s$-channel $(\mathcal{M}_{H})$ terms.
The first thee diagrams are determined entirely by EW couplings while the last is set by the $H^0-H^0-k-k$ coupling $\lambda_{kZBh}$. For the $h^+h^-$ channel, the final diagram is set by the $H^0-H^0-h-h$ coupling $\lambda_{hZBh}$.

Similarly, the $W^+W^-$ channels are mediated each by the three diagrams (not shown) with $s$-channel  $\gamma/Z/H^0$ appearing as intermediaries. The $\gamma/Z$ diagrams are determined entirely by EW couplings while the third is set by $\lambda_{kZBh}$ and $\lambda_{hZBh}$ for $k^{--}k^{++}$ and $h^-h^+$, respectively.

Notably, the helicity amplitude for $Z_0 Z_0 \to k^{--} k^{++}$, where $Z_0$ is a longitudinally polarized $Z$ boson, undergoes strong cancellations. Without any approximations, one finds the scaling
\begin{align}
\left[\mathcal{M}_4+\mathcal{M}_t+\mathcal{M}_u\right]_{(\lambda_{Z_A},\lambda_{Z_B})=(0,0)}\
\sim\ 
M_Z^2\ \times\ \frac{(\text{powers of }s,\ m_k^2,\ \text{and}\ M_Z^2)}{(\text{higher powers of }s,\ m_k^2,\ \text{and}\ M_Z^2)}.
\end{align}
This means that in the high-energy limit, where $s\gg M_Z^2, m_k^2$, the pure gauge contribution vanishes. Taking the $(M_Z^2/s), (m_k^2/s)\to0$ limit before summing diagrams leads to separately divergent terms. Some of this behavior can be attributed to the structure of longitudinal polarization vectors, which scale like $\varepsilon^\mu(q,\lambda=0)\sim q^\mu/M_Z + \mathcal{O}(M_Z/q^0)$.  Intuitively, scalars in the Zee-Babu model carry only hypercharge, not weak isospin. Therefore, in the unbroken phase, they should decouple from the weak sector. 
The same behavior is found in the other channels.

Summing over all diagrams, followed by taking the high-energy limit
\begin{align}
 (M_V^2/s),\ (m_H^2/s)\ (m_k^2/s),\ (m_h^2/s) \to\ 0\ ,
\end{align}
leads to the simple expressions:
\begin{subequations}
\label{eq:partial_wave_me}
 \begin{align}
 -i\mathcal{M}(Z_0 Z_0\ \to\ k^{--}k^{++}) \quad     &= \quad
  -i\mathcal{M}(W^+_0 W^-_0\ \to\ k^{--}k^{++})  &= -i \lambda_{kH}\ ,
  \\
 -i\mathcal{M}(Z_0 Z_0\ \to\ h^{-}h^{+}) \quad       &= \quad
 -i\mathcal{M}(W^+_0 W^-_0\ \to\ h^{-}h^{+})    &= -i \lambda_{hH}\ .
\end{align}
\end{subequations}
For completeness, the partonic cross sections in this kinematic limit simplify to the expression
\begin{align}
 \hat{\sigma}(V_0 V_0^\dagger \to S S^\dagger) = \frac{1}{16\pi}\frac{1}{M_{VV}^2} \vert \{ \lambda \} \vert^2 \ , 
\end{align}
where $M_{VV}$ is the invariant mass of the $(V_0V_0)$ system, and $\{ \lambda \}$ is either $\lambda_{kZBh}$ or $\lambda_{ZBh}$.
(No spin averaging is needed as $V_0V_0$ are polarized.)
Following the procedure of Ref.~\cite{BuarqueFranzosi:2019boy}, we checked that our implementation of the Zee-Babu model (see Sec.~\ref{sec:setup_mc}) reproduces this cross section.

The $J=0$ partial-wave amplitudes are obtained from Eq.~\eqref{eq:partial_wave_me} using 
\begin{align}
 a_{J=0} &= \frac{1}{32\pi}\ \int_{-1}^{1}d\cos\theta_k\ \mathcal{M}(V_0 V_0\ \to\ S S^\dagger) \ ,
\end{align}
where $V_0\in\{W_0,Z_0\}$ and $S\in\{k,h\}$.
This results in the following partial-wave amplitudes:
\begin{subequations}
\begin{align}
 a_{J=0}(Z_0Z_0\to k^{--}k^{++})\quad &=\quad 
 a_{J=0}(W^+_0W^-_0\to k^{--}k^{++}) &= \frac{\lambda_{kH}}{16\pi}\ ,
 \\
 a_{J=0}(Z_0Z_0\to h^{-}h^{+})\quad &=\quad 
 a_{J=0}(W^+_0W^-_0\to h^{-}h^{+}) &= \frac{\lambda_{hH}}{16\pi}\ .
\end{align}
\end{subequations}
The perturbative condition of $\vert a_{J}\vert < 1/2$ constrains the $H^0-H^0-S-S$ couplings to be
\begin{align}
 \lambda_{kH},\ \lambda_{hH} < 8\pi.
\end{align}
While these bounds are relatively weak, more aggressive restrictions on $\vert a_{J}\vert$ translate into more aggressive limits on $ \lambda_{kZBh}, \lambda_{hZBh}$. Further considerations from VBS is left to future work.


\subsection[kk and hh pairs at the LHC]{$k^{++}k^{--}$ and $h^+h^-$ pairs at the LHC}\label{sec:pheno_lhc}

\begin{figure}[!t]
\includegraphics[width=\textwidth]{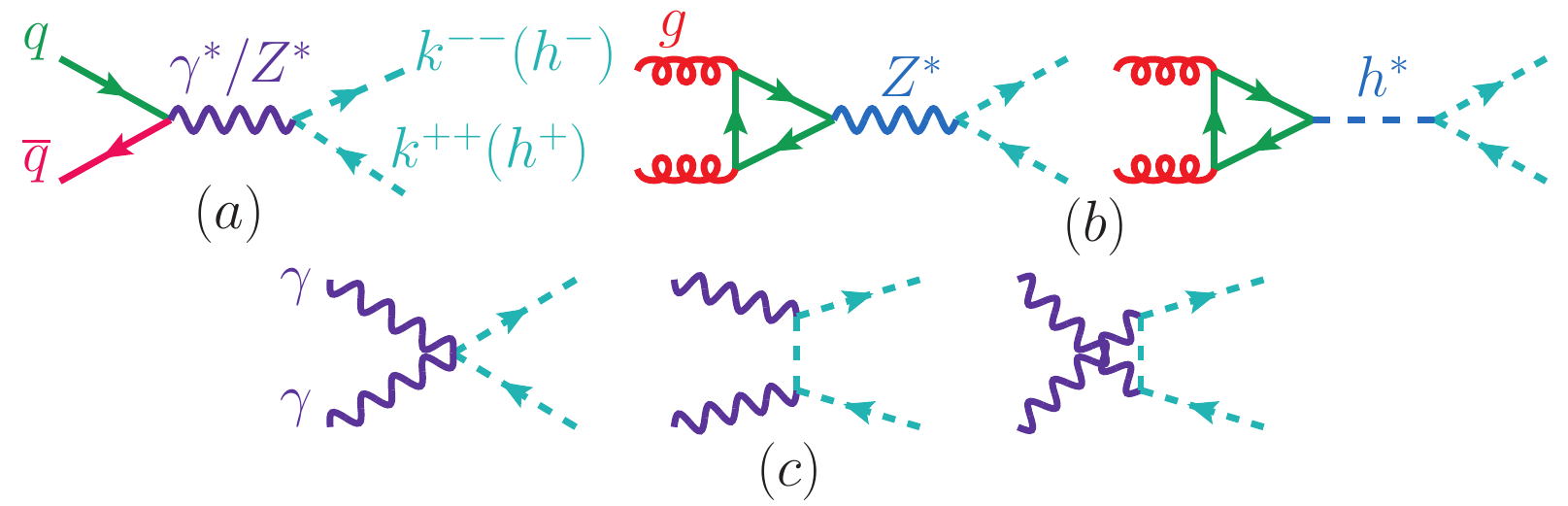} 
\caption{
Born-level partonic graphs depicting the production of $k^{--}k^{++}$ pairs (or $h^-h^+$ pairs) in the Zee-Babu model from the (a) Drell-Yan process, (b) gluon fusion, (c) photon fusion. 
}
\label{fig:diagram_ZeeBabu_MultiProd_LHC}
\end{figure}

\begin{figure}
\begin{center}
\subfigure[]{\includegraphics[width=0.485\textwidth]{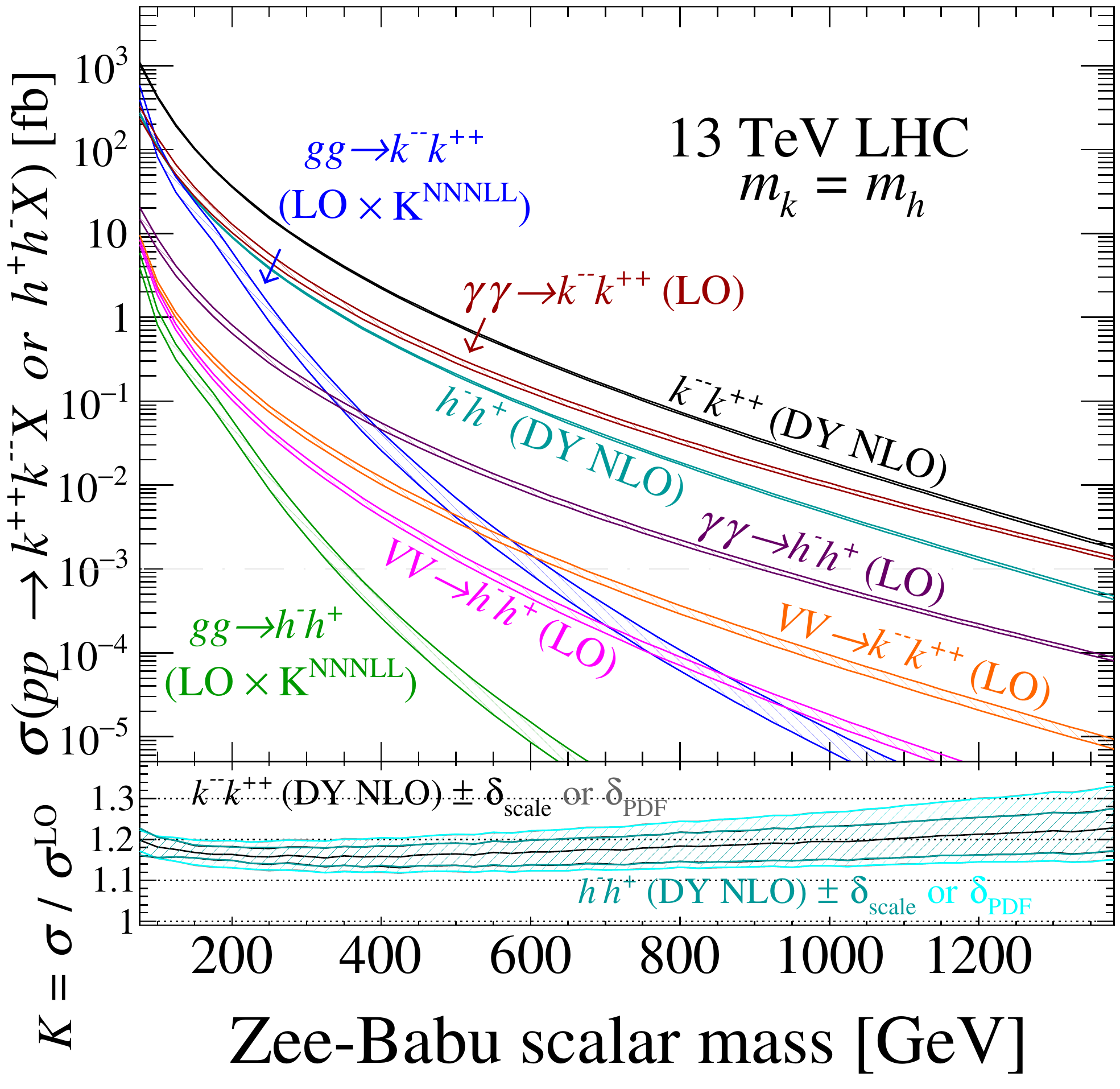}		\label{fig:zeeBabu_XSec_vs_Mass_LHCX13}}
\subfigure[]{\includegraphics[width=0.485\textwidth]{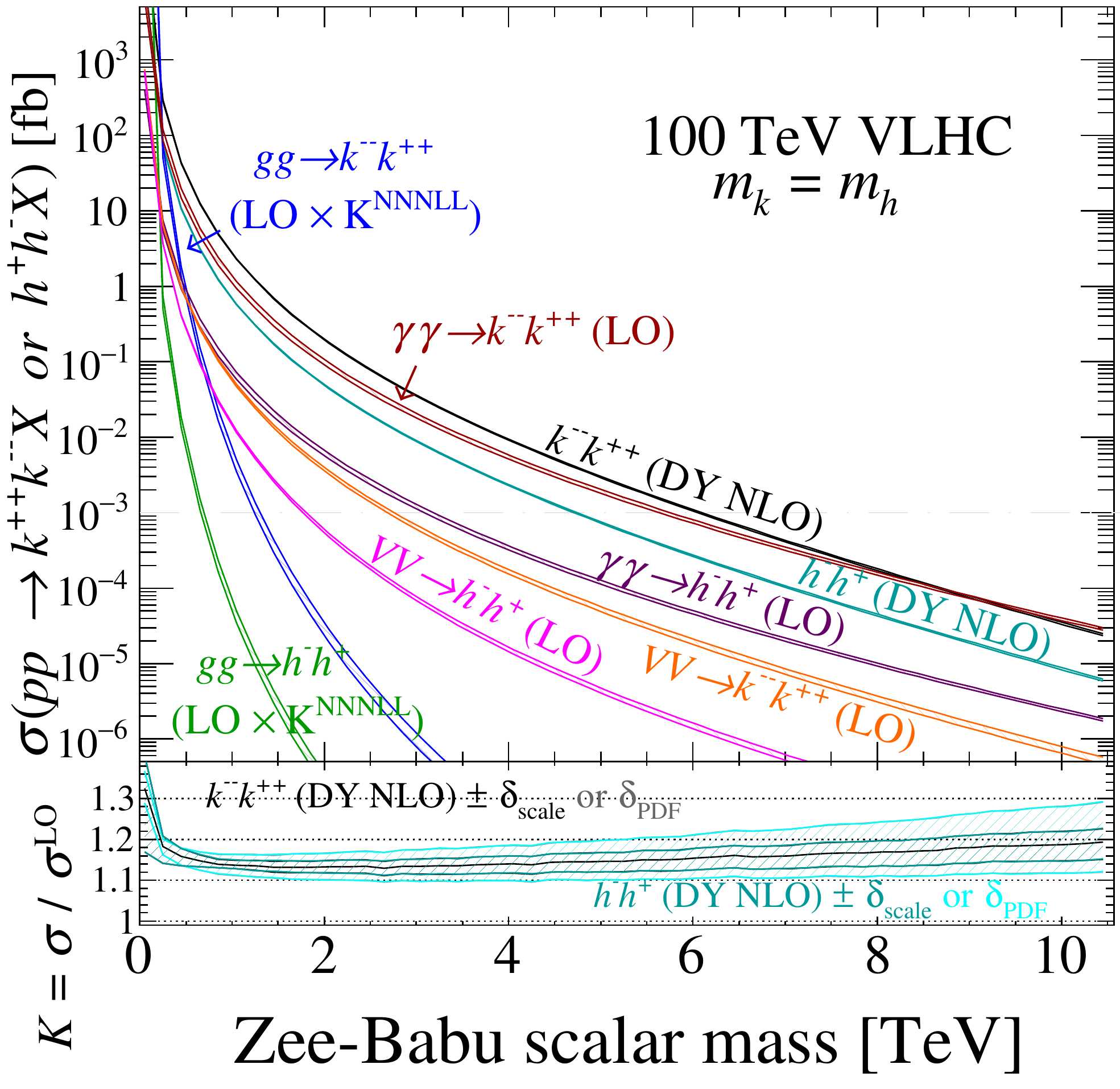}		\label{fig:zeeBabu_XSec_vs_Mass_LHC100}}
\end{center}
\caption{
(a) Upper panel: As a function of scalar mass, production-level cross sections at residual scale uncertainties at the $\sqrt{s}=13\TeV$ LHC of $k^{--}k^{++}$ and $h^-h^+$ pairs, as predicted by the Zee-Babu model, via the Drell-Yan mechanism (DY) at NLO in QCD, $\gamma\gamma$ fusion (AF) at LO, loop-induced $gg$ fusion (GF) channel scaled by an N$^3$LL $K$-factor~\cite{Fuks:2019clu}, and EW vector boson scattering $(VV)$ at LO with phase space cuts.
Lower panel: the ratio of the NLO and LO DY cross section with scale uncertainty (inner band) and PDF uncertainty (outer band).
(b) Same as (a) but for $\sqrt{s}=100\TeV$.
Cross sections 
assume the SM inputs of Sec.~\ref{sec:setup_sm}
as well as the Zee-Babu inputs 
of Eqs.~\eqref{eq:benchmark_inputs} and \eqref{eq:benchmark_inputs_update}.
}
\label{fig:zeeBabu_XSec_vs_Mass}
\end{figure}

We now turn to the production Zee-Babu scalars at the $\sqrt{s}=13\TeV$ LHC. We focus on  $k^{--}k^{++}$ and $h^{-}h^{+}$ pair production through a variety of processes that are depicted at the Born level in Fig.~\ref{fig:diagram_ZeeBabu_MultiProd_LHC}.
As an outlook, we also consider a hypothetical Very Large Hadron Collider (VLHC) at $\sqrt{s}=100\TeV$.
Our results are summarized in Fig.~\ref{fig:zeeBabu_XSec_vs_Mass}, where we plot as a function of scalar mass $(m_k=m_h)$ the total inclusive cross section $(\sigma)$ for these processes. 
For some channels these are the first LHC predictions that have been made in the context of the Zee-Babu model.

In high-$p_T$ hadron collisions, 
the inclusive production rate of final-state $\mathcal{F}$
is given by~\cite{Collins:1984kg,Collins:1985ue,Collins:2011zzd}
\begin{align}
\label{eq:factorTheorem}
 \sigma(pp \to \mathcal{F} + \text{anything}) &= 
 \sum_{i,j\in\{q,\overline{q},g,
 \gamma\}} \ 
 f_{i/p} \ \otimes \ 
 f_{j/p} \ \otimes \ 
 \Delta_{ij} \ \otimes \ 
 \hat{\sigma}_{ij\to\mathcal{F}}\ ,
\end{align}
where $\hat{\sigma}_{ij\to\mathcal{F}}$ is the partonic   
$ij\to \mathcal{F}$ scattering cross section as obtained from  the  formula
\begin{align}
\label{eq:partonxsec}
 \hat{\sigma}_{ij\to\mathcal{F}} = \int dPS_{n_f}\ \frac{d\hat{\sigma}_{ij\to\mathcal{F}}}{dPS_{n_f}},
 \qquad
\frac{d\hat{\sigma}_{ij\to\mathcal{F}}}{dPS_{n_f}}  
 = \frac{1}{2Q^2}\frac{1}{\mathcal{S}_i\mathcal{S}_j}\frac{1}{N_c^i N_c^j} \sum_{\rm dof}\vert\mathcal{M}_{ij\to\mathcal{F}}\vert^2 \ . 
\end{align}
Here, $\mathcal{M}_{ij\to\mathcal{F}}$ is the partonic matrix element calculable using perturbative methods and the Feynman rules of Sec.~\ref{sec:theory_model}; $N_c$ and $\mathcal{S}$ are, respectively, the color and spin multiplicities of $i$ and $j$; the summation is over all discrete dof / multiplicities;
and $Q^2=(p_i+p_j)^2 > M^2(\mathcal{F})$ is the (squared) hard scattering scale. $Q$ must exceed the invariant mass of $\mathcal{F}$ in order for the process to proceed.  The phase space volume element for an $n_f$-body final state is defined in Eq.~\eqref{eq:def_phase_space}.

In Eq.~\eqref{eq:factorTheorem}, the $f$ are the collinear PDFs that represent the likelihood of finding partons $i, j\in\{q,\overline{q},g,\gamma\}$, for light quark species $q$, in proton $p$ carrying particular longitudinal momentum fractions. $\Delta_{ij}$ describes the likelihood of soft radiation emitted in the $ij\to\mathcal{F}$ scattering process.
The symbol $\otimes$ denotes the convolution of these probabilities. 
Cross sections throughout this section 
assume the SM inputs of Sec.~\ref{sec:setup_sm}
and the Zee-Babu inputs 
of Eq.~\eqref{eq:benchmark_inputs}.
For select (gluon fusion) computations, the Zee-Babu inputs of Eq.~\eqref{eq:benchmark_inputs_update} are used and will be discussed below.

\paragraph*{Drell-Yan:} 
We begin with pair production via the Drell-Yan (DY) mechanism, i.e., quark-antiquark annihilation, which at the Born level is depicted in Fig.~\ref{fig:diagram_ZeeBabu_MultiProd_LHC}(a)
and given by
 \begin{align}
  q\overline{q}\ \to\ \gamma^* / Z^*\ \to\ k^{--}k^{++}
  \quad\text{or}\quad h^- h^+ \quad \text{at}\quad \mathcal{O}(\alpha^2)\ .
 \end{align}
To simulate this at NLO with the \libName libraries and \texttt{mg5amc}, we use the commands\footnote{See also Ref.~\cite{Alwall:2014hca} for instructions on operating \texttt{mg5amc}.}:
 \begin{verbatim}
set acknowledged_v3.1_syntax true
import model SM_ZeeBabu_NLO
define kk = k++ k--
define hh = h+  h-
define qq = u c d s b u~ c~ d~ s~ b~
generate qq qq > kk kk QED=2 QCD=0 [QCD]
output DirName1
generate qq qq > hh hh QED=2 QCD=0 [QCD]
output DirName2
 \end{verbatim}
 
For both $k^{--}k^{++}$ (black) and $h^-h^+$ (teal) production, we show in Fig.~\ref{fig:zeeBabu_XSec_vs_Mass_LHCX13} the inclusive production cross sections for $\sqrt{s}=13\TeV$  at NLO in QCD and with residual scale uncertainties (band thickness).
While predictions at NLO in QCD for Type II scalars have been available for some time~\cite{Muhlleitner:2003me,Fuks:2019clu}, this is the first for the Zee-Babu model.
Over the mass range $m_k,m_h = 50\GeV-1400\GeV$, we find that the cross sections for the two processes span approximately
\begin{subequations}
\begin{align}
 \sigma_{13\TeV}^{\rm DY~(NLO)}(k^{--}k^{++}) &\approx 
 6\pb-1.6\ab\ ,
 \\
 \sigma_{13\TeV}^{\rm DY~(NLO)}(h^{-}h^{+}) &\approx 
 1.5\pb-0.4\ab\ ,
\end{align}
\end{subequations}
with residual scale uncertainties spanning about $\delta \sigma_{13\TeV}^{\rm DY~(NLO)}\approx\pm2\%-\pm5\%$ for both channels.

An interesting observation is that ratio of the two DY rates is constant and equal to four.
This can be attributed to the couplings of $k$ and $h$ to $\gamma$ and $Z$.
As shown in the Lagrangian of Eq.~\eqref{eq:lag_kin_ewsb}, the three-point $S-S^\dagger-V$ vertices  are proportional to the electric charge of the scalar but are otherwise the same for $k$ and $h$. Hence, the rates differ by the
square of the charges:
\begin{equation}
\frac{\sigma_{13\TeV}^{\rm DY~(NLO)}(k^{--}k^{++})}{\sigma_{13\TeV}^{\rm DY~(NLO)}(h^-h^+)} = \left(\frac{Q_k}{Q_h}\right)^2 = 4.
\end{equation}
Further investigation of the matrix element shows that in the large $m_k, m_h$ limit, i.e., where $\mathcal{O}(M_Z^2/Q^2)$ terms can be neglected, a strong destructive cancellation occurs between the $\gamma$ and $Z$ diagrams. For both the $u\overline{u}$ and $d\overline{d}$ parton channels, the $\gamma-Z$ interference term is negative and    slightly larger in magnitude than the $Z$ channel. (In the $d\overline{d}$ channel, the $\gamma$, $Z$, and interference terms are actually all comparable in size.) In essence, the DY channel is driven by the $\gamma$ diagram.

To quantify the size of QCD corrections, we show the NLO in QCD $K$-factor $(K^{\rm NLO})$
\begin{align}
 K^{\rm NLO} &= \sigma^{\rm NLO}\ /\ \sigma^{\rm LO}\ .
\end{align}
in the lower panel of Fig.~\ref{fig:diagram_ZeeBabu_MultiProd_LHC}(a) as a function of mass. 
For both DY channels, $K^{\rm NLO}$ spans roughly
\begin{align}
 K^{\rm NLO} \approx 1.15 - 1.22\ .
\end{align}
These numbers are comparable to exotically charged scalar  production in the Type II Seesaw~\cite{Muhlleitner:2003me,Fuks:2019clu} when adjusted for PDF and scale choices. In the lower panel, we also show the residual scale uncertainty at NLO (darker inner bands) and PDF uncertainty (lighter outer bands). PDF uncertainties roughly span $\delta \sigma_{13\TeV}^{\rm DY~(NLO)}\approx\pm2\%-\pm8\%$. 
The curves and bands for $k^{--}k^{++}$ and $h^-h^+$ pair production overlap almost perfectly. This follows from the DY rates for $k$ and $h$ differing by a constant at both LO and NLO, which cancel when taking the respective ratios.

\paragraph*{Photon Fusion:}
Next we consider inclusive  pair production from $\gamma\gamma$ fusion (AF), given by
 \begin{align}
  \gamma\gamma\ \to\ k^{--}k^{++}
  \quad\text{or}\quad h^- h^+ \quad \text{at}\quad \mathcal{O}(\alpha^2)\ ,
 \end{align}
 and shown diagrammatically in Fig.~\ref{fig:diagram_ZeeBabu_MultiProd_LHC}(c).
This process can be simulated at LO  using the syntax
 \begin{verbatim}
generate a a > kk kk QED=2 QCD=0
output DirName3
generate a a > hh hh QED=2 QCD=0
output DirName4
 \end{verbatim}
 Over the mass range investigated, the cross sections for the two processes approximately span
\begin{subequations}
\begin{align}
 \sigma_{13\TeV}^{\rm AF~(LO)}(k^{--}k^{++}) &\approx 
 1\pb-1.2\ab\ ,
 \\
 \sigma_{13\TeV}^{\rm AF~(LO)}(h^{-}h^{+}) &\approx 
 60\fb-75\zb\ ,
\end{align}
\end{subequations}
with scale uncertainties reaching $\delta \sigma_{13\TeV}^{\rm AF~(LO)}\approx\pm20\%~(\pm5\%)$ at low (high) masses for both channels.
These moderate scale uncertainties are QED uncertainties.
They are common to $\gamma$-induced processes at LO (when the photon is inelastic)
and can generically~\cite{Alva:2014gxa,Degrande:2016aje} be attributed to 
logarithmic and 
power-law terms in real-radiation matrix elements at NLO in QED, i.e., logarithmic and
power-law terms in the tree-level $q\gamma \to q SS^\dagger$ matrix element.
More specifically, the corrections 
are associated with tree-level $q\to q\gamma$ splittings
and, after phase space integration, have the forms 
$\mathcal{O}\left(\log(M_{\gamma\gamma}/p_T^\gamma)\right)$ and
$\mathcal{O}(p_T^\gamma/M_{\gamma\gamma})$.
Like in QCD, real radiation diagrams can alternatively be described by employing so-called multi-leg matching (MLM) techniques.
These would systematically combine, for example, the matrix elements for $\gamma\gamma\to k^{++}k^{--}$ at $\mathcal{O}(\alpha^2)$ and $q\gamma\to q k^{++}k^{--}$ at $\mathcal{O}(\alpha^3)$, capture corrections through $\mathcal{O}(\alpha^3)$, but remain LO accurate.

PDF uncertainties are about  $\delta \sigma_{13\TeV}^{\rm AF~(LO)}\approx\pm2\%$ over the mass range. 

 As in the DY case,  
 the ratio of the two AF rates are  proportional to ratio of electric charges:
\begin{equation}
\frac{\sigma_{13\TeV}^{\rm AF~(LO)}(k^{--}k^{++})}{\sigma_{13\TeV}^{\rm AF~(LO)}(h^-h^+)} = \left(\frac{Q_k}{Q_h}\right)^4 = 16.
\end{equation}
This follows from the fact that the three point $S-S^\dagger-\gamma$ vertex and the four point $S-S^\dagger-\gamma-\gamma$ vertex are each proportional to the charge of $k$ and $h$.
Due to (a) the destructive interference between the $\gamma$ and $Z$ contributions in the DY channel, (b) the charge enhancement in the AF channel, and (c) the fact that the $\gamma\gamma$ luminosity is sourced from valance quark scattering whereas the DY process is sourced by valance-sea scattering, the $\gamma\gamma\to k^{--}k^{++}$ cross section consistently sits between the $k^{--}k^{++}$ and $h^-h^+$ DY rates. The $\gamma\gamma\to h^-h^+$ channel sits below all three curves. This suggests that a second $k^{--}k^{++}$ could be seen at the LHC before the $h^-h^+$ channel.

We caution that the similarity of the DY and AF rates for $k^{--}k^{++}$ production in the Zee-Babu model is mostly a consequence of the suppressed DY rate, not the ``largeness'' of the photon PDF.
This is an uncommon occurrence but also not an artifact of the photon PDF.
As discussed in Sec.~\ref{sec:typeii}, doubly charged scalars in other models carry different gauge quantum numbers; this often leads to larger DY rates in those models~\cite{Gunion:1996pq}.
Claims that the cross sections for photon-induced processes readily exceed DY rates can usually be attributed to mis-modeling of photon PDFs or discounting large uncertainties. For dedicated discussions, see Refs.~\cite{Han:2007bk,Ball:2013hta,Alva:2014gxa,Fuks:2019clu}.

 \begin{table*}[t!]
\begin{center}
\resizebox{\textwidth}{!}{
\begin{tabular}{c c | c | r r c || r r c}
\hline\hline
\multicolumn{3}{c}{} & \multicolumn{3}{c}{$\sqrt{s} = 13\TeV$ LHC} & \multicolumn{3}{c}{$\sqrt{s} = 100\TeV$ LHC}
\\
\multicolumn{2}{c|}{Process} & mass [GeV] & 
$\sigma^{\rm LO}$ [fb] & $\sigma^{\rm NLO}$ [fb] & $K$ & 
$\sigma^{\rm LO}$ [fb] & $\sigma^{\rm NLO}$ [fb] & $K$
\\ 
\hline\hline
\multirow{5}{*}{$k^{--}k^{++}$}	 & DY	 & \multirow{5}{*}{450} 	& $1.15\cdot10^0\ ^{+7\%}_{-6\%}	\ ^{+4\%}_{-3\%}$ 	& $1.34\ ^{+2\%}_{-2\%}	\ ^{+4\%}_{-4\%}$	& $1.16$
& $37.1\cdot10^0\ ^{+3\%}_{-4\%}	\ ^{+2\%}_{-2\%}$ 	& $43.0\cdot10^0\ ^{+2\%}_{-2\%}	\ ^{+2\%}_{-2\%}$	& $1.16$
\\
& AF	 &  	& $492\cdot10^{-3}\ ^{+9\%}_{-9\%}	\ ^{+2\%}_{-2\%}$ &	&
& $17.0\cdot10^0\ ^{+15\%}_{-14\%}	\ ^{+1\%}_{-1\%}$ &	&
\\
& GF	 &  	& $4.16\cdot10^{-3}\ ^{+31\%}_{-22\%}	\ ^{+5\%}_{-5\%}$ &	& $3.10$
& $599\cdot10^{-3}\ ^{+17\%}_{-14\%}	\ ^{+1\%}_{-1\%}$ &	& $2.60$
\\
& VBF	 &  	& $6.53\cdot10^{-3}\ ^{+11\%}_{-9\%}	\ ^{+2\%}_{-2\%}$ &	&
& $972\cdot10^{-3}\ ^{+2\%}_{-2\%}	\ ^{+1\%}_{-1\%}$ &	&
\\
\hline
\multirow{5}{*}{$h^{-}h^{+}$}   & DY	 & \multirow{5}{*}{450} 	& $288\cdot10^{-3}\ ^{+7\%}_{-6\%}	\ ^{+4\%}_{-4\%}$ 	& $334\cdot10^{-3}\ ^{+2\%}_{-2\%}	\ ^{+4\%}_{-4\%}$	& $1.16$
& $9.28\cdot10^0\ ^{+3\%}_{-4\%}	\ ^{+2\%}_{-2\%}$ 	& $10.8\cdot10^0\ ^{+2\%}_{-2\%}	\ ^{+2\%}_{-2\%}$	& $1.16$
\\
& AF	 &  	& $30.8\cdot10^{-3}\ ^{+9\%}_{-9\%}	\ ^{+2\%}_{-2\%}$ &	&
& $1.06\cdot10^0\ ^{+15\%}_{-14\%}	\ ^{+1\%}_{-1\%}$ &	&
\\
& GF	 &  	& $41.6\cdot10^{-6}\ ^{+31\%}_{-22\%}	\ ^{+5\%}_{-5\%}$ &	& $3.10$
& $5.99\cdot10^{-3}\ ^{+17\%}_{-14\%}	\ ^{+1\%}_{-1\%}$ &	& $2.60$
\\
& VBF	 &  	& $2.50\cdot10^{-3}\ ^{+11\%}_{-9\%}	\ ^{+2\%}_{-3\%}$ &	&
& $411\cdot10^{-3}\ ^{+2\%}_{-2\%}	\ ^{+1\%}_{-1\%}$ &	&
\\
\hline
\multirow{5}{*}{$k^{--}k^{++}$}	 & DY	 & \multirow{5}{*}{1250} 	& $3.24\cdot10^{-3}\ ^{+13\%}_{-11\%}	\ ^{+7\%}_{-6\%}$ 	& $3.94\cdot10^{-3}\ ^{+3\%}_{-4\%}	\ ^{+7\%}_{-6\%}$	& $1.22$
& $1.07\cdot10^0\ ^{+1\%}_{-1\%}	\ ^{+2\%}_{-2\%}$ 	& $1.21\cdot10^0\ ^{+1\%}_{-1\%}	\ ^{+2\%}_{-2\%}$	& $1.13$
\\
& AF	 &  	& $2.59\cdot10^{-3}\ ^{+5\%}_{-5\%}	\ ^{+2\%}_{-2\%}$ &	&
& $564\cdot10^{-3}\ ^{+10\%}_{-10\%}	\ ^{+2\%}_{-2\%}$ &	&
\\
& GF	 &  	& $271\cdot10^{-9}\ ^{+38\%}_{-26\%}	\ ^{+18\%}_{-15\%}$ &	& $3.10$
& $433\cdot10^{-6}\ ^{+21\%}_{-16\%}	\ ^{+2\%}_{-2\%}$ &	& $2.60$
\\
& VBF	 &  	& $17.5\cdot10^{-6}\ ^{+15\%}_{-12\%}	\ ^{+3\%}_{-3\%}$ &	&
& $24.8\cdot10^{-3}\ ^{+4\%}_{-4\%}	\ ^{+2\%}_{-2\%}$ &	&
\\
\hline
\multirow{5}{*}{$h^{-}h^{+}$}	 & DY	 & \multirow{5}{*}{1250} 	& $811\cdot10^{-6}\ ^{+13\%}_{-11\%}	\ ^{+7\%}_{-6\%}$ 	& $986\cdot10^{-6}\ ^{+3\%}_{-4\%}	\ ^{+8\%}_{-6\%}$	& $1.22$
& $267\cdot10^{-3}\ ^{+1\%}_{-1\%}	\ ^{+2\%}_{-2\%}$ 	& $303\cdot10^{-3}\ ^{+1\%}_{-1\%}	\ ^{+2\%}_{-2\%}$	& $1.13$
\\
& AF	 &  	& $162\cdot10^{-6}\ ^{+5\%}_{-5\%}	\ ^{+2\%}_{-2\%}$ &	&
& $35.2\cdot10^{-3}\ ^{+10\%}_{-10\%}	\ ^{+2\%}_{-2\%}$ &	&
\\
& GF	 &  	& $2.71\cdot10^{-9}\ ^{+38\%}_{-26\%}	\ ^{+18\%}_{-15\%}$ &	& $3.10$
& $4.33\cdot10^{-6}\ ^{+21\%}_{-16\%}	\ ^{+2\%}_{-2\%}$ &	& $2.60$
\\
& VBF	 &  	& $2.68\cdot10^{-6}\ ^{+15\%}_{-12\%}	\ ^{+4\%}_{-4\%}$ &	&
& $5.35\cdot10^{-3}\ ^{+4\%}_{-4\%}	\ ^{+2\%}_{-1\%}$ &	&
 \\
\hline\hline
\end{tabular}
} 
\caption{
For representative masses $m_k,\ m_h$ [GeV] (column 2), 
the predicted cross sections [fb] at LO and/or NLO in QCD for inclusive $pp\to k^{--}k^{++}+X$ (rows 1 and 3) and $pp\to h^-h^++X$ (rows 2 and 4) production at $\sqrt{s}=13\TeV$ (column 3) and 100 TeV (column 4), for the Drell-Yan process (DY), photon fusion (AF), gluon fusion (GF), and EW vector boson scattering (VBS).
Also shown are scale uncertainties [\%], PDF uncertainties [\%], and QCD $K$-factor (if present).
Cross sections assume the SM inputs of Sec.~\ref{sec:setup_sm}
as well as the Zee-Babu inputs 
of Eqs.~\eqref{eq:benchmark_inputs} and \eqref{eq:benchmark_inputs_update}.
}
\label{tab:xsec}
\end{center}
\end{table*}

\paragraph*{Gluon Fusion:}
We now consider $k$ and $h$ pair production from gluon fusion (GF):
\begin{align}
\label{eq:proc_gf}
  gg\ \to\ H^{0*} / Z^*\ \to\ k^{--}k^{++}
  \quad\text{or}\quad h^- h^+ \quad \text{at}\quad \mathcal{O}(\alpha_s^2 y_t \lambda,\ \alpha_s^2\alpha \lambda)\ .
 \end{align}
 This loop-induced process is mediated by
 $s$-channel $H^0$ and $Z$ bosons, as depicted in Fig.~\ref{fig:diagram_ZeeBabu_MultiProd_LHC}(b).
With \libName\texttt{\_NLO}, we simulate the channel at LO, i.e., at one loop in $\alpha_s$, in \texttt{mg5amc} using 
 \begin{verbatim}
generate g g > kk kk QED=2 QCD=2 [noborn=QCD]
output DirName5
generate g g > hh hh QED=2 QCD=2 [noborn=QCD]
output DirName6
 \end{verbatim}
 
 We first note that the $Z$ contribution vanishes due to two mechanisms: (a) Angular momentum conservation in the $gg\to Z^*$  sub-graph causes the transverse component of the $Z$'s propagator $\Pi_{\rho\sigma}^Z(q)$, i.e., the $g_{\rho\sigma}$ term, to vanish when contracted with the quark loops. (b) The longitudinal component of the $Z$'s propagator, i.e., the $q_\rho q_\sigma/M_Z^2$ term in the Unitary gauge, 
 vanishes when contracted with the $Z-S-S^\dagger$ vertex $\Gamma^{\sigma}_{Z-S-S}$. Hence for $q=p_S+p_{S^\dagger}$, one has
 \begin{align}
 \Pi_{\rho\sigma}^Z(q)\ \Gamma^{\sigma}_{Z-S-S} &\sim q_\rho q_\sigma (p_S^\sigma - p_{S^\dagger}^\sigma)
  = q_\rho (p_S+p_{S^\dagger})\cdot (p_S-p_{S^\dagger}) 
= q_\rho (m_S^2 - m_S^2) =0\ .
 \end{align}
Ultimately, this can attributed to $Z^*$ behaving as if it is a pseudoscalar in the $gg\to Z^*\to X$ process~\cite{Ruiz:2017yyf}, which is at odds with the parity conserving nature of the $S-S^\dagger-Z$ coupling.

A second comment is that the $S-S-H^0$ vertices differ only by a normalization. More specifically, from the Lagrangian in Eq.~\eqref{eq:lag_scalar}, the $\Gamma_{H^0-k-k}$ and $\Gamma_{H^0-h-h}$ vertices are
\begin{subequations}
 \begin{align}
  \Gamma_{H^0-k-k} = -i\ v\ \lambda_{kH}\ ,
  \quad\text{and}\quad
  \Gamma_{H^0-h-h} = -i\ v\ \lambda_{hH}\ .
 \end{align}
\end{subequations}
The inputs of Eq.~\eqref{eq:benchmark_inputs} result in identical cross sections. To make things interesting, we set 
\begin{align}
\label{eq:benchmark_inputs_update}
 \lambda_{kH} = 1.0 \quad\text{and}\quad \lambda_{hH} = 0.1\ .
\end{align}
For all other channels in this section, we keep the $\{\lambda\}$ couplings as specified in Eq.~\eqref{eq:benchmark_inputs}.

A third comment is that the QCD corrections to GF are typically large. This follows from both positive, virtual corrections and the opening of partonic channels in real corrections.
To account for these, we apply a $K$-factor derived for the next-to-next-to-next-to-leading logarithmic threshold corrections (N$^3$LL(thresh.)) to exotic scalar production in the Type II Seesaw~\cite{Fuks:2019clu}
\begin{align}
K^{\rm N^3LL} &= \sigma^{\rm N^3LL(thresh.)}\ /\ \sigma^{\rm LO} = 3.10\ . 
\end{align}
This scale factor captures the leading contributions to the GF channel at NNLO~\cite{Ahrens:2009cxz}.
Using this $K$-factor is justified by the following:
The processes in Eq.~\eqref{eq:proc_gf} and the analogous processes in the Type II Seesaw contain the same sub-graphs that are susceptible to QCD corrections;
that is,
the $gg\to H^*/Z^*$ sub-graphs are the same for all four processes.
Therefore, the QCD corrections to the production cross sections are the same.
Moreover, in Ref.~\cite{Fuks:2019clu}, the $K$-factors at N$^3$LL are reported to span 
$K^{\rm N^3LL}=3.04-3.15$ 
for scalar masses between 100 GeV and 800 GeV,
with scale uncertainties reaching $\mathcal{O}(5\%-10\%)$,
when using the scale choices stipulated in Sec.~\ref{sec:setup_sm}. 
Approximating $K^{\rm N^3LL}\approx3.1$ for this entire mass range leads to at most a $\pm2\%$ over/underestimation of QCD corrections, which is within scale uncertainties.

Over the mass range $m_{k},m_{h}=100\GeV-1\TeV$, the N$^3$LL-corrected cross sections for $k^{--}k^{++}$ and $h^-h^+$ pair production from GF span about
\begin{subequations}
\begin{align}
 K^{\rm N^3LL}\times\sigma_{13\TeV}^{\rm GF~(LO)}(k^{--}k^{++}) &\approx 
 100\fb-9\zb\ ,
 \\
 K^{\rm N^3LL}\times\sigma_{13\TeV}^{\rm GF~(LO)}(h^{-}h^{+}) &\approx 
 1\fb-0.09\zb\ .
\end{align} 
\end{subequations}
At low masses, the $k^{--}k^{++}$ channel exhibits a cross section that is comparable to those of DY and AF. The rates for both GF channels quickly fall as masses increase. Beyond $m_k~(m_h) \sim 800~(400)\GeV$, the cross sections are negligible for the LHC, unless the couplings in Eq.~\eqref{eq:benchmark_inputs_update} are as large as $\{\lambda\}\sim\mathcal{O}(\pi)$. The difference between $k$ and $h$ reflects the inputs of Eq.~\eqref{eq:benchmark_inputs_update}. 

While the scale uncertainties of GF at LO reach about $\mathcal{O}(30\%)$, these are known to underestimate the actual uncertainty. At N$^3$LL(thresh.), uncertainties can reach $\delta\sigma^{\rm N^3LL(thresh.)}_{13\TeV}\sim\pm20\%$ due to the absence of real corrections at $\mathcal{O}(\alpha_s)$~\cite{Ruiz:2017yyf,Fuks:2019clu}.
PDF uncertainties in the LO rates are as low as $\delta\sigma^{\rm LO}_{13\TeV}\sim\pm2\%$ for low masses and as large as $\delta\sigma^{\rm LO}_{13\TeV}\sim\pm20\%$ for high  masses.

\paragraph*{Vector boson scattering:} 
Finally, we consider VBS, which is given by the permutations of 
  \begin{align}
  VV^\dagger \to\ k^{--}k^{++}
  \ \text{or}\ h^- h^+ \ ,\ V\in\{W^\pm,Z,\gamma^*\}\ .
 \end{align}
The diagrams for $Z\gamma$ scattering are similar to those given for $\gamma\gamma$ and $ZZ$ in Figs.~\ref{fig:diagram_ZeeBabu_zz_kk_scatt} and \ref{fig:diagram_ZeeBabu_MultiProd_LHC}. We include full interference by considering at the full, tree-level $2\to 4$ process
\begin{align}
\label{eq:procDef_vbs}
 q_1\ q_2\ \to\ q_1'\ q_2'\ S\ S^\dagger\quad \text{at}\quad \mathcal{O}(\alpha_s^0\alpha^4)\ .
\end{align}
To simulate this at LO, the corresponding syntax in our Monte Carlo setup is
\begin{verbatim}
generate qq qq > kk kk qq qq QED=4 QCD=0
output DirName7
generate qq qq > hh hh qq qq QED=4 QCD=0
output DirName8
\end{verbatim}

While the photons in VBS are virtual and never on-shell, there is some phase space overlap with the AF channel due to PDF fitting and evolution. Moreover, as we include interference from all gauge-invariant diagrams, Eq.~\eqref{eq:procDef_vbs} formally includes
diboson production, e.g., $q\overline{q} \to SS^\dagger \gamma^* \to SS^\dagger q\overline{q}$, and associated Higgs processes, e.g., $q\overline{q} \to Z H^* \to q\overline{q} SS^\dagger$. To minimize these additions and to regulate infrared divergences, we apply the following kinematic restrictions:
\begin{align}
\label{eq:cuts_vbs}
 p_T^q >\ & 30\GeV\ ,\quad   
 \vert\eta^q\vert > 5\ ,\quad M(q,q) > 1\TeV,\quad
 \ M(S,S^\dagger) > 150\GeV\ .
\end{align}

 Over the mass range investigated, the cross sections for the two processes roughly span
\begin{subequations}
\begin{align}
 \sigma_{13\TeV}^{\rm VBS~(LO)}(k^{--}k^{++}) &\approx 
 17\fb-7\zb\ ,
 \\
 \sigma_{13\TeV}^{\rm VBS~(LO)}(h^{-}h^{+}) &\approx 
 16\fb-1\zb\ .
\end{align}
\end{subequations}
The cross sections for both VBS channels sit well below those for AF but also differ qualitatively. At lower masses, the VBS rates are comparable to each other but bifurcate at higher masses. ($VV\to k^{--}k^{++}$ is always larger than $VV\to h^{-}h^{+}$.)
Since $\{\lambda\}=1$, the Higgs-mediated sub-processes likely play a bigger role for smaller masses and gauge-mediated sub-processes likely play a bigger role for larger masses. 
For both channels, we report that scale uncertainties range from about $\delta\sigma_{13\TeV}^{\rm VBS~(LO)}\sim\pm7\%$ at low masses to about $^{+15\%}_{-13\%}$ at high masses. For both channels, PDF uncertainties remain stable at $\delta\sigma_{13\TeV}^{\rm VBS~(LO)}\sim\pm2\%-\pm3\%$ for all masses.

\paragraph*{Summary:} For representative scalar masses, we summarize the above cross sections, scale uncertainties, PDF uncertainties, and QCD $K$-factors at the $\sqrt{s}=13\TeV$ LHC for all processes in Table~\ref{tab:xsec}. As an outlook for future  experiments, we present the same results at a hypothetical $\sqrt{s}=100\TeV$ VLHC in Fig.~\ref{fig:zeeBabu_XSec_vs_Mass_LHC100} and also in Table~\ref{tab:xsec}. (For the GF channel, we use $K^{\rm N^3LL}=2.60$.) For brevity, we do not comment much on cross sections at higher energies. Aside from the obvious jump in parton luminosities, which manifests as higher production rates, the cross section hierarchy does not qualitatively change  from $\sqrt{s}=13\TeV$.
Likewise, uncertainties at 100 TeV do not qualitatively differ from 13 TeV outside extreme values of $m_k$ and $m_h$.

In these figures and table, we quantify for various production mechanisms the dependence on scalar masses, collider energy, PDF choice, factorization/renormalization scale choice, and some QCD corrections. For the DY and AF channels at their orders of perturbation theory, there are no additional dependencies in their cross sections  on Zee-Babu model parameters since they are controlled entirely by SM gauge couplings. 
The GF channels, $gg\to k^{++}k^{--}$ and $gg\to h^{+}h^{-}$,
are controlled additionally by the top quark mass, which is well-measured, as well as the scalar couplings 
$\lambda_{kH}$ and $\lambda_{hH}$, respectively.
These couplings appear quadratically in cross sections, i.e., $\sigma_{\rm GF}\sim \vert \lambda\vert^2$.
And aside from these, the two GF cross sections are actually identical at LO in EW theory. (This degeneracy is broken at NLO in the EW theory since $k$ and $h$ have different weak hypercharges.)
Therefore, in some sense, the inputs of  Eq.~\eqref{eq:benchmark_inputs_update} and the subsequent differences in cross sections illustrate the dependence on the scalar couplings $\lambda_{kH}$ and $\lambda_{hH}$.
Similarly, the differences in the VBF channels in the high-mass limit can be understood as varying $\lambda_{kH}$ and $\lambda_{hH}$.
As discussed in Sec.~\ref{sec:pheno_unitarity}, the two VBF channels are driven by $\lambda_{kH}$ and $\lambda_{hH}$ in the high-energy limit; this limit is partially triggered by taking the masses of $k$ and $h$ to be much larger than those of the $W$ and $Z$.
Moreover, as only the production of $k^{++}k^{--}$ and $h^{+}h^{-}$ pairs are being considered (and not, say, same-sign $k^{\pm\pm}k^{\pm\pm}$ pairs),  
$\lambda_{kH}$ and $\lambda_{hH}$ are the only scalar couplings to appear.
Other processes must be considered to explore other scalar couplings.


\section{Distinguishing the Zee-Babu and Type II Seesaw models at the LHC}\label{sec:typeii}

Doubly and singly charged scalars are not unique predictions of any one model. They are in fact integral to several scenarios~\cite{Konetschny:1977bn, Cheng:1980qt,Lazarides:1980nt,Schechter:1980gr, Mohapatra:1979ia,Mohapatra:1980yp,Zee:1985rj,Zee:1985id,Babu:1988ki}. Therefore, if doubly charged scalars are discovered at the LHC, work must be done to discern their nature, e.g., what are their gauge quantum numbers, decay rates, and coupling strengths. In this section, we explore several ways one can potentially distinguish exotic scalars in the Zee-Babu model from those in the Type II Seesaw.

This section contains the main results of our study.
In summary, we find that one must rely on decay correlations of exotic scalars to distinguish the models; most (but not all) production-level observables are too similar to be of use.
We start with Sec.~\ref{sec:typeii_normalization}, where we compare production cross sections of exotic scalars. (Similar results at LO have been reported before~\cite{Gunion:1996pq}; however, our improved numerical analysis permits us to make new statements.)
In Sec.~\ref{sec:typeii_kinematics}, we compare predictions for differential distributions of doubly charged scalar pairs up to NLO+LL(PS) accuracy.
We then reinterpret constraints on the Type II Seesaw from the LHC~\cite{ATLAS:2022yzd} in terms of the Zee-Babu model in Sec.~\ref{sec:typeii_limits}.
We discuss in Sec.~\ref{sec:typeii_decay} correlations in exotic scalar decays.
And in Sec.~\ref{sec:typeii_lnv}, we discuss some criteria for establishing LNV in the two scenarios.


\subsection{Total Cross Section}\label{sec:typeii_normalization}

\begin{figure}[!t]
\begin{center}
\subfigure[]{\includegraphics[width=0.485\textwidth]{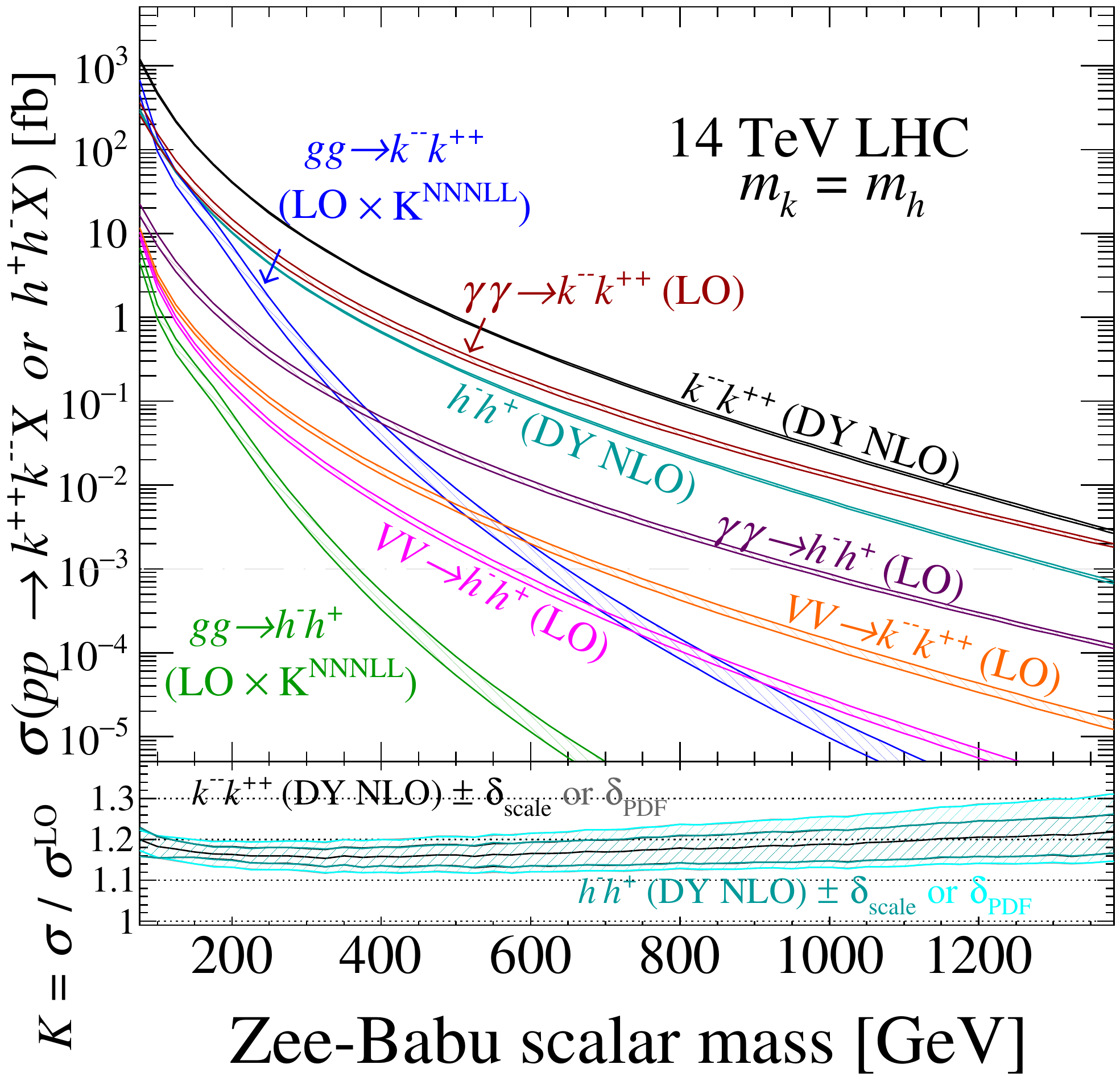}		\label{fig:zeeBabu_XSec_vs_Mass_LHCX14}}
\subfigure[]{\includegraphics[width=0.485\textwidth]{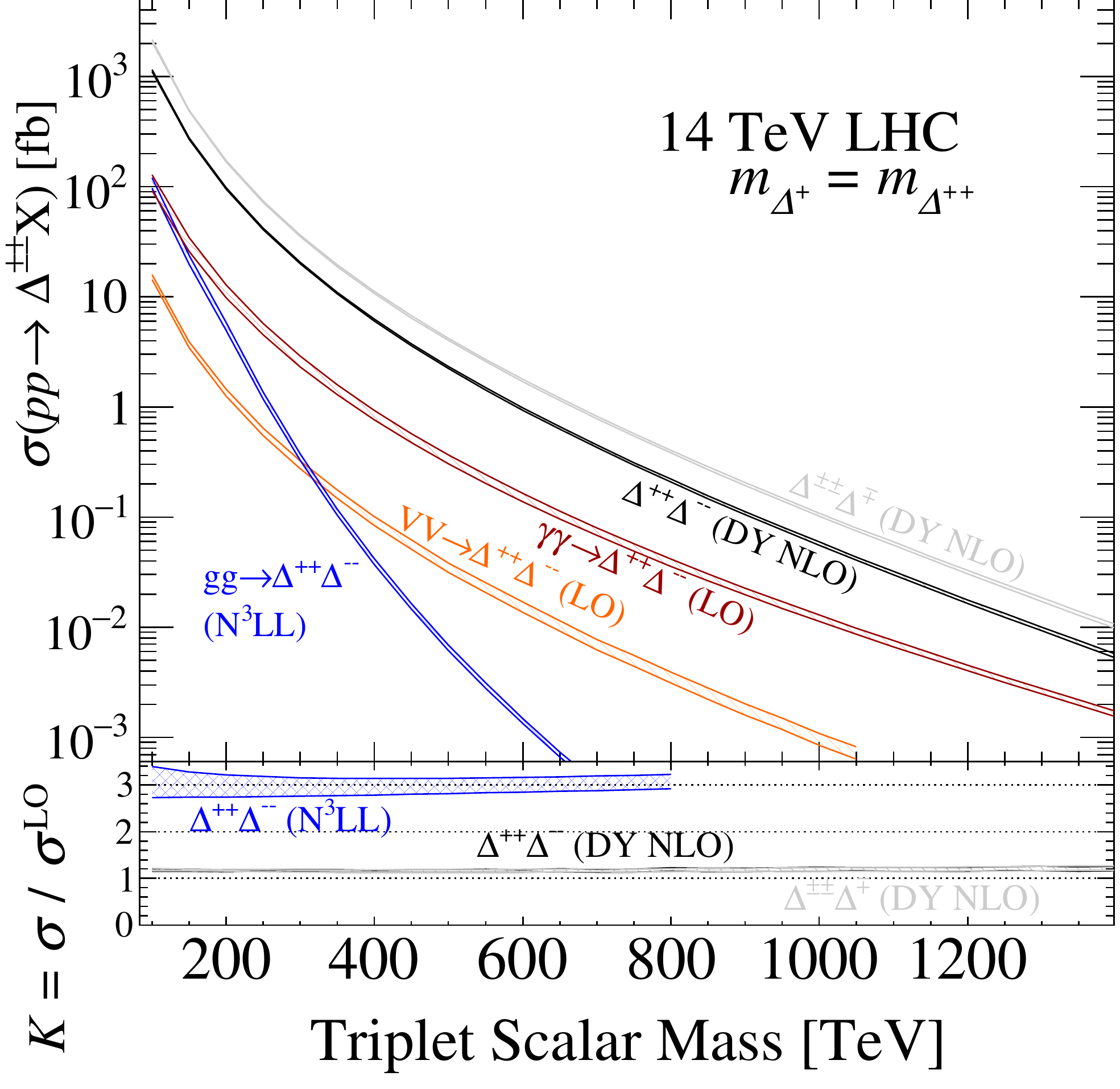}		\label{fig:typeII_XSec_vs_Mass_LHCX14_Update}}
\end{center}
\caption{
(a) Same as Fig.~\ref{fig:zeeBabu_XSec_vs_Mass_LHCX13} but for $\sqrt{s}=14\TeV$.
(b) Analogous to (a) but for the production of doubly $(\Delta^{\pm\pm})$ and singly $(\Delta^{\pm})$ charged scalars  in the Type II Seesaw; figure adapted from Ref.~\cite{Fuks:2019clu}.
}
\label{fig:zeeBabu_vs_typeII_XSec_vs_Mass_LHCX14}
\end{figure}

We start by briefly summarizing the relevant principles of the Type II Seesaw model.
In the notation of Ref.~\cite{Fuks:2019clu}, the model extends the SM by a single complex scalar multiplet $\hat{\Delta}$ that carries the quantum number assignment $(\mathbf{1},\mathbf{3},+1)$  under the SM gauge group $\mathcal{G}_{\rm SM}$ (see Sec.~\ref{sec:theory_model}). In terms of U$(1)_{\rm EM}$ states, $\hat{\Delta}$ and its vev $v_\Delta$ are given by 
\begin{align}
\hat\Delta =   
\begin{pmatrix}
\frac{1}{\sqrt{2}} \hat\Delta^+ & \hat\Delta^{++}\\
\hat\Delta^0 & -\frac{1}{\sqrt{2}} \hat\Delta^+
\end{pmatrix}
\ , 
\quad \text{with}\quad 
\langle\hat\Delta\rangle = 
\frac{1}{\sqrt{2}}\ 
\begin{pmatrix}
0 & 0 \\
v_\Delta & 0 
\end{pmatrix}
\ .
\end{align}
Conventionally, $\hat{\Delta}$ is assigned lepton number $L_{\hat{\Delta}} = -2$, making its Yukawa couplings to SM leptons LN conserving. The kinematic term and covariant derivative of $\hat{\Delta}$ are given by
\begin{align}
 \Delta\mathcal{L}_{\rm Kin.}^{\rm Type~II} = 
 {\rm Tr}\big[D_\mu \hat\Delta^\dag D^\mu\hat\Delta\big],\
 \text{with}\ 
 D_\mu \hat\Delta =  \partial_\mu\hat\Delta
   - \frac{i}{2} g W_\mu^k \big[\sigma_k\hat\Delta-\hat\Delta\sigma_k\big]
   - i g' B_\mu \hat\Delta \ .
\end{align}
After EWSB, one obtains the mass eigenstates $\Delta^{\pm\pm}$ and $\Delta^\pm$. The state $\Delta^{\pm\pm}$ is 
completely aligned with the gauge state $\hat\Delta^{\pm\pm}$, while $\Delta^{\pm}$ is partially misaligned with the gauge state $\hat\Delta^{\pm}$.  $\Delta^{\pm}$ is an admixture of $\hat\Delta^{\pm}$ and the  Goldstone boson $G^\pm$, which decouples when $v_\Delta$ vanishes.

Regarding quantum number assignments, $\hat{\Delta}$ is defined with hypercharge $Y_{\hat{\Delta}} = +1$. This means that the particles $\Delta^{++}$ and $\Delta^{+}$ carry weak isospin charges $(T_L)^3 = +1$ and $0$, respectively, and antiparticles carrying the opposite charges. The gauge states $k$ and $h$ in the Zee-Babu model are defined with $Y_{k} = -2$ and $Y_{h} = -1$, so the particles $k^{--}$ and $h^{-}$ carry negative electric charges,  antiparticles carry positive electric charges, and all states have $(T_L)^3 = 0$.
This distinction is why $\hat{\Delta}$ is assigned $L_{\hat{\Delta}} = -2$ while $k$ and $h$ are assigned $L_k,\ L_h = +2$.

Given this, we first show in Fig.~\ref{fig:zeeBabu_XSec_vs_Mass_LHCX14} the production rate for $k^{--}k^{++}$ and $h^-h^+$ pairs in the Zee-Babu model as a function of mass through various mechanisms at the $\sqrt{s}=14\TeV$ LHC, i.e., the 14 TeV analogue of Fig.~\ref{fig:zeeBabu_XSec_vs_Mass}.
In comparison, we show in Fig.~\ref{fig:typeII_XSec_vs_Mass_LHCX14_Update}
the $\sqrt{s}=14\TeV$ LHC cross sections for $\Delta^{\pm\pm}$ associated and $\Delta^{++}\Delta^{--}$ pair production in the Type II Seesaw at various accuracies, as adapted from Ref.~\cite{Fuks:2019clu}.
Qualitatively, there are several similarities between the two models:
(i) Both models feature pair production of doubly and singly (not shown) charged scalars through the  neutral current DY mechanism.
(ii) Both models feature pair production of charged scalars through AF.
(iii) Both models feature pair production of charged scalars through GF.
(iv) Both models feature pair production of charged scalars through VBF.

There are also three qualitative distinctions:
(i) Unlike the Zee-Babu model, the Type II Seesaw admits associated production via the charged current DY process, i.e., $q\overline{q'}\to W^{\pm*} \to \Delta^{\pm\pm}\Delta^\mp$, since the $\Delta^{\pm(\pm)}$ couple directly to the $W$ boson.
(ii) As the $\Delta^{\pm(\pm)}$ belong to a multiplet in the Type II Seesaw, the existence of CP-even and -odd states $\Delta^0$ and $\xi^0$   implies additional production channels (not shown).
(iii) Due to the different gauge couplings and multiplet states, many more sub-processes are present in VBS in the Type II Seesaw than in the Zee-Babu case.

The qualitative similarities listed above are also quantitative similarities. Explicit computation shows that the AF and GF cross sections for doubly and singly charged in the two scenarios are the same. (This assumes that masses $S-S^\dagger-H^0$ couplings, etc., are set equal.) Likewise, QCD corrections for these processes are the same; EW corrections are anticipated to be different due to the different gauge quantum numbers of the scalars but this must be investigated.

\begin{figure}[!t]
\begin{center}
\subfigure[]{\includegraphics[width=0.485\textwidth]{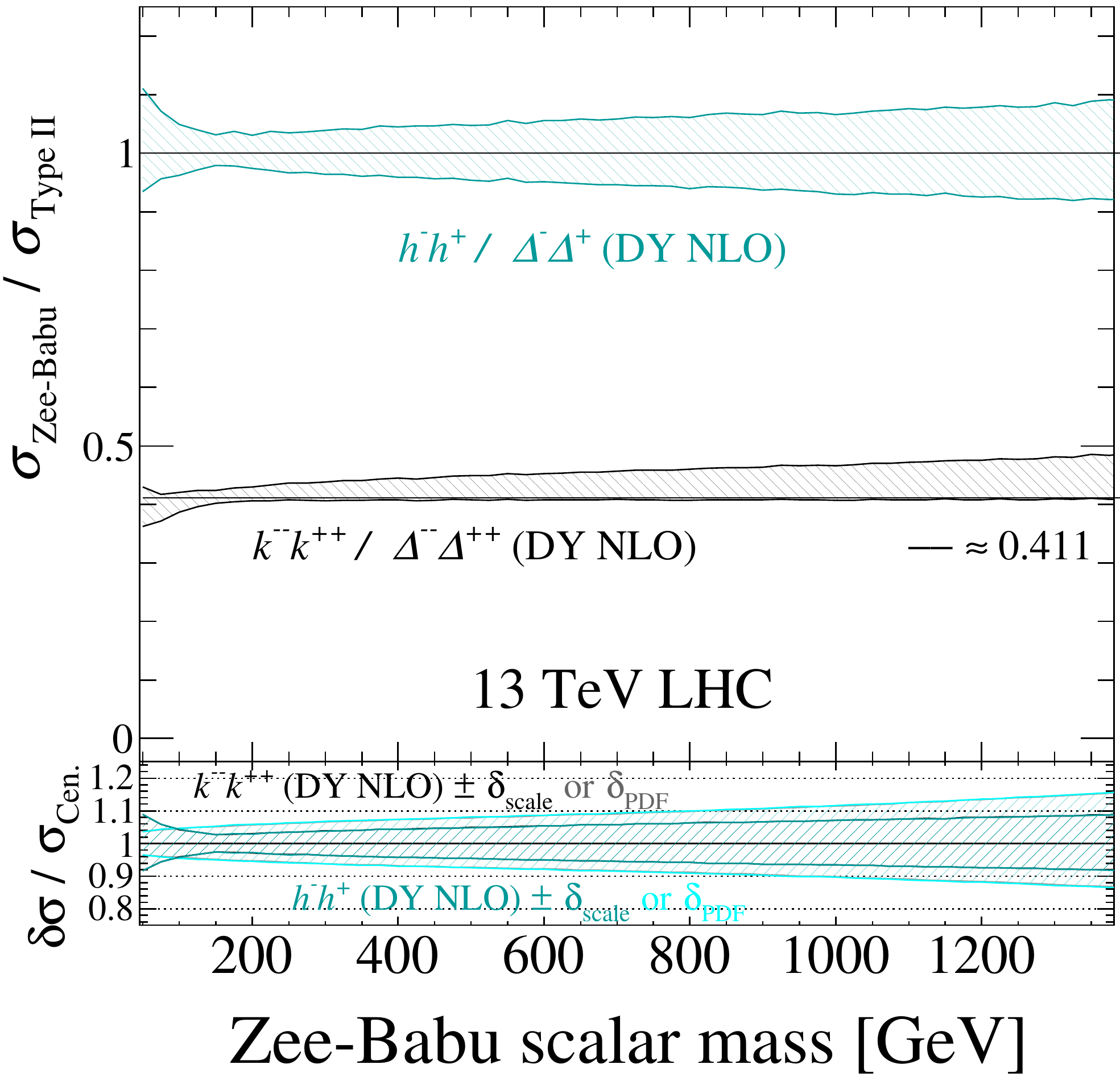}		\label{fig:zeeBabu_XSecRatio_wTypeII_LHCX13}}
\subfigure[]{\includegraphics[width=0.485\textwidth]{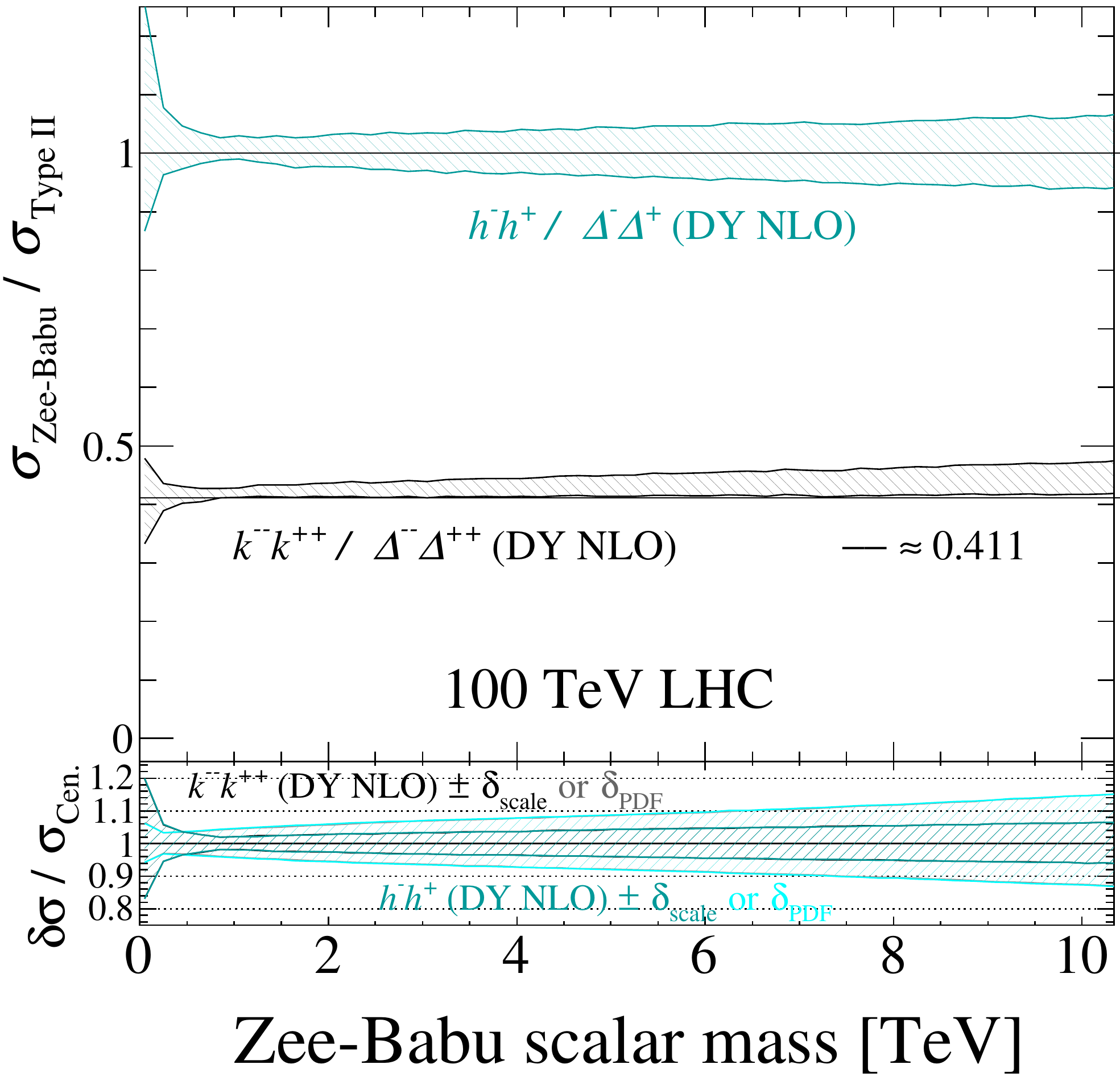}		\label{fig:zeeBabu_XSecRatio_wTypeII_LHC100}}
\end{center}
\caption{
Upper panel:
At (a) $\sqrt{s}=13\TeV$ and (b) 100\TeV, and as a function of scalar mass, the ratio of hadronic production cross sections, 
  ${\sigma}^{\rm DY~(NLO)}_{\rm Zee-Babu}(pp\to SS^\dagger X) / {\sigma}^{\rm DY~(NLO)}_{\rm Type~II}(pp\to SS^\dagger X)$,
  at NLO in QCD and with scale uncertainties (band)
  for $S\in\{k^{--},h^-\}$ in the Zee-Babu model 
  and $S\in\{\Delta^{++},\Delta^+\}$ in the Type II Seesaw model.
  Lower panel: scale (darker band) and PDF (lighter band) uncertainties.
}
 \label{fig:zeeBabu_XSecRatio_wTypeII}
\end{figure}

To further study the quantitative similarities of the neutral current DY channels, we define the scale factors $\xi$ and $\chi$ as the ratio of the Zee-Babu and Type II hadronic cross sections:
\begin{align}
\xi = 
  \frac{{\sigma}^{\rm DY~(NLO)}_{\rm Zee-Babu}(pp\to k^{--}k^{++}X)}{{\sigma}^{\rm DY~(NLO)}_{\rm Type~II}(pp\to \Delta^{++}\Delta^{--}X)}
  \quad\text{and}\quad
   \chi = 
  \frac{{\sigma}^{\rm DY~(NLO)}_{\rm Zee-Babu}(pp\to h^{-}h^{+}X)}{{\sigma}^{\rm DY~(NLO)}_{\rm Type~II}(pp\to \Delta^{+}\Delta^{-}X)}
  \ .
\end{align}
For all cross sections, we use the total rate at NLO in QCD.
Conservative scale uncertainties are obtained by taking the ratio of extrema as allowed by scale variation, e.g., $\max(\xi) = \max({\sigma}^{\rm DY~(NLO)}_{\rm Zee-Babu}) / \min({\sigma}^{\rm DY~(NLO)}_{\rm Type~II})$. In Fig.~\ref{fig:zeeBabu_XSecRatio_wTypeII}, we plot the scale factors $\xi$ (black band) and $\chi$ (teal band) with scale uncertainties (band thickness) as a function of scalar mass, fixing all equal, at (a) $\sqrt{s}=13\TeV$ and (b) $100\TeV$. Guidelines (black) are given at $\xi_* \approx 0.411$ and $\chi_* =1$. 
While the ratio for doubly charged scalars is situated around $\xi \approx 0.4-0.45$ and is largely independent of mass and $\sqrt{s}$, the ratio for singly charged scalars sits universally at unity.

To understand Fig.~\ref{fig:zeeBabu_XSecRatio_wTypeII}, note that the generic cross section for producing $SS^\dagger$ pairs of mass $m_S$, electric charge $Q_S$, and weak isospin charge $(T_L^S)^3$, 
from massless SM fermions $f\overline{f}$ is~\cite{Gunion:1996pq}
\begin{subequations}
\label{eq:dy_generic}
\begin{align}
 \hat{\sigma}^{\rm DY~(LO)}(f\overline{f}\to SS^\dagger) &= 
 \left(\frac{\pi \alpha_{\rm EM}^2 \beta^3 Q^2}{6 N_c}\right) 
 \left[P_{\gamma\gamma} + P_{\gamma Z} + P_{ZZ} \right], \quad \beta = \sqrt{1-\frac{m_S^2}{Q^2}},
 \\
 P_{\gamma\gamma} &= \frac{2 Q_S^2 Q_f^2}{Q^4},
 \\
 P_{\gamma Z} &= 
 \frac{2 Q_S Q_f A\ \left(a_L^f + a_R^f\right)}{\sin^2\theta_W\ \cos^2\theta_W}
 \frac{(Q^2-M_Z^2)}{Q^2\left[(Q^2-M_Z^2)^2 + (M_Z\Gamma_Z)^2\right]^2}
 , 
 \\
 P_{ZZ} &= \frac{A^2\ \left((a_L^f)^2 + (a_R^f)^2 \right)}{\sin^4\theta_W\ \cos^4\theta_W}
 \frac{1}{\left[(Q^2-M_Z^2)^2 + (M_Z\Gamma_Z)^2\right]^2}
 , 
 \\
 A = (T_L^S)^3 - Q_f\sin^2\theta_W, & 
 \quad
 a_L^f = (T_L^f)^3 - Q_f\sin^2\theta_W,\
 \quad 
 a_R^f = - Q_f\sin^2\theta_W\ .
\end{align}
\end{subequations}
Here, $Q_f$ and $(T_L^f)^3$ are, respectively, the electric and weak isospin charges of $f$.
The terms $P_{\gamma\gamma}$, $P_{\gamma Z}$, and $P_{ZZ}$ denote the pure photon, interference, and pure $Z$ contributions to $SS^\dagger$ production. 
In the following we work in the limit $(M_Z^2/Q^2)\to0$, and implicitly take $M_Z=0$.

First, note that the entire $m_S$ dependence in the partonic cross section is contained in the momentum/phase-space factor $\beta$. For fixed flavor $f$, scattering scale $Q$, and mass, the ratio of partonic cross sections, i.e.,  
$\hat{\sigma}^{\rm DY~(LO)}(f\overline{f}\to k^{--}k^{++})/\hat{\sigma}^{\rm DY~(LO)}(f\overline{f}\to \Delta^{++}\Delta^{--})$, is independent of $m_S$. 
At the hadronic level, this holds under reasonable approximations:
Assuming the production of TeV-scale $SS^\dagger$ is driven by valence-sea annihilation, that at large momentum fractions the up-flavor PDF is twice as large as the down-flavor PDF, i.e., $f_{u/p}\approx 2f_{d/p}$, and sea densities are equal, i.e., $f_{\overline{u}/p}\approx f_{\overline{d}/p}$, then the ratio of hadronic cross section is proportional to
\begin{align}
\xi = 
 \frac{{\sigma}^{\rm DY~(LO)}(pp\to k^{--}k^{++}X)}{{\sigma}^{\rm DY~(LO)}(pp\to \Delta^{++}\Delta^{--}X)} 
 \propto
    \frac{f_{u/p}\otimes f_{\overline{u}/p} \otimes \left(\beta^3/Q^2\right)}{f_{u/p}\otimes f_{\overline{u}/p} \otimes \left(\beta^3/Q^2\right)} \ .
\end{align}
Nominally, the dependence on $m_S$ in the ratio cancels. 
(There is a small dependence on $m_S$ for the doubly charged case that we address below.)
The ratio is also stable at NLO in QCD.
For instance: at NLO in QCD, 
virtual and soft corrections factorize (see, for example, Ref.~\cite{Ruiz:2015zca}) and cancel. Under our assumptions, PDF subtraction terms also follow this behavior.

Moving onto the value of the ratios themselves, we note that the quantities $(Q^4\ P_{VV'})$, for $V\in\{\gamma,Z\}$, depend only on gauge quantum number when $M_Z$ can be neglected.
For the $u\overline{u}\to k^{--}k^{++}$ channel, the $\gamma-Z$ interference $(Q^4\ P_{\gamma Z})$ and pure $Z$ $(Q^4\ P_{ZZ})$ contributions strongly cancel, resulting in an $\mathcal{O}(-10\%)$ correction to the pure $\gamma$ contribution $(Q^4\ P_{\gamma\gamma})$. (This is a gauge-dependent statement and can be interpreted differently.) For the $d\overline{d}\to k^{--}k^{++}$ channel, the cancellation is stronger since $Q^4 P_{\gamma\gamma} < Q^4 \vert P_{\gamma Z} \vert < Q^4 P_{ZZ}$. The cancellation between the $\gamma-Z$ interference and pure $Z$ contributions is an $\mathcal{O}(+6\%)$ addition to the pure $\gamma$ term. Essentially, $f\overline{f}\to k^{--}k^{++}$ can be treated as a QED process, which is consistent with the $k-k-Z$ coupling being Weinberg angle-suppressed (see Sec.~\ref{sec:theory_model}). For $f\overline{f}\to \Delta^{++}\Delta^{--}$, the nonzero weak isospin charge induces constructive interference among the three terms for both up and down flavors.

All inputs equal, the ratio of partonic cross sections 
for doubly charged scalars simplify to ratios of ``$P_{VV'}$-terms.'' For flavor combinations $(u\overline{u})$ and $(d\overline{d})$, these are given by
\begin{align}
\mathcal{U} \equiv
\frac{\hat{\sigma}^{\rm DY~(LO)}(u\overline{u}\to k^{--}k^{++})}{\hat{\sigma}^{\rm DY~(LO)}(u\overline{u}\to \Delta^{++}\Delta^{--})} 
&\approx
\frac{\left[P_{\gamma\gamma}+P_{\gamma Z}+P_{ZZ}\right]\Big\vert_{\rm ZB}}{\left[P_{\gamma\gamma}+P_{\gamma Z}+P_{ZZ}\right]\Big\vert_{\rm Type~II}}
\\
& = 
\frac{136\sin^4\theta_W}{21-8\cos(2\theta_W)+5\cos(4\theta_W)}
\approx 0.499\ ,
\\
\mathcal{D} \equiv
\frac{\hat{\sigma}^{\rm DY~(LO)}(d\overline{d}\to k^{--}k^{++})}{\hat{\sigma}^{\rm DY~(LO)}(d\overline{d}\to \Delta^{++}\Delta^{--})} 
&\approx
\frac{40\sin^4\theta_W}{9-4\cos(2\theta_W)+5\cos(4\theta_W)}
\approx 0.237\ .
\end{align}
We find good numerical agreement between this and the full matrix element calculation. Parameterizing the relative $(u\overline{u})$ and $(d\overline{d})$ contribution na\"ively as $(2/3)$ $\mathcal{U}$ and $(1/3)$ $\mathcal{D}$ 
gives
\begin{align}
\xi_\star = 
 \frac{2}{3}\mathcal{U} + \frac{1}{3}\mathcal{D} \approx 0.411\ .
\end{align}
This agrees remarkably well with the numerical results reported in Fig.~\ref{fig:zeeBabu_XSecRatio_wTypeII}.

A closer inspection shows that the central value of the ratio $\xi$ sits just below (above) the $\xi_\star\approx 0.411$ guideline at the lowest (highest) masses. 
This follows from PDF dynamics, namely deviations from our crude assumptions that $f_{{u}/p}=2f_{{d}/p}$
and $f_{\overline{u}/p}=f_{\overline{d}/p}$.
For instance: masses that are $\mathcal{O}(100-300)\GeV$ require momentum fractions that are $\mathcal{O}(0.01-0.05)$ at $\sqrt{s}=13\TeV$. This is where an asymmetry occurs between the $\overline{u}$ and $\overline{d}$ PDFs, with $f_{\overline{d}/p} > f_{\overline{u}/p}$. Also in this range, the $u$ PDF is only $\mathcal{O}(20\%-40\%)$ larger $(\mu_f = 2m_s)$ than the $d$.
This means that $(d\overline{d})$ annihilation occurs more frequently than na\"ively argued and pulls down the scale factor $\xi$.
At larger momentum fractions, i.e., beyond $\mathcal{O}(0.1)$, the $\overline{u}-\overline{d}$ asymmetry closes and the $u/d$ ratio increases.  Subsequently, $(u\overline{u})$ annihilation occurs more frequently and pulls up $\xi$. The estimate $\xi_\star$ remains within or at the edge of the scale uncertainty band. PDF uncertainties for individual cross sections (lower panel) are at the $\pm5\%-\pm15\%$ level for the masses investigated. Therefore, while deviations from $\xi_\star$ can evolve with improved PDF fits, the change will not be significant. 

At 100 TeV (Fig.~\ref{fig:zeeBabu_XSecRatio_wTypeII_LHC100}), this behavior is unchanged for the masses under investigation since the same regions of individual PDFs are being probed. For example: the momentum fraction
\begin{align}
 x \sim 2m_S/\sqrt{s} = 130\ {\rm GeV}/ 6.5\ {\rm TeV}~(1300\ {\rm GeV} / 6.5\ {\rm TeV}) = 0.02~(0.2)
\end{align}
 probes the same parts of a PDF at $\sqrt{s}=13\TeV$ as the momentum fraction
 \begin{align}
 x \sim 2m_S/\sqrt{s} =1\ {\rm TeV}/ 50\ {\rm TeV}~(10\ {\rm TeV} / 50\ {\rm TeV}) = 0.02~(0.2)
 \end{align}
 at $\sqrt{s}=100\TeV$. 
This is a manifestation of Bjorken scaling.
Therefore, the (small) differences between the 
curves at 13 and 100 TeV are due to DGLAP evolution, which is logarithmic.
(As stipulated in Eq.~\eqref{eq:scale_choice}, PDFs are evolved up to factorization scales that scale as $\mu_f \sim 2m_S$.)

Turning to singly charged scalars, note that the gauge charges for $h^-$ and $\Delta^-$ are both $Q_S = -1$, $Y_S = -1$,  and $(T_L^S)=0$.
Therefore, for fixed $f$ and $m_S$, the partonic  cross sections for $f\overline{f}\to SS^\dagger$ are the same. It follows that the ratio of hadronic rates is unity for all masses:
\begin{align}
\label{eq:scale_factor_hh}
 \chi = 
  \frac{{\sigma}^{\rm DY~(NLO)}(pp\to h^{-}h^{+}X)}{{\sigma}^{\rm DY~(NLO)}(pp\to \Delta^{+}\Delta^{-}X)}  = 1.
\end{align}
This behavior is reflected in Fig.~\ref{fig:zeeBabu_XSecRatio_wTypeII} at both collider energies. A caveat of this result is that cross sections are obtained at LO in the EW theory. Since $h^\mp$ is an SU$(2)_L$ singlets but $\Delta^\pm$ belongs to a triplet representation, it is likely that EW corrections  can break this degeneracy. 

\paragraph*{Discussion:}
If $k^{--}k^{++}$ and/or $h^-h^+$ pairs are discovered at the LHC, one could arguably use the observed cross section to extract gauge quantum numbers. However, $k$ and $h$ are shortly lived (see Sec.~\ref{sec:pheno_decay}) and readily decay to SM particles. Therefore, what is actually measured is the combination of production and decay rates. As shown in Table~\ref{tab:widths}, Eq.~\eqref{eq:decay_rate_k}, and Eq.~\eqref{eq:decay_rate_h}, the branching rates of $k$ and $h$ are sensitive to the relative sizes of masses and $\mu_{\not L}$, not just Yukawa couplings. Factors of $\xi\sim0.4$ can easily be absorbed by a branching rate.
This last statement is also true for the Type II Seesaw. 
Decay rates of charged scalars in the Type II Seesaw are also correlated with neutrino oscillation parameters~\cite{FileviezPerez:2008jbu}.
However,  the uncertainty oscillation parameters remain sufficiently large to effectively absorb factors of $\xi$~\cite{Fuks:2019clu}.


\subsection{Kinematic distributions of doubly charged scalars}\label{sec:typeii_kinematics}

Beyond total cross sections, it is also possible to compare Zee-Babu and Type II scalars at a differential level. One argument goes that since the $k^{\mp\mp}$ and $\Delta^{\mp\mp}$ (or $h^\mp$ and $\Delta^\mp$) carry different gauge quantum numbers, and hence couple to the intermediate $\gamma/Z$ with different strengths, then one may anticipate differences in differential distributions.
We report that this argument does not work in the present case. 
Even at the differential level, we find that kinematic distributions of scalars produced by the DY process in the Zee-Babu and Type II models have the same shape and differ by only a normalization; the normalization is given by the ratio of hadronic cross sections.
This finding also holds for the GF and AF channels.

\begin{figure}[!t]
\begin{center}
\subfigure[]{\includegraphics[width=0.4\textwidth]{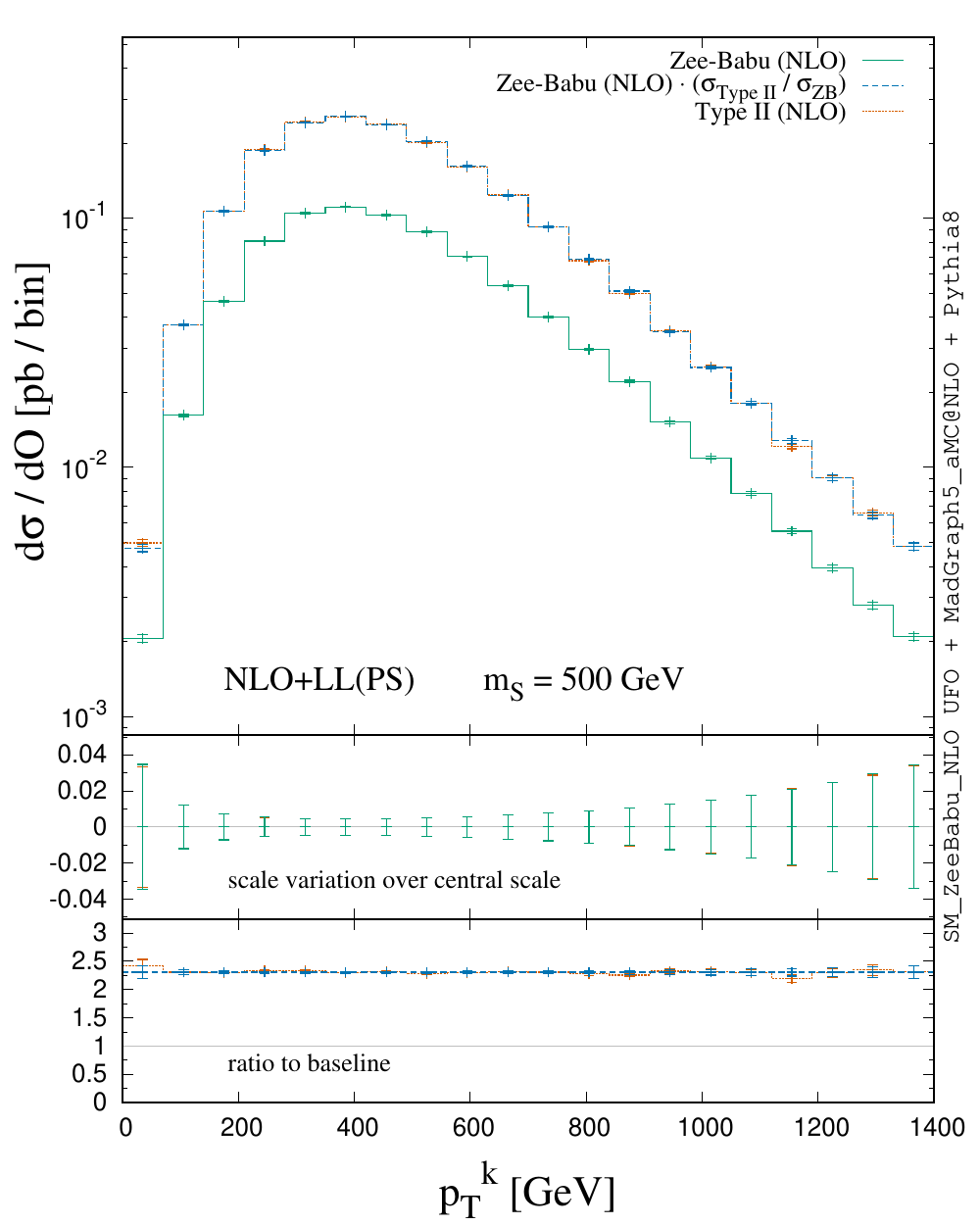}		\label{fig:zeeBabu_kinematics_LHCX13_MX500GeV_pTk}}
\subfigure[]{\includegraphics[width=0.4\textwidth]{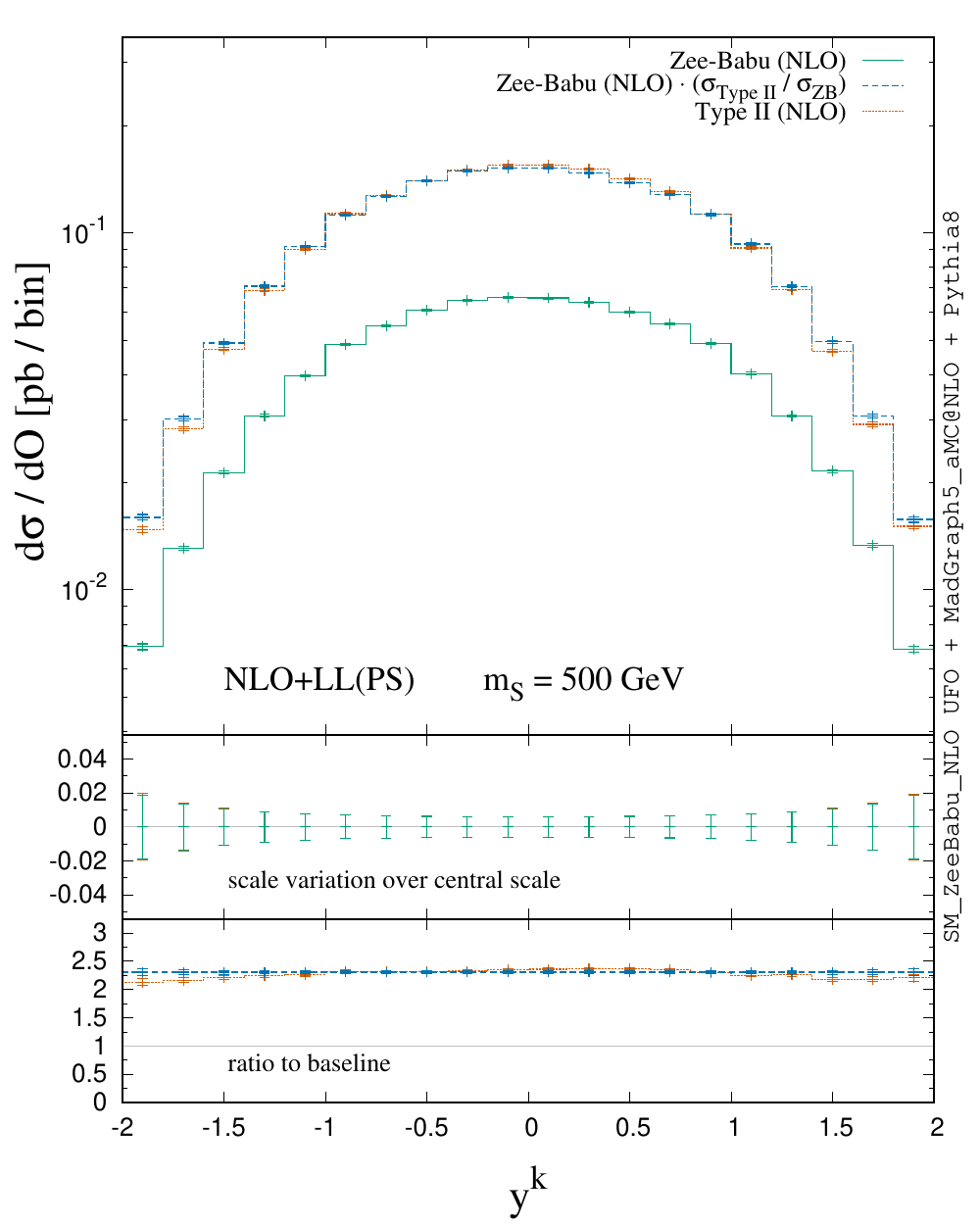}		\label{fig:zeeBabu_kinematics_LHCX13_MX500GeV_rapk}}
\\
\subfigure[]{\includegraphics[width=0.4\textwidth]{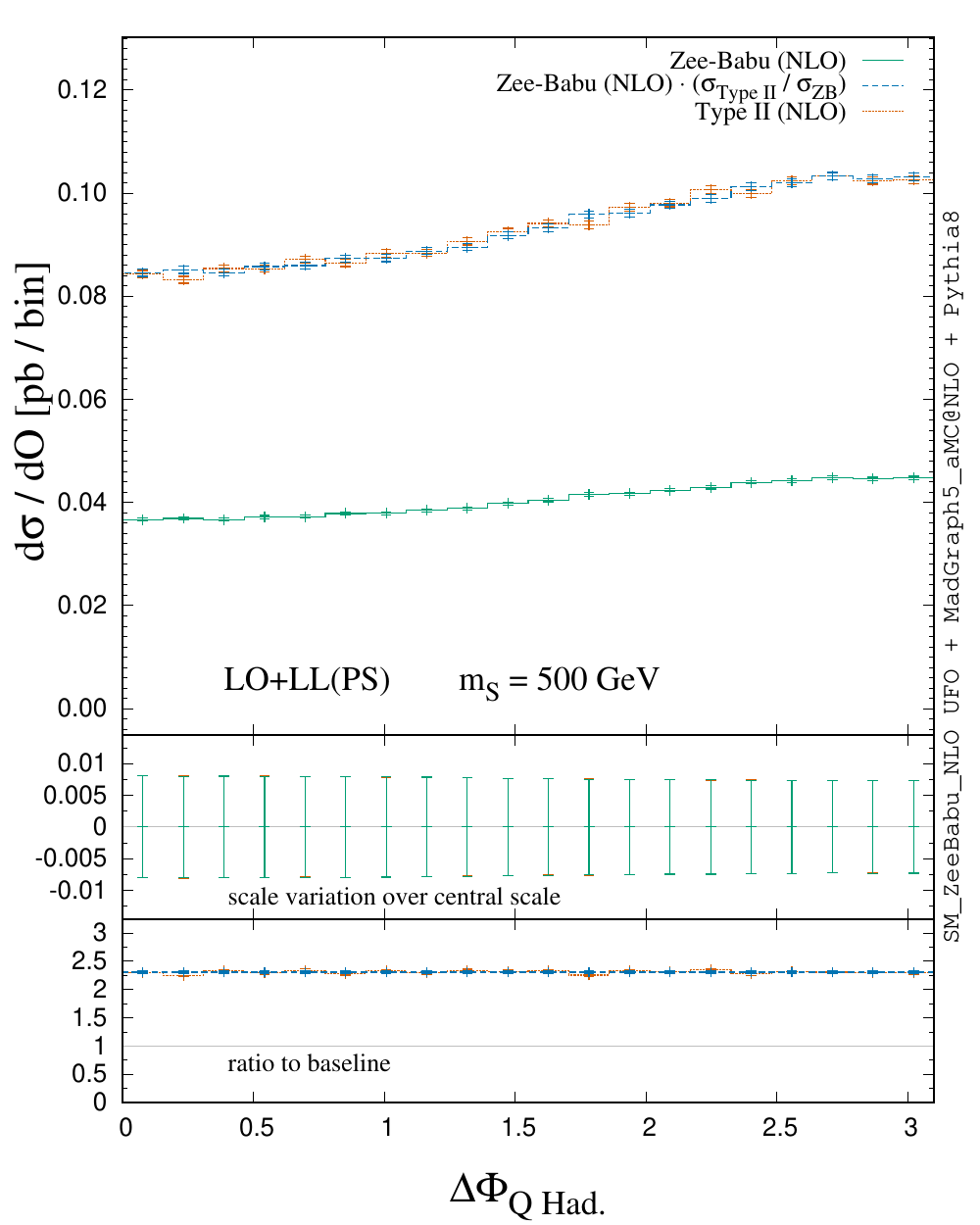}		\label{fig:zeeBabu_kinematics_LHCX13_MX500GeV_dPhi}}
\subfigure[]{\includegraphics[width=0.4\textwidth]{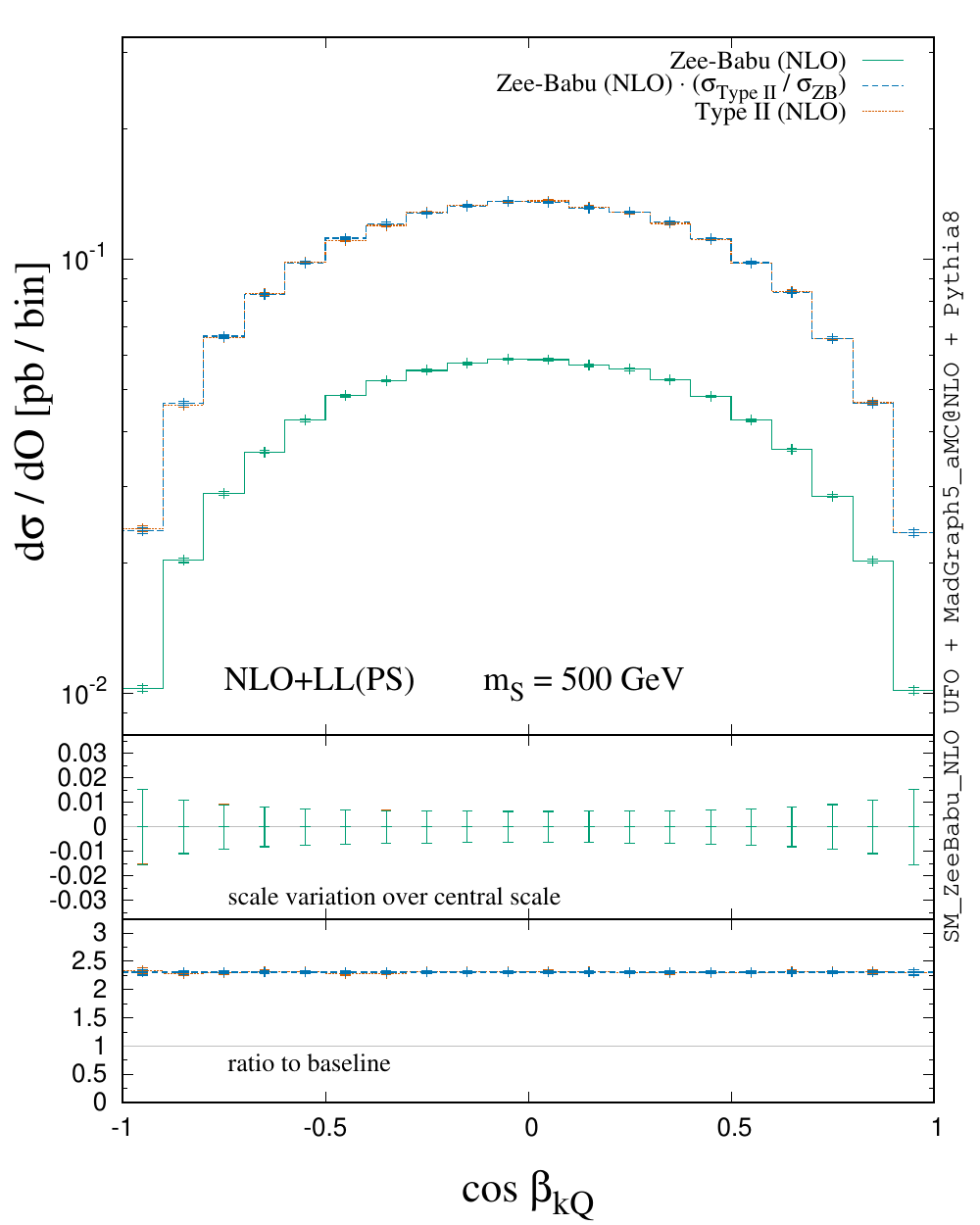}		\label{fig:zeeBabu_kinematics_LHCX13_MX500GeV_cosBoost}}
\end{center}
\caption{
Top panel: 
In Drell-Yan production of scalar pairs at the $\sqrt{s}=13\TeV$ LHC, $pp\to SS^\dagger+X$, the differential cross sections at NLO+LL(PS) or LO+LL(PS) with respect to (a) $p_T^S$, (b) $y^S$, (c) $\Delta \Phi_{S\rm Had.}$, (d) $\cos\beta_{SQ}$ for scalar $S\in\{k^{++}, \Delta^{++}\}$
from:
(a) the Zee-Babu model ({\color{teal}solid teal});
(b) the Type II model ({\color{red}dotted red}); and
(c) the Zee-Babu model normalized by the scale factor $\xi(m_s)$ ({\color{blue}dashed blue}),
assuming a benchmark mass of $m_S = 500\GeV$.
See text for observable definitions.
Middle panel: scale variation relative to the central scale choice.
Bottom panel: ratio of curves (b) and (c) with respect to (a).
}
 \label{fig:zeebabu_kinematics_loMass}
\end{figure}

To demonstrate this, we simulate at NLO+LL(PS) for $\sqrt{s}=13\TeV$ the two DY processes\footnote{
While not shown, we have checked that the behavior and trend reported throughout this section also hold for $h^-h^+$ and $\Delta^+\Delta^-$ pairs. In this instance, the scale factor $\chi=1$ is used, in accordance with Eq.~\eqref{eq:scale_factor_hh}. 
}
\begin{subequations}
\begin{align}
 \text{Zee-Babu} :\quad& p p \to \gamma^*/Z^* X\to k^{--}k^{++}X, \\
\text{Type II}   :\quad& p p \to \gamma^*/Z^* X\to \Delta^{--}\Delta^{++}X,
\end{align}
\end{subequations}
 following the methodology in Sec.~\ref{sec:setup}.
This is the first time that kinematic distributions of the Zee-Babu model beyond LO+LL(PS) have been reported.
We normalize the total Zee-Babu cross section to Type II cross section using the mass-dependent scale factor $\xi(m_s)$:
\begin{align}
\label{eq:scale_factor_def}
\xi^{-1}(m_s) &= 
  \frac{\sigma_{13\TeV}^{\rm DY~(NLO)}(pp\to \Delta^{--}\Delta^{++}X)\vert_{m_\Delta=m_s}}{\sigma_{13\TeV}^{\rm DY~(NLO)}(pp\to k^{--}k^{++}X)\vert_{m_k=m_s}} \ .
\end{align}
For representative masses
$m_k,\ m_\Delta = m_s = 500\GeV$ and $1250\GeV$,  we obtain the scale factors
\begin{align}
\xi(m_S = 500\GeV) \approx 2.31
\quad\text{and}\quad
\xi(m_S = 1250\GeV) \approx 2.23\ .
\end{align}
From Fig.~\ref{fig:zeeBabu_vs_typeII_XSec_vs_Mass_LHCX14},
the scale uncertainty in this ratio is below 10\%.

\begin{figure}[!t]
\begin{center}
\subfigure[]{\includegraphics[width=0.45\textwidth]{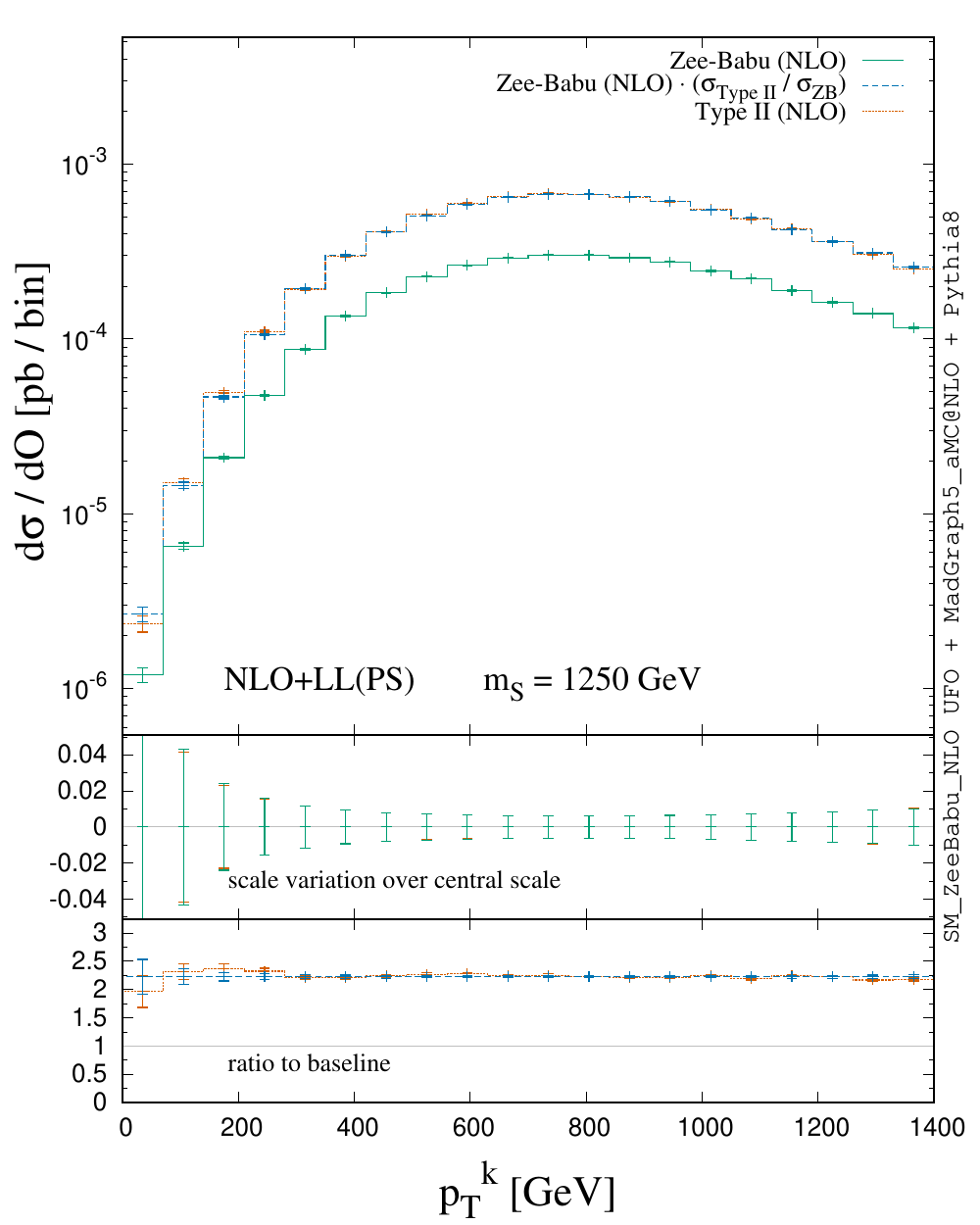}		\label{fig:zeeBabu_kinematics_LHCX13_M1250GeV_pTk}}
\subfigure[]{\includegraphics[width=0.45\textwidth]{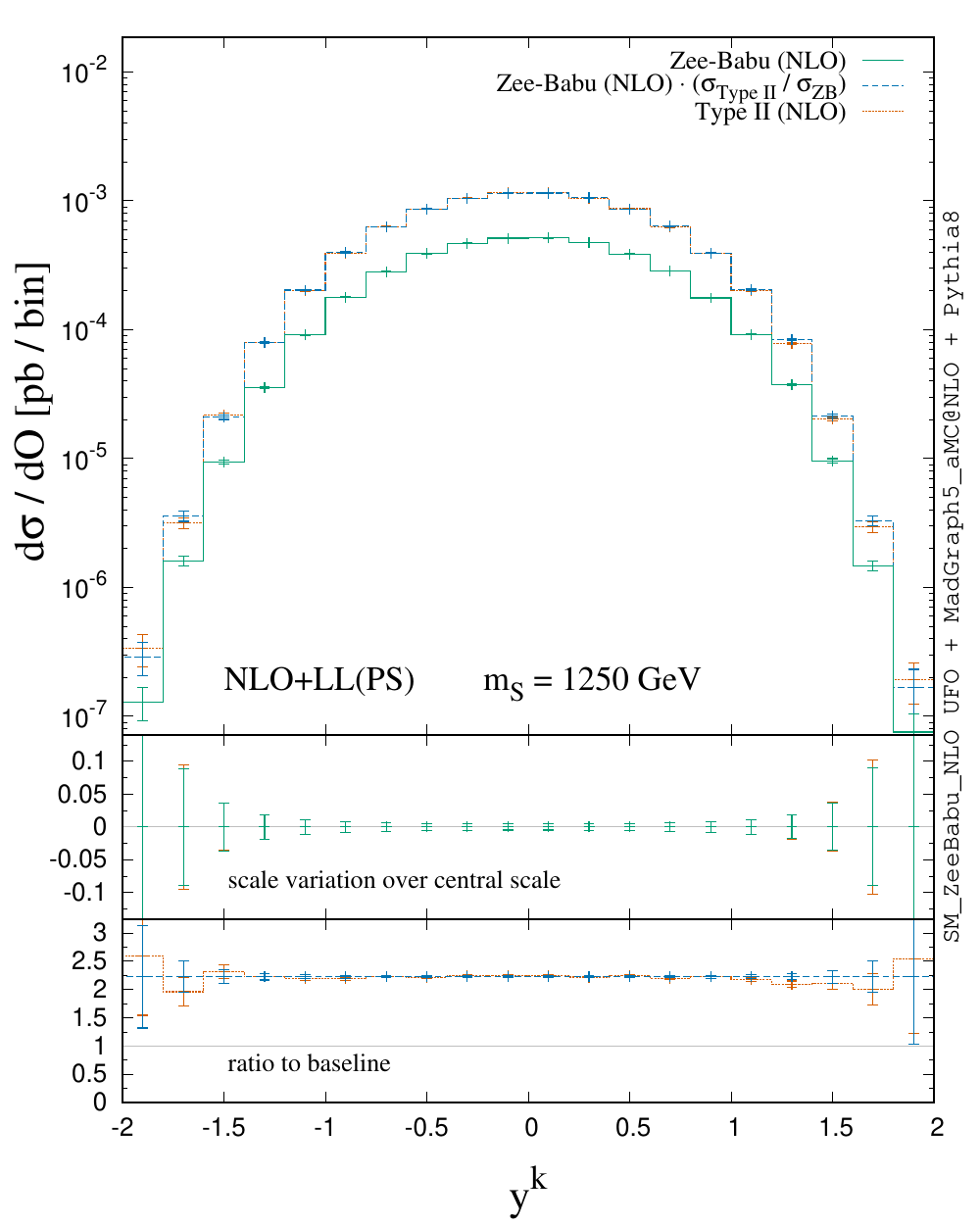}		\label{fig:zeeBabu_kinematics_LHCX13_M1250GeV_rapk}}
\\
\subfigure[]{\includegraphics[width=0.45\textwidth]{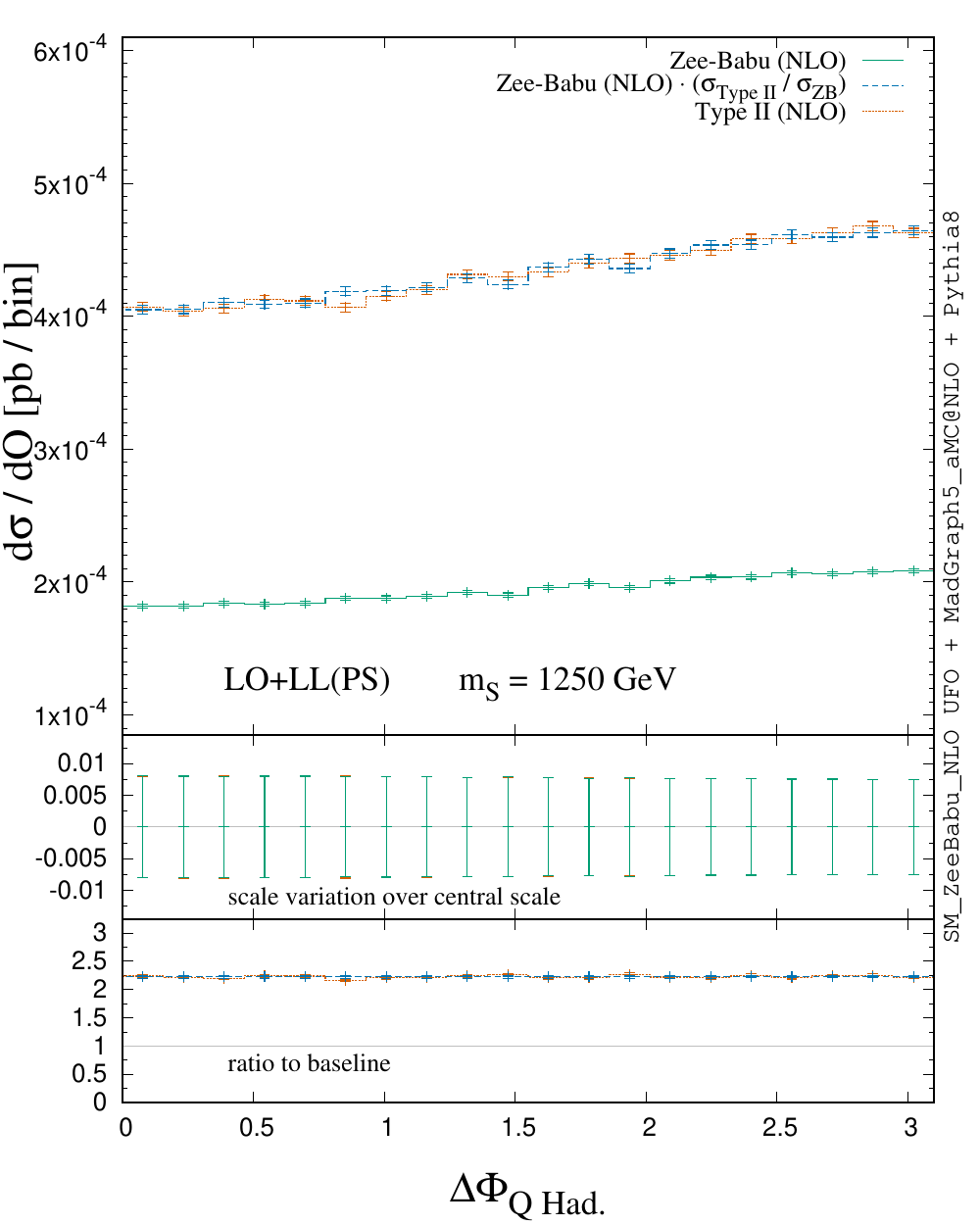}		\label{fig:zeeBabu_kinematics_LHCX13_M1250GeV_dPhi}}
\subfigure[]{\includegraphics[width=0.45\textwidth]{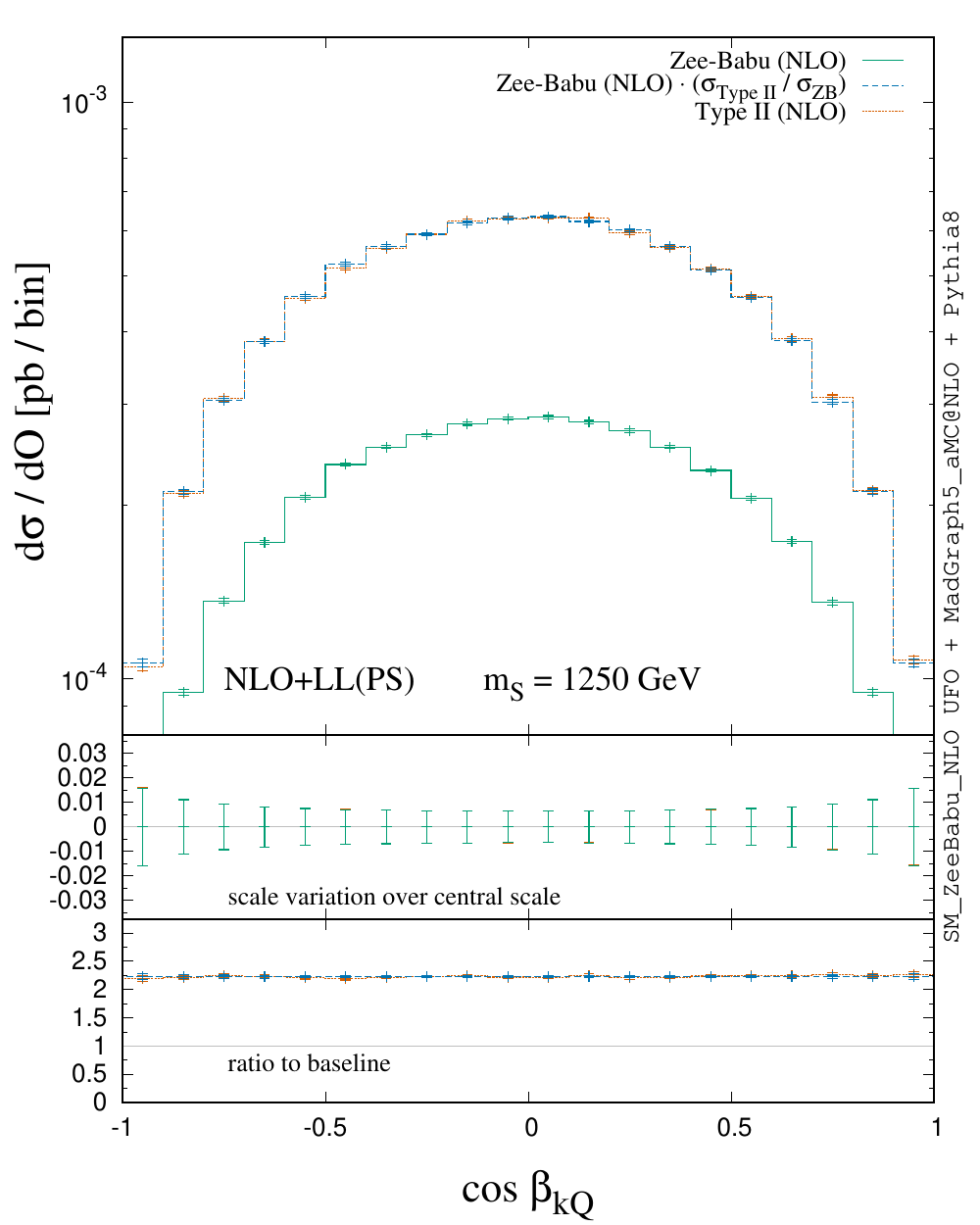}		\label{fig:zeeBabu_kinematics_LHCX13_M1250GeV_cosBoost}}
\end{center}
\caption{Same as Fig.~\ref{fig:zeebabu_kinematics_loMass} but for $m_S = 1250\GeV$.
}
 \label{fig:zeebabu_kinematics_hiMass}
\end{figure}

In the following distributions (upper panel), we plot three quantities:
(a) the Zee-Babu prediction ({\color{teal}solid teal});
(b) the Type II prediction ({\color{red}dotted red});
(c) the Zee-Babu prediction normalized by the scale factor $\xi^{-1}(m_s)$ ({\color{blue}dashed blue}).
For all three curves and for a given observable $\mathcal{O}$, we also show (middle panel) the bin-by-bin scale variation of the differential cross section with respect to the central scale choice (see Sec.~\ref{sec:setup_mc}).
Symbolically, this is given by
\begin{align}
 \text{scale variation at}\ \mathcal{O}^* 
 = \cfrac{\frac{d\sigma^{\rm NLO+LL(PS)}}{d\mathcal{O}}\Big\vert_{\mathcal{O}=\mathcal{O}^*}^{\zeta\in\{0.5,1.0,2.0\}}}{\frac{d\sigma^{\rm NLO+LL(PS)}}{d\mathcal{O}}\Big\vert_{\mathcal{O}=\mathcal{O}^*}^{\zeta=1}}\ .
\end{align}
We also show (lower panel) the bin-by-bin ratio of the Type II and scaled Zee-Babu differential rates 
to the unscaled Zee-Babu prediction. The unscaled rate is our baseline.
This is given by
\begin{align}
 \text{ratio to baseline at }\ \mathcal{O}^* &= \frac{1}{\sigma^{\rm DY~(NLO)}}\Big\vert_{\rm Zee-Babu}\
 \times\ \frac{d\sigma^{\rm NLO+LL(PS)}}{d\mathcal{O}}\Big\vert_{\mathcal{O}=\mathcal{O}^*}\ .
\end{align}
We report results for $m_S = 500\GeV$ in Fig.~\ref{fig:zeebabu_kinematics_loMass}  and for $m_S = 1250\GeV$ in Fig.~\ref{fig:zeebabu_kinematics_hiMass}.
The results are based on samples with $N=400k$ events each; statistical uncertainties are denoted by crosses $+$.
Closed bars $][$ denote scale uncertainties.
We focus on $k^{++}$ and $\Delta^{++}$ but by momentum conservation the kinematics of $k^{--},\ \Delta^{--}$  mirror those for the positively charged states.

We start with Fig.~\ref{fig:zeeBabu_kinematics_LHCX13_MX500GeV_pTk}. There we show (top) the transverse momentum distribution $p_T^S$ of positively charged scalars $S\in\{k^{++},\ \Delta^{++}\}$. 
As a function of increasing $p_T$, the distributions rise to a maximum at about $p_T\sim 400\GeV$, or $(p_T/m_S)\sim 0.8$, and then fall with a power-law-like behavior. The spectra become small at small $(p_T/m_S)$ since $SS^\dagger$ pairs are not produced precisely at threshold but rather with modest momenta.
The qualitative behavior of all three distributions is the same but the unscaled Zee-Babu curve sits below the scale Zee-Babu curve and the Type II curve. 
Scale uncertainties (middle) for all three curves reach about \confirm{$\pm4\%$} at small and large $p_T$; 
at intermediate $p_T$, scale uncertainties reduce to the sub-percent level.
In comparison to the baseline (bottom), the bin-by-bin normalization of the Zee-Babu and Type II models are statistically indistinguishable. This is the first indication that kinematic distributions of doubly charged scalar production in the two models are identical, up to an overall normalization.

In Fig.~\ref{fig:zeeBabu_kinematics_LHCX13_M1250GeV_rapk} we show the rapidity $y$ of $S$, defined by
\begin{align}
 y = \frac{1}{2}\ \log\left(\frac{E+p_z}{E - p_z}\right)\ .
\end{align}
One sees that a bulk of the phase space sits in the range $\vert y \vert< 1.5$, indicating a longitudinal momentum small compared to the total energy carried by $S$. In other words,  $SS^\dagger$ pairs produced at the LHC are largely high-$p_T$ objects  with only moderate longitudinal momentum. The scale uncertainty ranges from about \confirm{$\pm2\%$} at large rapidities $(\vert y\vert>1.75)$ but reduce to the sub-percent level at small rapidities $(\vert y\vert<0.5)$.
In comparison to the baseline, the bin-by-bin normalization between the scaled Zee-Babu distribution and the Type II distribution are statistically indistinguishable. At low (high) rapidity, the Type II curve slightly juts above (under) the scaled Zee-Babu curve. However, the statistical uncertainty bars overlap for all $y$.

To explore angular correlations, we consider in Fig.~\ref{fig:zeeBabu_kinematics_LHCX13_M1250GeV_dPhi} the azimuth separation $(\Delta \Phi)$ between $S$ and the underlying hadronic environment. This is given by
\begin{align}
 \Delta \Phi_{S\ \rm Had} = 
  \cos^{-1}\left[\
 \hat{p}_T^S\ \cdot\ \hat{p}_T^{\rm Had.} \ 
 \right]
 = 
 \cos^{-1}\left[
 \frac{\vec{p}_T^{\ S}\ \cdot\ \vec{p}_T^{\ \rm Had.} }{p_T^S\ p_T^{\rm Had.}}\right]\ ,
\end{align}
where  $p^{\rm Had.}$ is the vector sum over all hadrons within a maximum rapidity of $y^{\max} =10$, i.e.,
\begin{align}
p^{\mu\ \rm Had.} = \sum_{i\in\{\text{Had.}\}} p_i^\mu, 
\quad\text{with}\quad \vert y^i \vert < y^{\max}\ .
\end{align}
We impose $\vert y^i\vert < y^{\max}$ to exclude beam remnant 
and simplify our generator-level analysis.
(Generally, such objects have little-to-no impact on transverse kinematics.)
We observe that $\Delta\Phi$ is a mostly flat distribution, with a slight monotonic increase as one goes from a parallel orientation $(\Delta\Phi=0)$ to a back-to-back orientation $(\Delta\Phi=\pi)$. The means that it is more (less) likely for $S$ and the hadronic activity to propagate in opposite (same) transverse direction.
To understand this, note that the transverse part of $p^{\rm Had.}$ is also the recoil of the $(SS^\dagger)$ system:
\begin{equation}
 \vec{p}_T^{\ \rm Had.} = - \vec{q}_T,\ \text{where}\ q^\mu = (p_S + p_{S^\dagger})^\mu \ .
\end{equation}
Hence, the angle $\Delta \Phi_{S\ \rm Had}$ can be interpreted as an azimuthal rotation of $S$ relative to the $(SS^\dagger)$ system in the transverse plane. In the limit that $\vert \vec{q}_T\vert / Q\to0$, where $Q=\sqrt{q^2}$ is the invariant mass of the $(SS^\dagger)$ system, Born-like kinematics are recovered and the $\Delta \Phi_{S\ \rm Had}$ distribution is flat, i.e., there is no dependence on $\Delta \Phi_{S\ \rm Had}$ in $2\to2$ scattering. As  $\vert \vec{q}_T\vert$ grows, the $(SS^\dagger)$ system, and hence $S$, recoils more against the hadronic activity. 
Nonzero $\vert \vec{q}_T\vert$ induces back-to-back separation in the transverse plane 
and simultaneously suppresses same-direction propagation.

\begin{figure}[!t]
\begin{center}
\subfigure[]{\includegraphics[width=0.45\textwidth]{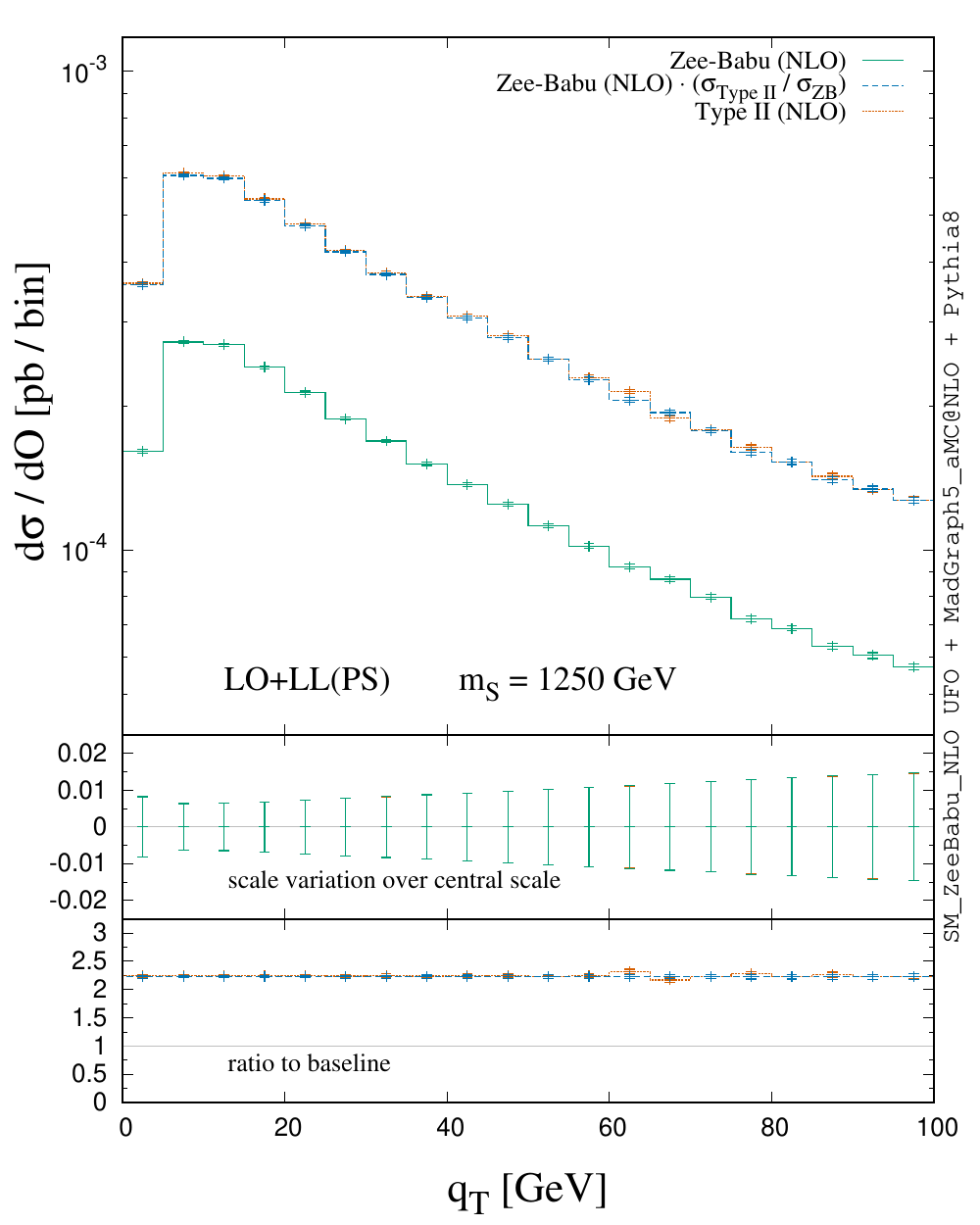}		\label{fig:zeeBabu_kinematics_LHCX13_MX500GeV_qT}}
\subfigure[]{\includegraphics[width=0.45\textwidth]{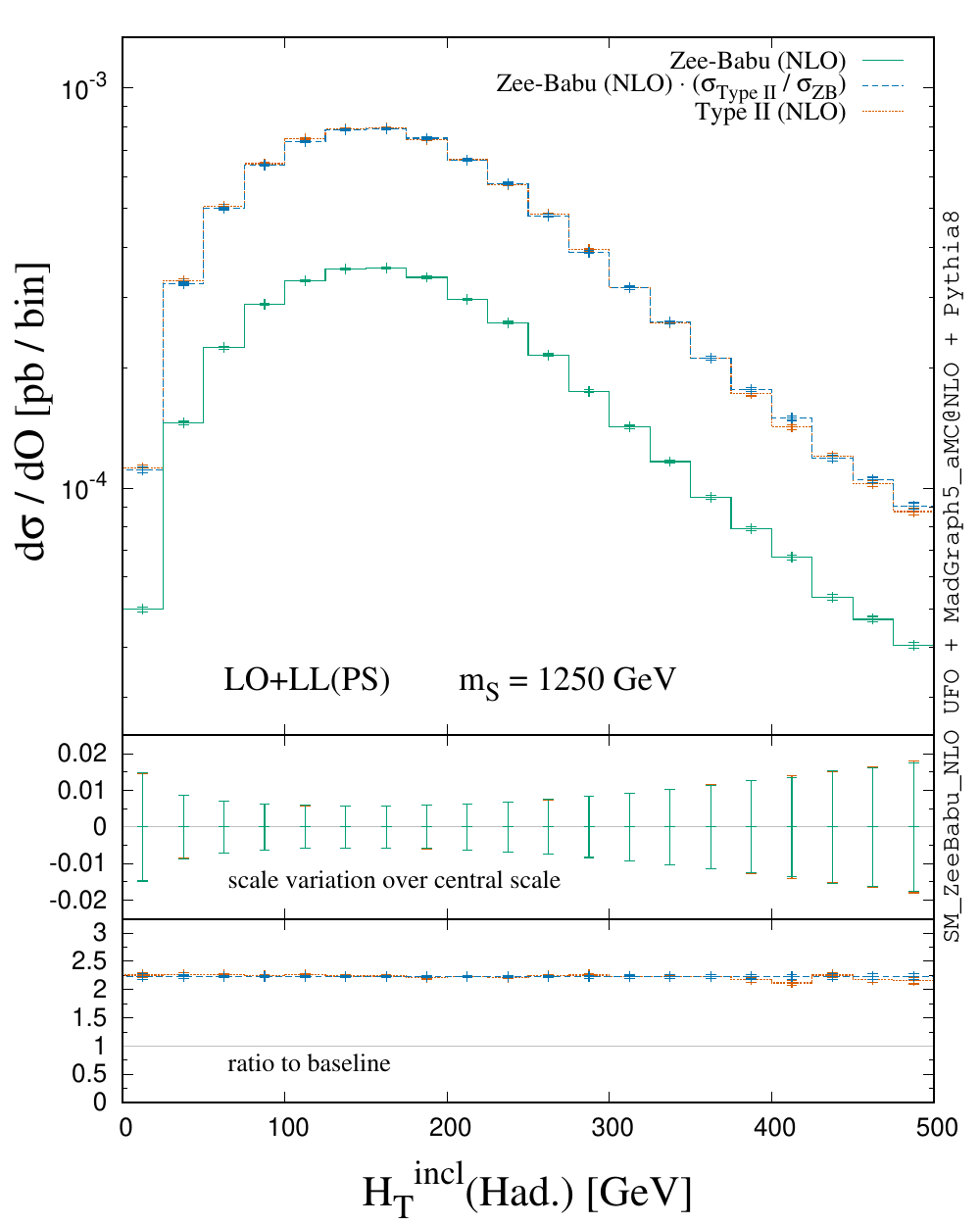}		\label{fig:zeeBabu_kinematics_LHCX13_MX500GeV_HTinclHad}}
\\
\subfigure[]{\includegraphics[width=0.45\textwidth]{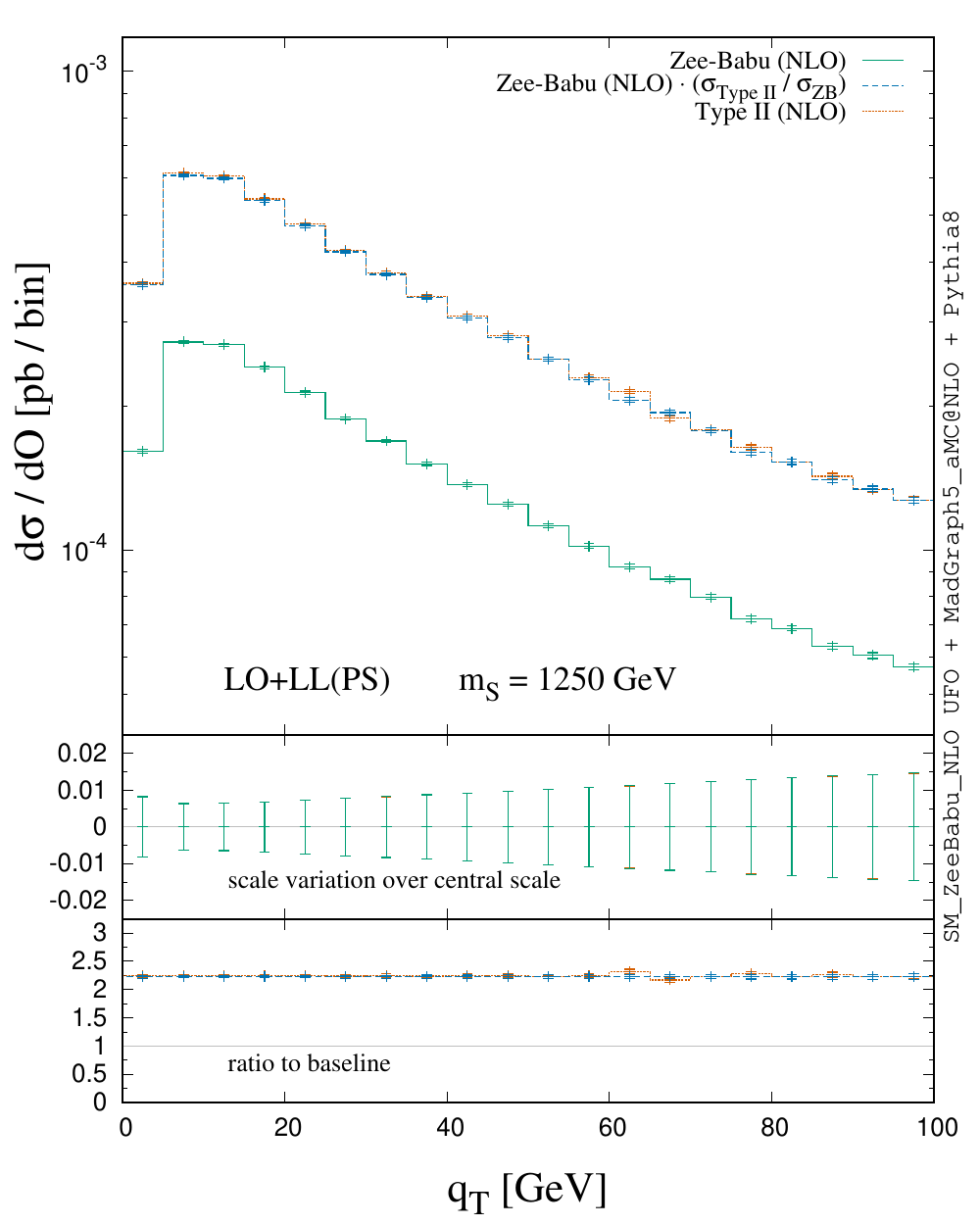}		\label{fig:zeeBabu_kinematics_LHCX13_M1250GeV_qT}}
\subfigure[]{\includegraphics[width=0.45\textwidth]{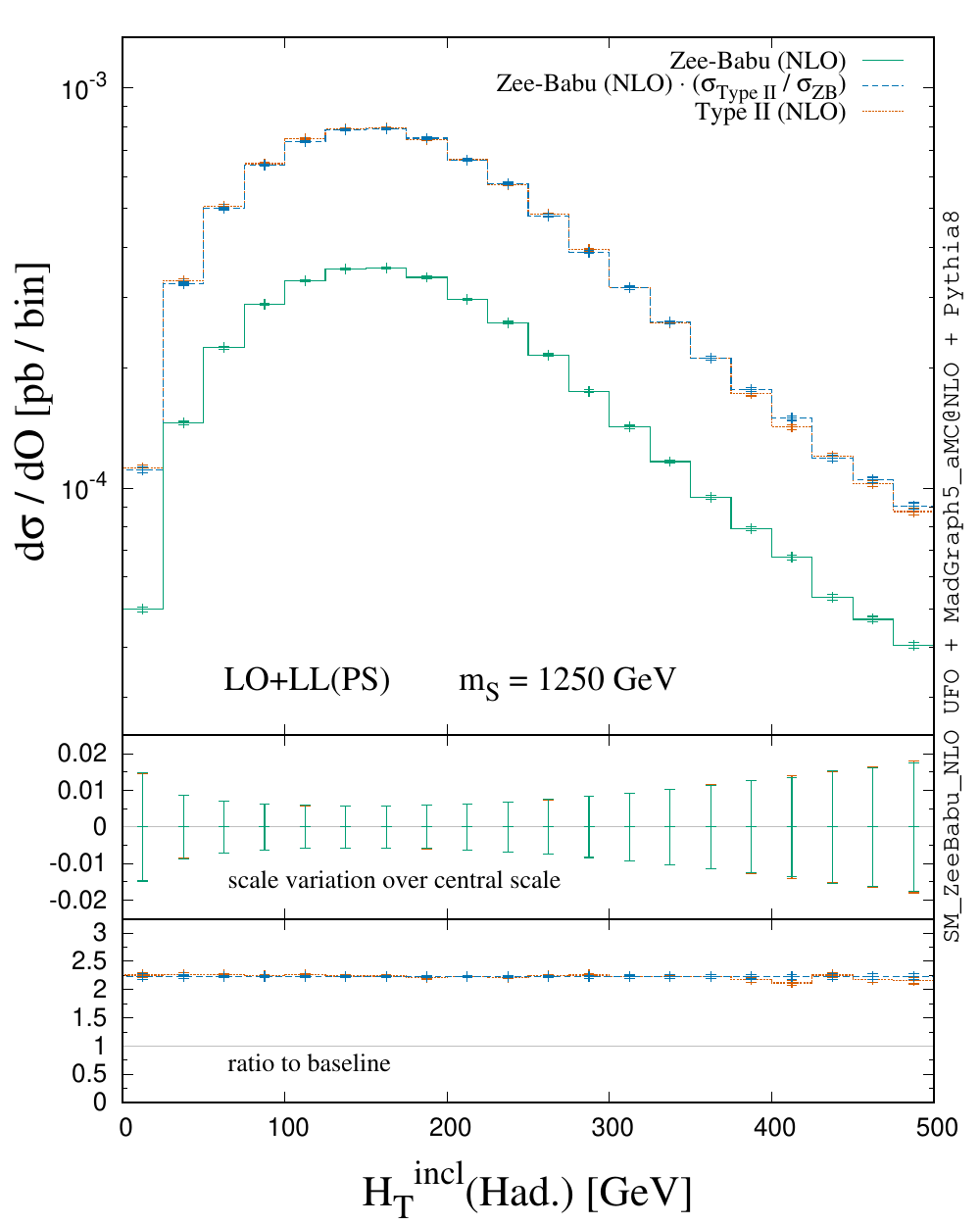}		\label{fig:zeeBabu_kinematics_LHCX13_M1250GeV_HTinclHad}}
\end{center}
\caption{
(a,b) Same as Fig.~\ref{fig:zeebabu_kinematics_loMass} but for shapes $Q_T$ and $H_T^{\rm incl}$(Had.) at LO+LL(PS);
(c,d) Same as (a,b) but for $m_S = 1250\GeV$.
}
 \label{fig:zeebabu_kinematics_ue}
\end{figure}

Notably, $\Delta\Phi_{S\ \rm Had.}$ is only accurate to LO+LL(PS) in our simulations. When the $q\overline{q}\to SS^\dagger$ matrix element is known at LO, $\Delta \Phi_{S\ \rm Had}$ is ill-defined as the $(SS^\dagger)$ system carries no $p_T$. The $p_T$ of the $(SS^\dagger)$ system is generated first at LL accuracy by the parton shower and eventually at LO accuracy by the real radiation correction at NLO in QCD.
Despite this formally lower accuracy, the scale uncertainties  are at the sub-percent level. Again, the bin-by-bin normalizations of the scaled Zee-Babu distribution and Type II distribution are statistically indistinguishable.

In Fig.~\ref{fig:zeeBabu_kinematics_LHCX13_MX500GeV_cosBoost}, 
we plot the polar distribution $(\cos\beta_{SQ})$ 
of $S$ in the frame of the $(SS^\dagger)$ system  relative to the propagation direction of the $(SS^\dagger)$ system in the lab frame. 
Defining $p_S^{\ (SS^\dagger)}$ to be the momentum of $S$ in the $(SS^\dagger)$ frame, the observable is given symbolically by
\begin{align}
 \cos(\beta_{SQ})\ =\ \hat{p}_S^{(SS^\dagger)}\ \cdot \ \hat{q}
 =\  \frac{\vec{p}_S^{\ (SS^\dagger)}\ \cdot \ \vec{q}}{\vert \vec{p}_S^{\ (SS^\dagger)}\vert\ \vert \vec{q} \vert }\ ,
\end{align}
The distributions exhibit NLO+LL(PS) accuracy since $q$ has longitudinal momentum in the lab frame, event at LO. We observe that all three curves obey a $d\sigma\sim(1-\cos^2(\beta_{SQ}))$ distribution. This follows from angular momentum conservation: Imagining the decay of massive, virtual photon $\gamma^*\to SS^\dagger$, the Feynman rules of Eq.~\eqref{eq:lag_kin_ewsb}  indicate that the corresponding helicity amplitude describes a $p$-wave process with $-i\mathcal{M}(\gamma^*_\lambda\to SS^\dagger)\sim \sin(\beta_{SQ})$ for either transverse polarization of $\gamma^*_\lambda$, and where the $\hat{z}$-axis is aligned  with $\hat{q}$.  At the square level, one obtains 
\begin{align}
\sum_{\lambda=\pm}\vert \mathcal{M}_\lambda\vert^2\sim \sin^2(\beta_{SQ}) = 1-\cos^2(\beta_{SQ})\ .
\end{align}
Scale uncertainties reach as large as \confirm{$\pm1.5\%$} in backward $(\cos\beta_{SQ}=-1)$ and forward $(\cos\beta_{SQ}=+1)$ regions, and are below \confirm{$\pm1\%$} in the central region $(\cos\beta_{SQ}=0)$. The bin-by-bin normalizations of the scaled Zee-Babu and Type II distributions are statistically indistinguishable.

In Fig.~\ref{fig:zeebabu_kinematics_hiMass}, we plot the same observables as in Fig.~\ref{fig:zeebabu_kinematics_loMass} for the benchmark $m_S = 1250\GeV$. 
Qualitatively, we find strong similarities between the two mass choices. 
Quantitatively, the absolute normalizations of the differential distributions are smaller than the previous case due to naturally the smaller production cross section. Beyond this, shape broadening or narrowing can be attributed to the larger mass scale. 
For the $p_T$ and $y$ distributions, we find a slightly larger residual scale uncertainty, but also find that uncertainties stay below \confirm{$\pm4\%$}.
Importantly, as in the low-mass case, the Zee-Babu and Type II distributions are statistically indistinguishable.

For completeness, we consider two measures of the hadronic activity in $SS^\dagger$ pair production to demonstrate that the underlying event also remains unchanged between the Zee-Babu and Type II scenarios.
For (a,b) $m_S = 500\GeV$ and (c,d) $m_S=1250\GeV$, we show in Fig.~\ref{fig:zeebabu_kinematics_ue}(a,c) the transverse momentum of the $(SS^\dagger)$ system, $q_T$, and (c,d) the
inclusive $H_T$, defined by
\begin{align}
H_T^{\rm incl}(\text{Had.}) = \sum_{i\in\{\text{Had.}\}} p_{T}^i, 
\quad\text{with}\quad \vert y^i \vert < y^{\max}\ ,
\end{align}
which is built directly from hadrons. Despite being LO+LL(PS) accurate, scale uncertainties reach only $\pm2\%$. As before, the scaled Zee-Babu and Type II distributions are  indistinguishable.


\subsection{Limits and projections for the LHC}\label{sec:typeii_limits}

As demonstrated above, total and differential cross sections for pair production of charged scalars  in the Types II and Zee-Babu models differ at most by an overall normalization. Consequentially, their decay products will inherit this sameness and also exhibit nearly identical kinematics.

\begin{figure}[t!]
\begin{center}
\subfigure[]{\includegraphics[width=0.485\textwidth]{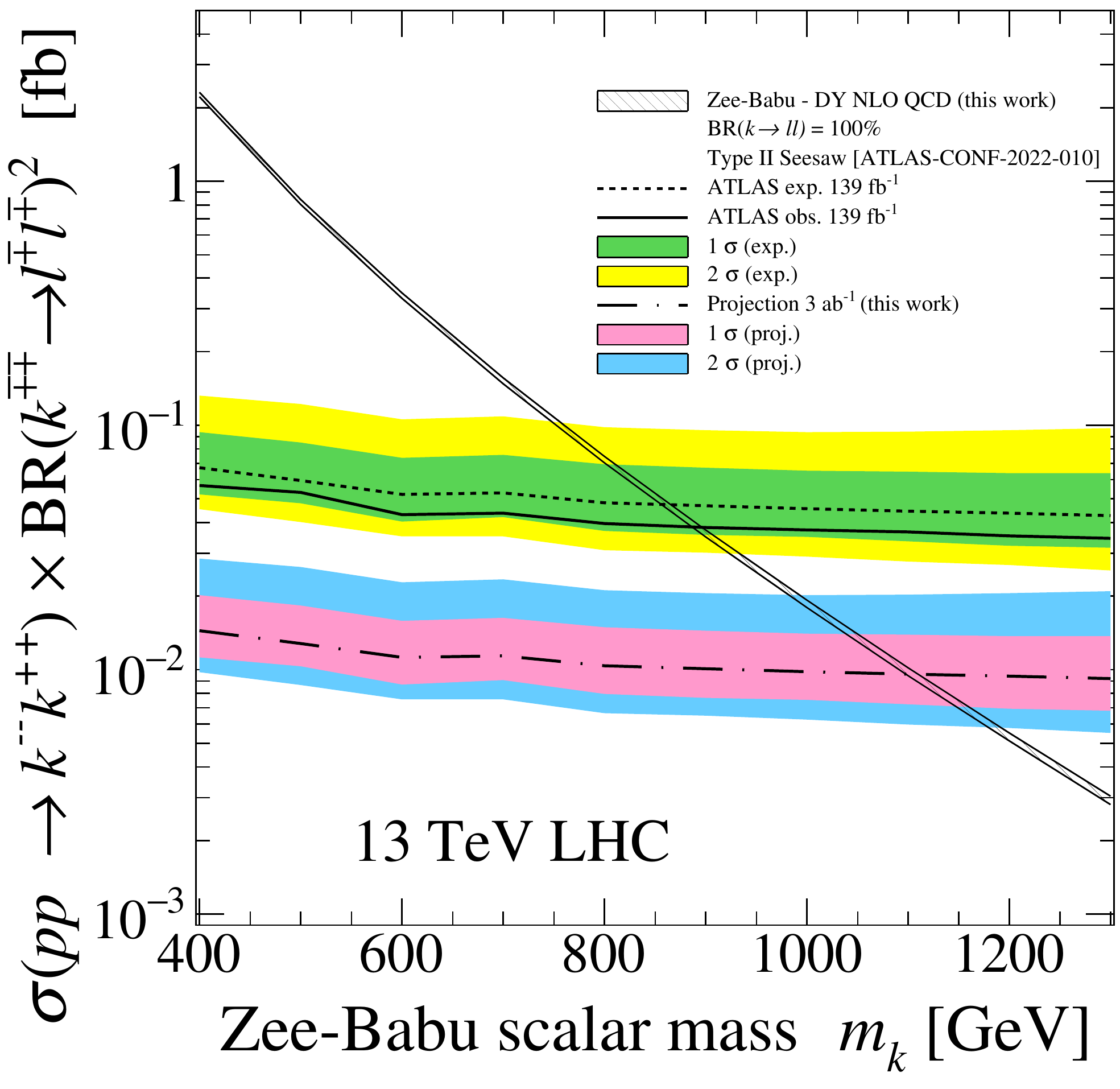} \label{fig:zeeBabu_ATLAS_Excl_vs_Mass}}
\subfigure[]{\includegraphics[width=0.485\textwidth]{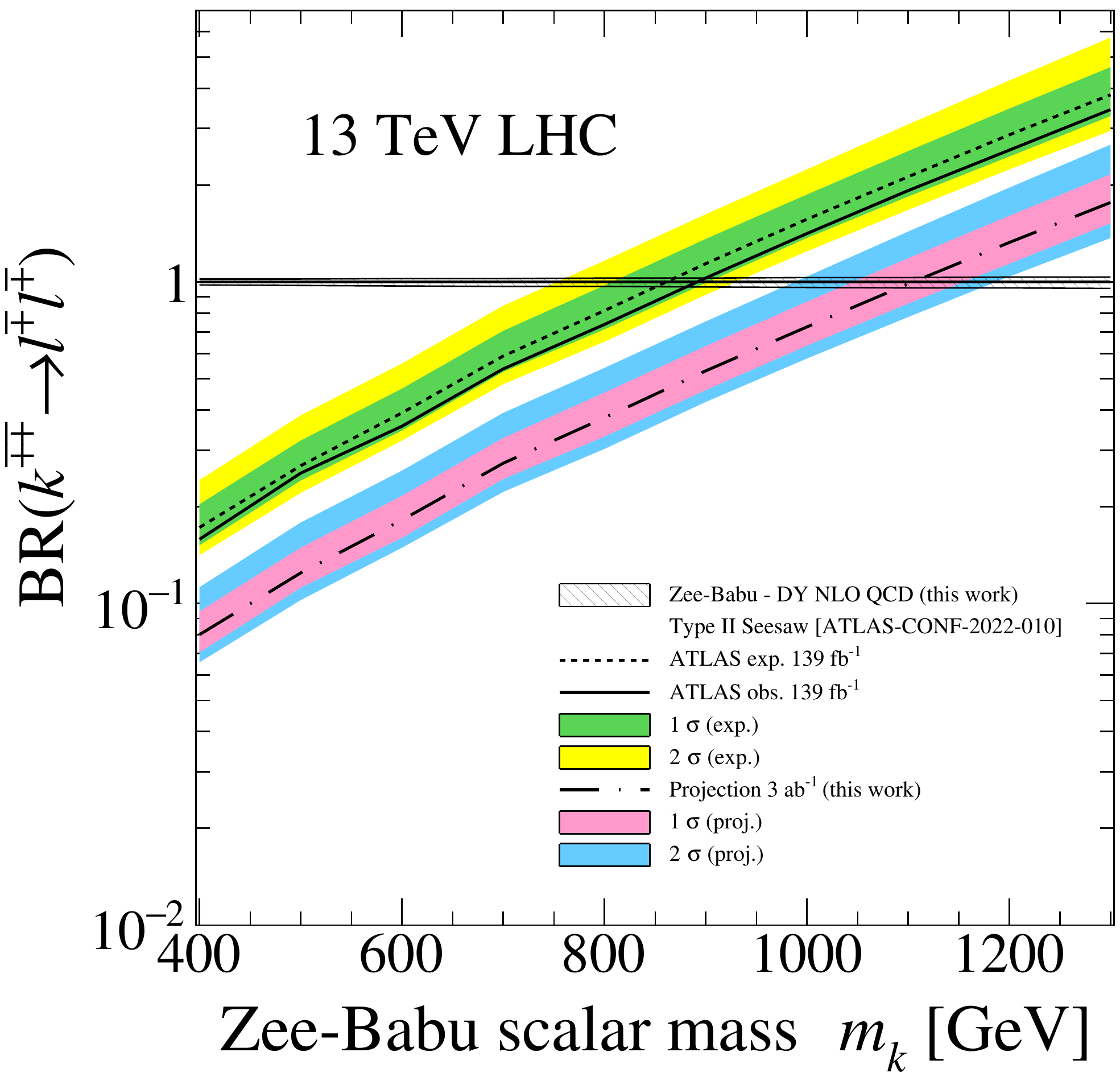} \label{fig:zeeBabu_ATLAS_BRexcl_vs_Mass}}
\end{center}
\caption{
(a) Estimated and projected cross section limits at 95\% CL on $k^{--}k^{++}\to4\ell$ production, $\ell\in\{e,\mu\}$, in the Zee-Babu model at the $\sqrt{s}=13\TeV$ LHC with $\mathcal{L}\approx139\invfb$ and $\mathcal{L}\approx3\invab$ of data, respectively, as derived from constraints on $\Delta^{++}\Delta^{--}\to 4\ell$ production in the Type II Seesaw 
by the ATLAS experiment with $\mathcal{L}\approx139\invfb$~\cite{ATLAS:2022yzd}. 
(b) Same as (a) but for the $k^{\mp\mp}\to\ell^\mp\ell'^\mp$ branching rate. 
}
 \label{fig:ZeeBabu_ATLAS_Excl_vs_Mass}
 \end{figure}

Despite this hardship, there a silver lining of this sameness: the selection $(\varepsilon)$ and acceptance $(\mathcal{A})$ efficiencies obtained by LHC experiments in searches for charged scalars in the Type II Seesaw are automatically applicable to the Zee-Babu model. This is a nontrivial conclusion. It implies that for common final states the two models can be tested simultaneously at the LHC without additional event generation or additional signal/control/validation-region modeling.

Normally, new signal events must be simulated to reinterpret or recast a collider analysis for one scenario in terms of a second scenario. That is not needed for the charged scalars in the Zee-Babu and Type II models. For example: suppose that after all acceptance and selection requirements the total number of Type II events at a given integrated luminosity $\mathcal{L}$ is
\begin{align}
 n_{\rm Type~II}\ =\ \varepsilon\ \times \mathcal{A}\ \times \mathcal{L}\ \times \sigma_{\rm Type~II}^{\rm DY~(NLO)}\ .
\end{align}
Since the selection and acceptance efficiencies are the same, which follows from final states having identical kinematics, the corresponding number of Zee-Babu events is 
\begin{align}
    n_{\rm ZB}\ = \varepsilon\ \times \mathcal{A}\ \times \mathcal{L}\ \times \sigma_{\rm ZB}^{\rm DY~(NLO)}
    = 
    n_{\rm Type~II}\ \times \left(\frac{\sigma_{\rm ZB}^{\rm DY~(NLO)}}{\sigma_{\rm Type~II}^{\rm DY~(NLO)}}\right) \ .
    \label{eq:nsig_ratio}
\end{align}
One can identify the cross-section ratio in Eq.~\eqref{eq:nsig_ratio} as the 
scale factor $\xi(m_S)$ in Eq.~\eqref{eq:scale_factor_def}.
Since the final states are assumed to be the same, the backgrounds are the same. And after acceptance and selection cuts, the number of background events are the same. Subsequently, the upper limit on a cross section derived for one scenario is the same as for the other. The specific parameter space excluded is, of course, different for the two scenarios.

\begin{table}[t!]
\begin{center}
\resizebox{\columnwidth}{!}{
\begin{tabular}{c | c c c c c c c c c c}
\hline\hline
\multicolumn{10}{c}{BR$(k^{\mp\mp}\to \ell^\mp\ell'^\mp)$}
\\
\hline
mass of $k^{\mp\mp}$ [GeV]
	& $400$		& $500$		& $600$		& $700$		& $800$		& $900$		& $1000$		& $1100$		& $1200$		& $1300$	\\ 
 \hline 
 BR$^{\rm obs.}_{\rm 95\%}(139\invfb)$
	& 0.158	& 0.254	& 0.356	& 0.536	& 0.739	& 1.028	& 1.413	& 1.926	& 2.574	& 3.434\\ 
 \hline 
 BR$^{\rm proj.}_{\rm 95\%}(3\invab)$
	& 0.080	& 0.125	& 0.182	& 0.273	& 0.378	& 0.528	& 0.726	& 0.985	& 1.329	& 1.770\\ 
\hline\hline
\end{tabular}
} 
\caption{
For representative $m_k$ (top), estimated (middle) and projected (bottom) limits on the decay rate of $k^{\mp\mp}\to \ell^\mp\ell'^\mp$ set by ATLAS at 95\% CL using $\mathcal{L}\approx139\invfb$~\cite{ATLAS:2022yzd}
and $\mathcal{L}=3\invab$ of data.
}
\label{tab:zeeBabuATLAS}
\end{center}
\end{table}

To demonstrate this we consider the search  for $\Delta^{--}\Delta^{++}$ pairs in the $4\ell^\pm$ channel,
\begin{align}
pp\ \to\ \Delta^{++}\Delta^{--}\  \to\ \ell^+ \ell'^+ \ell''^- \ell'''^-\ 
\end{align}
where $\ell\in\{e,\mu\}$, 
by the ATLAS experiment using the full Run II data set~\cite{ATLAS:2022yzd}. In Ref.~\cite{ATLAS:2022yzd}, a mass-dependent 95\% confidence level (CL) limit $(\sigma^{\rm ATLAS}_{95\%\ \rm CL})$ is reported on the (unfolded) quantity
\begin{align}
  \sigma(pp\to \Delta^{--}\Delta^{++})& \times \text{BR}^2(\Delta^{\pm\pm} \to \ell^\pm\ell'^\pm) \ .
  \label{eq:def_xsec_limit_typeII}
\end{align}
Applying the same limit to the (unfolded) quantity in the Zee-Babu model
\begin{align}
  \sigma(pp\to k^{--}k^{++})& \times \text{BR}^2(k^{\pm\pm} \to \ell^\pm\ell'^\pm) \ ,
  \label{eq:def_xsec_limit_zb}
\end{align}
we obtain the limit shown in Fig.~\ref{fig:ZeeBabu_ATLAS_Excl_vs_Mass}.
Assuming branching rates of unity, $k^{\mp\mp}$ with masses
\begin{align}
m_k < {890\GeV} 
\end{align}
are excluded by ATLAS at 95\% CL using approximately 
$\mathcal{L}\approx139\invfb$ of data at $\sqrt{s}=13\TeV$.
Assuming that sensitivity $\sigma^{\rm ATLAS}_{95\%\ \rm CL}$ scales with the square root of integrated luminosity, i.e.,
\begin{align}
 \sigma^{\rm ATLAS}_{95\%\ \rm CL} \Big\vert_{\mathcal{L}_2} = \sqrt{\frac{\mathcal{L}_1}{\mathcal{L}_2}} \times \sigma^{\rm ATLAS}_{95\%\ \rm CL} \Big\vert_{\mathcal{L}_1},
\end{align}
and unchanged selection and acceptance efficiencies, we also show the projected sensitivity with $\mathcal{L}\approx3\invab$ of data at $\sqrt{s}=13\TeV$. 
For branching rates of unity, $k^{\mp\mp}$ with masses 
\begin{align}
m_k < {1110\GeV} 
\end{align}
can be excluded at 95\% CL.
This sensitivity can be improved with higher collider energies,
larger data sets, and improved analysis techniques.
(This would equally benefit searches for $\Delta^{\mp\mp}$.)

Invert Eq.~\eqref{eq:def_xsec_limit_zb}, we obtain limits on the $k^{\mp\mp}\to \ell^\mp\ell'^\mp$ branching rate.
This is given by
\begin{align}
  \text{BR}(k^{\pm\pm} \to \ell^\pm\ell'^\pm)\ <\ 
  \sqrt{
  \frac{\sigma^{\rm ATLAS}_{95\%\ \rm CL}}{\sigma(pp\to k^{--}k^{++})}
  }\ .
\end{align}
In Fig.~\ref{fig:zeeBabu_ATLAS_BRexcl_vs_Mass} we show the estimated and projected limits 
for  $\mathcal{L}\approx139\invfb$ and $\mathcal{L}=3\invab$.
As summarized in Table~\ref{tab:zeeBabuATLAS}, decay rates as small as $16\%~(74\%)$ can be probed for $m_k = 400~(800)\GeV$ with the full Run II data set. Tentatively, this can be improved two-fold at the HL-LHC.


\subsection{Correlations in flavor-violating decays of singly charged scalars}\label{sec:typeii_decay}

We now discuss a consequence of the relationships between Yukawa couplings $f_{\ell\ell'}$ and oscillation parameters as 
summarized in Eqs.~\eqref{eq:yukawa_ratio_no} and \eqref{eq:yukawa_ratio_io}.
As discussed in Secs.~\ref{sec:pheno_decay} and \ref{sec:typeii_normalization}, $k$ and $h$ are shortly lived and decay readily to SM particles. In principle, production rates themselves are not observed at the LHC but rather the combination of production and decay rates. For the production of $h^-h^+$ pairs, which preferably decay via the $h^\mp\to\ell^\mp\nu_{\ell'}$ channel, neutrinos cannot be flavor tagged at the LHC. One can only identify the charged lepton $\ell^\mp$. Subsequently, measurements of the $pp \to h^-h^+ \to  \ell^-\ell'^+\nu_{\ell''}\nu_{\ell'''}$ process implicitly sum over all neutrino states.

Guided by this, we define the following pair of inclusive partial widths for $h$:
\begin{subequations}
 \begin{align}
  \Gamma(h^\mp \to e^\pm \nu_X)     &= \sum_{\ell=e}^\tau\ \Gamma(h^\mp \to e^\pm \nu_\ell) \ ,
  \\
  \Gamma(h^\mp \to \mu^\pm \nu_X)   &= \sum_{\ell=e}^\tau\ \Gamma(h^\mp \to \mu^\pm \nu_\ell) \ . 
 \end{align}
\end{subequations}
Recalling that $f_{ee}, f_{\mu\mu}=0$, and neglecting $\mathcal{O}(m_\ell^2/m_h^2)$ terms in expressions for total widths gives
\begin{align}
\label{eq:branching_ratio_full}
 \mathcal{R}^h_{e\mu} &= \frac{{\rm BR}(h^\mp \to e^\pm \nu_X)}{{\rm BR}(h^\mp \to \mu^\pm \nu_X)}
 = \frac{\vert f_{e\mu}\vert^2 + \vert f_{e\tau}\vert^2}{\vert f_{e\mu}\vert^2 + \vert f_{\mu\tau}\vert^2}
= \cfrac{\vert \frac{f_{e\mu}}{f_{\mu\tau}}\vert^2 + \vert \frac{f_{e\tau}}{f_{\mu\tau}}\vert^2}{\vert \frac{f_{e\mu}}{f_{\mu\tau}}\vert^2 + 1}\ ,
\end{align}
which is the $e$-over-$\mu$ branching ratio for $h$.
Inserting Eqs.~\eqref{eq:yukawa_ratio_no} and \eqref{eq:yukawa_ratio_io}, we obtain
\begin{subequations}
\label{eq:branching_ratio_by_ordering}
\begin{align}
  \mathcal{R}^h_{e\mu}\Big\vert_{\rm NO} &= \frac{2(A^2+B^2)}{1+A^2+B^2-2AB\cos(\delta_{\rm CP})}\ ,
  \\
 A &= \cos(\theta_{23}) \sec(\theta_{13}) \tan(\theta_{12}),
 \quad
 B = \sin(\theta_{23}) \tan(\theta_{13}),
  \\
  \mathcal{R}^h_{e\mu}\Big\vert_{\rm IO} &= \frac{2}{2+\cos(2\theta_{13})+\cos(2\theta_{23})} 
  \ ,
\end{align}
\end{subequations}
for the NO and IO of neutrino masses. These simple, analytic expressions are a direct consequence of Eqs.~\eqref{eq:yukawa_ratio_no} and \eqref{eq:yukawa_ratio_io} but have not previously been  reported in the literature.

The utility of the branching ratio $\mathcal{R}$, as oppose to the branching rate BR, is its independence of $h$'s total width. Unlike individual branching rates, the ratio $\mathcal{R}^h_{e\mu}$ is measurable at the LHC. It can be determined, for example, by comparing the event numbers $(N)$ in the processes:
\begin{subequations}
\begin{align}
 p p &\to \gamma^*/Z^* \to h^+ h^- \to e^+ e^- + \not\!\! E_T \ ,
 \\
 p p &\to \gamma^*/Z^* \to h^+ h^- \to \mu^+ \mu^- + \not\!\! E_T \ ,
\end{align}
\end{subequations}
where $\not\!\!\!E_T$ is the missing transverse momentum of the event.
Accounting for selection and acceptance efficiencies, integrated luminosity $(\mathcal{L})$, 
the hadronic cross section $\sigma_{\ell\ell'}$,
and assuming the narrow width approximation, 
then the branching ratio (squared) is obtained from the ratio
\begin{align}
 \frac{N_{ee}}{N_{\mu\mu}} = 
 \frac{\mathcal{L}\times  \sigma(pp \to h^+ h^- \to e^+ e^- + \not\!\! p_T)}{\mathcal{L}\times  \sigma(pp \to h^+ h^- \to \mu^+ \mu^- + \not\!\! p_T)}
 = 
 \frac{{\rm BR}(h^\mp\to e^\mp \nu_X)^2}{{\rm BR}(h^\mp\to \mu^\mp \nu_X)^2} 
 =
 \left(\mathcal{R}^h_{e\mu}\right)^2 
 \ .
\end{align}
$\mathcal{R}^h_{e\mu}$ can also be extracted from the ratio of $(e^+ \mu^-+ \not\!\! p_T)$ and $(\mu^+\mu^-+ \not\!\! p_T)$ events. 
The $e$-over-$\tau$ and $\mu$-over-$\tau$ branching ratios can also be constructed in the same manner.

\begin{table}[t!]
\begin{center}
\resizebox{.65\columnwidth}{!}{
\npdecimalsign{.}
\nprounddigits{1}
\begin{tabular}{c | c | n{3}{2} | n{2}{2} | n{3}{2} | n{3}{2} }
\hline\hline
mass ordering & extremum\ & 
\multicolumn{1}{c|}{$\theta_{12}$} & 
\multicolumn{1}{c|}{$\theta_{13}$} & 
\multicolumn{1}{c|}{$\theta_{23}$} & 
\multicolumn{1}{c}{$\delta_{\rm CP}$}
\\
\hline
\multirow{2}{*}{NO} & max & 35.86 & 8.96997 & 39.5 & 105
\\
                    & min & 31.27 & 8.20004 & 52.0 & 180
\\
\hline
\multirow{2}{*}{IO} & max &  & 8.97997 & 52.1 & 
\\
                    & min &  & 8.24007 & 39.8  &
\\
\hline\hline
\end{tabular}
\npnoround
} 
\caption{For the normal (NO) and inverse (IO) order of neutrino masses, the values of oscillation angles and phase needed to maximize or minimize the branching ratio $\mathcal{R}^h_{e\mu}$.
}
\label{tab:branchingRatio_nuInputs}
\end{center}
\end{table}

For the NO and IO scenarios, and using the central values and $\pm3\sigma$ ranges of Ref.~\cite{Esteban:2020cvm}
(NuFIT 5.1-without SK-atm), we find that neutrino oscillation data predict the branching ratios
\begin{align}
  \mathcal{R}^h_{e\mu}\Big\vert_{\rm NO} = 0.313\ ^{+55\%}_{-20\%}
  \quad\text{and}\quad
  \mathcal{R}^h_{e\mu}\Big\vert_{\rm IO} = 0.715\ ^{+3\%}_{-11\%}\ .
\end{align}
Uncertainties are obtained by varying each oscillation parameter over its
allowed $3\sigma$ range, as obtained by Ref.~\cite{Esteban:2020cvm}, and extracting the local maxima and minima. The $^{+55\%}_{-20\%}$ and $^{+3\%}_{-11\%}$ uncertainties for the NO and IO cases are the full $3\sigma$ windows. They correspond to the ranges:
 \begin{align}
  \mathcal{R}^h_{e\mu}\Big\vert_{\rm NO} \in \ [0.251,0.484] \quad\text{at}\ 3\sigma
  \quad\text{and}\quad
  \mathcal{R}^h_{e\mu}\Big\vert_{\rm IO} \in \ [0.637,0.739]\quad\text{at}\ 3\sigma\ .
 \end{align}

The relative smallness of the IO's uncertainty is due to its dependence on only two oscillation angles, one of which $(\theta_{13})$ is well measured. Fixing either $\theta_{12}$ or $\theta_{23}$ in the NO formula to its central value roughly halves the uncertainty.
Given the certainty in oscillation parameters, the predictions are sufficiently robust to discriminate between  the NO and IO.
For completeness, we report in Table~\ref{tab:branchingRatio_nuInputs} 
the oscillation angles and phase at which the $\mathcal{R}_{e\mu}^h$ are maximized/minimized.

\paragraph*{New correlations in low-energy transitions from oscillation data:}
As stipulated in Sec.~\ref{sec:pheno}, a comprehensive discussion on the phenomenology of the Zee-Babu model in low-energy processes is outside the scope of this study.
(These can be found in Refs.~\cite{Nebot:2007bc,Ohlsson:2009vk,Schmidt:2014zoa,Babu:2019mfe,Okada:2021aoi}.)
Nevertheless, in light of
Eqs.~\eqref{eq:branching_ratio_full} and 
\eqref{eq:branching_ratio_by_ordering}, 
which have not previously been reported, 
it is worth  commenting briefly that 
new connections can also be established between 
flavor-violating, low-energy transitions and 
oscillation data.

For instance: 
in the Zee-Babu model, $\ell\to\ell'\nu\nu$ decays are additionally mediated at tree-level by the charged scalar $h^\pm$ and the $f_{\ell\ell'}$ couplings.     Experimentally, this manifests as a Fermi constant $(G_F^{\rm ZB})$ that is shifted from its SM value $(G_F)$.
Explicitly, this shift is given by~\cite{Bertolini:1987kz,Nebot:2007bc}
\begin{align}
 \left(\frac{G_{F}^{\rm ZB}\vert_{\ell\to\ell'}}{G_F\vert_{\ell\to\ell'}}\right)^2 = 1 + \frac{\sqrt{2}}{m_h^2 G_F}\vert f_{\ell\ell'}\vert^2 + \mathcal{O}\left(\frac{1}{m_h^4 G_F^2}\right)
 \ ,
\end{align}
where $G_F$ is assumed to be extracted from, e.g., decays of hadrons.
This result holds so long as $m_h$ remains large compared to the vev of the SM Higgs. The extracted Fermi constant from various    $\ell\to\ell'\nu\nu$ decay channels then isolates the antisymmetric couplings:
\begin{subequations}
\begin{align}
 \left(\frac{G_{F}^{\rm ZB}\vert_{\tau\to\mu}}{G_F\vert_{\tau\to\mu}}\right)^2 
 -
 \left(\frac{G_{F}^{\rm ZB}\vert_{\tau\to e}}{G_F\vert_{\tau\to e}}\right)^2 &
 =
 \frac{\sqrt{2}}{m_h^2 G_F}
 \left(
 \vert f_{\mu\tau}\vert^2-\vert f_{e\tau}\vert^2
 \right) 
 + 
 \mathcal{O}\left(\frac{1}{m_h^4 G_F^2}\right)\ ,
 \\
  \left(\frac{G_{F}^{\rm ZB}\vert_{\tau\to\mu}}{G_F\vert_{\tau\to\mu}}\right)^2 
 -
 \left(\frac{G_{F}^{\rm ZB}\vert_{\mu\to e}}{G_F\vert_{\mu\to e}}\right)^2 &
 =
  \frac{\sqrt{2}}{m_h^2 G_F}
 \left(
 \vert f_{\mu\tau}\vert^2-\vert f_{e\mu}\vert^2
 \right) 
 + 
 \mathcal{O}\left(\frac{1}{m_h^4 G_F^2}\right)\ .
\end{align}
\end{subequations}

Taking this a step further, the ratio of these differences gives the ratio of $f_{\ell\ell'}$ couplings:
\begin{align}
 \mathbb{G}^{(\tau\to\mu),(\tau\to e)}_{(\tau\to\mu),(\mu\to e)} 
\equiv \frac{ \left(\cfrac{G_{F}^{\rm ZB}\vert_{\tau\to\mu}}{G_F\vert_{\tau\to\mu}}\right)^2 
 -
 \left(\cfrac{G_{F}^{\rm ZB}\vert_{\tau\to e}}{G_F\vert_{\tau\to e}}\right)^2 
}{ \left(\cfrac{G_{F}^{\rm ZB}\vert_{\tau\to\mu}}{G_F\vert_{\tau\to\mu}}\right)^2 
 -
 \left(\cfrac{G_{F}^{\rm ZB}\vert_{\mu\to e}}{G_F\vert_{\mu\to e}}\right)^2}
&=
\frac{\vert f_{\mu\tau}\vert^2-\vert f_{e\tau}\vert^2}{\vert f_{\mu\tau}\vert^2-\vert f_{e\mu}\vert^2}
 + 
 \mathcal{O}\left(\frac{1}{m_h^4 G_F^2}\right)
\\
&=
\frac{1-\vert \frac{f_{e\tau}}{f_{\mu\tau}}\vert^2}{1-\vert \frac{f_{e\mu}}{f_{\mu\tau}}\vert^2}
 + 
 \mathcal{O}\left(\frac{1}{m_h^4 G_F^2}\right)\ .
\end{align}
Once again, using Eqs.~\eqref{eq:yukawa_ratio_no} and \eqref{eq:yukawa_ratio_io} for NO and IO, respectively, one obtains correlations for flavor-violating transitions that are fixed by oscillation data.
Again, these expressions have not  previously been reported in the literature.
Such correlations can also be established in $\ell\to\ell'\gamma$ transitions. 
However, further investigations, including numerical studies, are left to future work.


\subsection{Establishing non-conservation of lepton number at the LHC}\label{sec:typeii_lnv}

If neutrino masses are generated through the Zee-Babu mechanism~\cite{Zee:1985rj,Zee:1985id,Babu:1988ki}, then neutrinos are Majorana fermions and LN is not conserved in scattering and decay processes. Whether this can be established through low-energy experiments, such as neutrinoless $\beta\beta$ decay, it is paramount to establish if LN is violated at other energies. Among other reasons, 
LNV is predicted in many neutrino mass models and
correlating LNV across the different scales can provide critical guidance on the underlying theory.
We now discuss a strategy for establishing LNV in the Zee-Babu model at the LHC. For reference, we start an analogous strategy in the Type II Seesaw.

\paragraph*{Establishing LNV with Type II scalars:}
Under conventional quantum number assignments, LN is broken explicitly 
in the scalar potential of both the Type II and Zee-Babu models by a dimensionful parameter $\mu_{\not L}$.
In both scenarios, the Yukawa couplings to exotically charged scalars are LN conserving. After EWSB, the triplet scalars of the Type II Seesaw acquire a vev $v_\Delta$ proportional to $\mu_{\not L}$, which manifests in the three-point $\Delta-\Delta-W$ coupling.

To establish LNV in the Type II Seesaw, one can search for the following LHC processes
\begin{subequations}
\begin{align}
 pp\ &\to\ \Delta^{++}\Delta^{--}\  \to\ \ell^+ \ell'^+ \ell''^- \ell'''^-\ ,
 \\
 pp\ &\to\ \Delta^{++}\Delta^{--}\  \to\ W^{+} W^{+}W^-W^-\ 
 \to\ 
 \ell^\pm \ell'^\pm j j j j\  +  \not\!\!E_T\ ,
\end{align}
\end{subequations}
and work by contradiction.
The argument goes as follows:
Assuming that LN is conserved, the four-lepton channel establishes that the $\Delta^{\mp\mp}$ states each carry $L=\pm2$ since each $\ell^\mp$ carries $L=\pm1$. 
The second channel is mostly reconstructable, particularly if employing techniques pioneered for extracting neutrino momenta in leptonic decays of top quark pairs~\cite{Dalitz:1991wa,Sonnenschein:2005ed,Sonnenschein:2006ud}. The channel establishes that the $\Delta^{\mp\mp}$ states carry of $L=0$ since $W^\pm$ carry $L=0$. This leads to a contradiction that LN is conserved. (Technically, the $v_\Delta$ in the $W-W-\Delta$ vertex carries away $\vert L\vert =2$.) Along these lines, the Zee-Babu model does not contain the $W-W-S$ vertex.
Therefore, observing the channel  would signal that the Zee-Babu model is not realized.

\paragraph*{Establishing LNV with Zee-Babu scalars:}
The above strategy does not apply to the Zee-Babu model since neither $k$ nor $h$ couples to the $W$. And based on the size of $\mu_{\not L}$, LN-violating processes may be inaccessibly at the LHC even if possible elsewhere.
Assuming relevant decay processes are accessible, one can establish LNV by 
searching for the following LHC processes
\begin{subequations}
 \begin{align}
 \label{eq:lnv_zb_4l}
  pp &\to k^{--}k^{++}\  \to\ \ell^+ \ell'^+ \ell''^- \ell'''^-\ ,
 \\
 \label{eq:lnv_zb_2lmet}
  pp &\to h^{-}h^{+}\ \quad \  \to\ \ell^+ \ell'^- + \not\!\!E_T\ ,
  \\
 pp &\to k^{--}k^{++}\  \to\ h^{-} h^{-}h^+h^+\ 
 \to\  \ell^- \ell'^- \ell''^+ \ell'''^+ +  \not\!\!E_T\ . \label{eq:lnv_zb_4lmet}
 \end{align}
\end{subequations}
The argument, which also works by contradiction, goes as follows:
Suppose LN is conserved.
Observing the four-lepton channel (Eq.~\eqref{eq:lnv_zb_4l}) establishes that each $k^{\mp\mp}$ carries $L=\pm2$ since each $\ell^\mp$ carries $L=\pm1$.
Observing the opposite-sign dilepton channel (Eq.~\eqref{eq:lnv_zb_2lmet}) establishes that $h^\mp$ carries $L=0$ or $L=\pm2$.
This assertion requires a few clarifying remarks.

First, the signature $pp\to\ell^+ \ell'^- \not\!\!\!E_T$ features a large multi-boson and top quark background, and will be difficult to observe. However, considering the extent to which SM simulations at NNLO+LL(PS) can describe data~\cite{Gehrmann:2014fva,Kallweit:2020gva,Lombardi:2021rvg,Mazzitelli:2021mmm}, 
the presumable mass difference between SM and Zee-Babu particles~\cite{Nebot:2007bc,Ohlsson:2009vk},
and prospective search strategies~\cite{delAguila:2013yaa,delAguila:2013mia,Fuks:2019iaj}, we premise this is attainable.
Second, it is possible to show that the signature's $\not\!\!\!E_T$  is driven by two neutrinos since
(a) the cross-section ratio of $pp\to\ell^+ \ell'^- \not\!\!E_T$ with respect to different lepton flavors is fixed by oscillation data
(see Sec.\ref{sec:typeii_decay}),
and (b) the kinematic distributions of $\ell$, $\ell'$, and $\not\!\!E_T$ are constrained since $h^\mp \to \ell^\mp\nu$ is a two-body decay involving (approximately) two massless states.
Therefore, extending NNLO+LL(PS) technology for $pp\to W^+W^-\to\ell^+ \ell'^- \not\!\!\!E_T$ in the SM~\cite{Gehrmann:2014fva,Kallweit:2020gva,Lombardi:2021rvg} to $h^-h^+$ pair production provides a (limited) means of checking whether the $\not\!\!E_T$ is driven heavier states, some light dark-sector fermion, or by more than two states. 
Failure to satisfy (a) and (b) would suggest that the Zee-Babu model is not realized. Even if satisfied, one can only assert that each $h^\mp$ carries $L=0$ or $L=\pm2$ since it is impossible to check the LN of outgoing neutrinos.

Finally, since each $h^\mp$ carries $L=0$ or $L=\pm2$, observing the four-lepton and $\not\!\!\!E_T$ channel (Eq.~\eqref{eq:lnv_zb_4lmet}), which again can be checked using kinematic distributions and ratios of cross sections, establishes that each $k^{\mp\mp}$ carries $L=0$ or $L=\pm4$. This is in contradiction with the four-lepton channel (Eq.~\eqref{eq:lnv_zb_4l}), which establishes $L_k=\pm2$, and implies that LN is not conserved. In the event that $m_h > m_k$, then the kinematically suppressed channel
\begin{align}
  pp &\ \to\ h^{-} h^+\ \to\ k^{--}h^{+*}h^{-*}k^{++}
   \to\ 6\ell +  \not\!\!\!E_T\ ,
\end{align}
shows that LN cannot be conserved since the $h^\mp \to k^{\mp\mp} h^{\pm*}$ splitting suggests $L_k = 0$ or $\pm 4$.


\section{Summary and Conclusions}\label{sec:conclude}

With widely felt impact in nuclear physics, astrophysics, and cosmology, the origin of neutrinos' tiny masses and large mixing is among the most pressing mysteries in particle physics today. Establishing whether neutrinos are their own antiparticles, implying that LN is not conserved, is also fundamental to model building. Naturally, there are numerous models of increasing complexity that answer these questions and are also testable at ongoing and near-future experiments.

Among these scenarios are the Type II Seesaw and Zee-Babu models for neutrino masses, which, less commonly, can reproduce oscillation data without invoking sterile neutrinos.
Both scenarios hypothesize the existence of exotically charged scalars that couple directly to the SM Higgs and SM gauge bosons, and therefore can be produced copiously at the LHC if kinematically accessible. In this study, we have revisited the phenomenology of the Zee-Babu model (Sec.~\ref{sec:pheno}) and focused (Sec.~\ref{sec:typeii}) on the ability to distinguish singly and doubly charged scalars from the two models at the LHC.
We conclude that this task is much more difficult than previously believed.

After reviewing the tenets of the Zee-Babu model (Sec.~\ref{sec:theory_model}),
and after presenting updated cross section predictions for $k^{--}k^{++}$ and $h^-h^+$ production at the LHC through various mechanisms up to NLO in QCD (Sec.~\ref{sec:pheno_lhc}),
we compared total (Sec.~\ref{sec:typeii_normalization}) and differential (Sec.~\ref{sec:typeii_kinematics}) predictions for the Zee-Babu and Type II Seesaw models.
All inputs equal, we find that total and differential rates for producing pairs of doubly and singly charged scalars are identical in shape and differ by a normalization equal to the ratio of hadronic cross sections, which can be unity. 
This holds for the Drell-Yan, $\gamma\gamma$ fusion, and $gg$ fusion, as well as observables at LO+LL(PS) and NLO+LL(PS) in QCD. Importantly, the differences in normalizations are sufficiently small that they can be hidden by unknown branching rates or unknown couplings to the SM Higgs. This similarity allows us to reinterpret LHC constraints and projected sensitivity on doubly charged scalars decaying to leptons from the Type II Seesaw in terms of the Zee-Babu model (Sec.~\ref{sec:typeii_limits}).

\paragraph*{Outlook:}
Despite potential hardships, there is some guidance on distinguishing the two models:
Unlike the Type II Seesaw, 
the Zee-Babu model predicts one massless neutrino,
and therefore features a clearer prediction for the rate of
neutrinoless $\beta\beta$ decay.
Aside from this, charged scalars in the Zee-Babu model do not couple to the $W$ boson at tree level. This means that the Type II Seesaw predicts several associated production channels at the LHC not found in the Zee-Babu model. 
If the Zee-Babu model is realized by nature, then these channels are absent.
Furthermore, neutrino oscillation parameters are now sufficiently precise to make clear predictions  
(Sec.~\ref{sec:typeii_decay}) for branching ratios, i.e., ratios of branching rates, of charged scalars in the Zee-Babu model. Such observables are less sensitive to unknown decay rates of charged scalars and are presented for the first time in Sec.~\ref{sec:typeii_decay}.
Similarly, new correlations between oscillation data and searches for lepton flavor violation at low-energy experiments, such as $\ell\to\ell'\nu\nu$ decays, can also be established. 
Finally, the inherent differences in the two models require different strategies for establishing LNV at the LHC (Sec.~\ref{sec:typeii_lnv}). Finally, it is also possible that NLO in EW corrections to the production rates of charged scalars at hadron colliders can help break degenerate predictions. In all these directions we encourage and anticipate future exploration.


\appendix


\section{Cross section normalizations at 13, 14, and 100 TeV}

In the following tables, we list cross sections at $\sqrt{s}=13$ (Tables~\ref{tab:app_xsec_LHCX13_kk} and \ref{tab:app_xsec_LHCX13_hh}), 14 (Tables~\ref{tab:app_xsec_LHCX14_kk} and \ref{tab:app_xsec_LHCX14_hh}), and 100 TeV
(Tables~\ref{tab:app_xsec_LHC100_kk} and \ref{tab:app_xsec_LHC100_hh}) for 
inclusive $pp\to k^{--}k^{++}$ production 
(Tables~\ref{tab:app_xsec_LHCX13_kk}, \ref{tab:app_xsec_LHCX14_kk}, and \ref{tab:app_xsec_LHC100_kk})
as well as 
inclusive $pp\to k^{--}k^{++}$ production 
(Tables~\ref{tab:app_xsec_LHCX13_hh}, \ref{tab:app_xsec_LHCX14_hh}, and \ref{tab:app_xsec_LHC100_hh})
via the Drell-Yan process.
For masses $m_k$ and $m_h$ [GeV] (column 1), the predicted cross sections [fb] for $\sqrt{s}=13\TeV$ at LO (column 2) and NLO (column 3) in QCD are provided. Also shown are scale uncertainties [\%], PDF uncertainties [\%], and the QCD $K$-factor (column 4). See Sec.~\ref{sec:setup_sm} for SM inputs.

\section*{Acknowledgments}

The author is grateful to 
Kaladi Babu,
Mikael Chala
Rupert Coy,
Benjamin Fuks,
Blaz Leban,
Miha Nemev{\v s}ek,
Miguel Nebot,
Jose Miguel No,
Nuria Rius,
Arcadi Santamaria, and
Carmona Tamarit 
for enlightening discussions.

The author acknowledges the support of Narodowe Centrum Nauki under Grant No. 2019/ 34/ E/ ST2/ 00186. The author also acknowledges the support of the Polska Akademia Nauk (grant agreement PAN.BFD.S.BDN. 613. 022. 2021 - PASIFIC 1, POPSICLE). This work has received funding from the European Union's Horizon 2020 research and innovation program under the Sk{\l}odowska-Curie grant agreement No.  847639 and from the Polish Ministry of Education and Science.

The author thanks the Pitt-PACC at the University of Pittsburgh for its hospitality during the progress of this work. The author would also like to thank the Instituto de Fisica Teorica (IFT UAM-CSIC) in Madrid for support via the Centro de Excelencia Severo Ochoa Program under Grant CEX2020- 001007-S, during the Extended Workshop ``Neutrino Theories,'' where this work developed.

\newpage 

\begin{table*}[t!]
\begin{center}
\resizebox{.95\textwidth}{!}{
\begin{tabular}{c | c c c | c c c | c}
\hline\hline
\multicolumn{8}{c}{$pp\to k^{--}k^{++}+X$}\\
\hline
\multicolumn{8}{c}{$\sqrt{s}=13\TeV$ LHC}
\\
mass [GeV] & $\sigma^{\rm LO}_{13\TeV}$ [fb] & $\delta_{\rm RG~scale}$ [\%] &
	 $\delta_{\rm PDF}$ [\%] & $\sigma^{\rm NLO}_{13\TeV}$ [fb] & $\delta_{\rm RG~scale}$ [\%] &
	 $\delta_{\rm PDF}$ [\%] & $K^{\rm NLO}$ \\
\hline
$50$	&	\texttt{4.959e+03} &	$^{+9\%}_{-10\%} $ &	$^{+1.7\%}_{-2.0\%} $	&	\texttt{6.102e+03} &	$^{+3.3\%}_{-5.4\%} $ & $^{+1.6\%}_{-1.9\%} $	&	$1.230$\\
$75$	&	\texttt{9.566e+02} &	$^{+4\%}_{-6\%} $ &	$^{+2.0\%}_{-2.3\%} $	&	\texttt{1.147e+03} &	$^{+2.5\%}_{-3.2\%} $ & $^{+2.0\%}_{-2.2\%} $	&	$1.199$\\
$125$	&	\texttt{1.661e+02} &	$^{+1\%}_{-2\%} $ &	$^{+2.4\%}_{-2.5\%} $	&	\texttt{1.953e+02} &	$^{+1.8\%}_{-1.6\%} $ & $^{+2.4\%}_{-2.5\%} $	&	$1.176$\\
$150$	&	\texttt{8.780e+01} &	$^{+0\%}_{-0\%} $ &	$^{+2.6\%}_{-2.6\%} $	&	\texttt{1.025e+02} &	$^{+1.5\%}_{-1.1\%} $ & $^{+2.6\%}_{-2.6\%} $	&	$1.167$\\
$200$	&	\texttt{3.097e+01} &	$^{+2\%}_{-2\%} $ &	$^{+2.9\%}_{-2.7\%} $	&	\texttt{3.603e+01} &	$^{+1.6\%}_{-1.3\%} $ & $^{+2.9\%}_{-2.8\%} $	&	$1.163$\\
$225$	&	\texttt{1.995e+01} &	$^{+2\%}_{-3\%} $ &	$^{+3.0\%}_{-2.8\%} $	&	\texttt{2.312e+01} &	$^{+1.7\%}_{-1.5\%} $ & $^{+3.0\%}_{-2.8\%} $	&	$1.159$\\
$275$	&	\texttt{9.174e+00} &	$^{+4\%}_{-3\%} $ &	$^{+3.3\%}_{-3.0\%} $	&	\texttt{1.065e+01} &	$^{+1.8\%}_{-1.6\%} $ & $^{+3.3\%}_{-3.0\%} $	&	$1.161$\\
$300$	&	\texttt{6.487e+00} &	$^{+4\%}_{-4\%} $ &	$^{+3.4\%}_{-3.1\%} $	&	\texttt{7.517e+00} &	$^{+1.9\%}_{-1.8\%} $ & $^{+3.4\%}_{-3.1\%} $	&	$1.159$\\
$350$	&	\texttt{3.443e+00} &	$^{+5\%}_{-5\%} $ &	$^{+3.5\%}_{-3.2\%} $	&	\texttt{3.993e+00} &	$^{+2.0\%}_{-2.0\%} $ & $^{+3.6\%}_{-3.3\%} $	&	$1.160$\\
$375$	&	\texttt{2.571e+00} &	$^{+6\%}_{-5\%} $ &	$^{+3.6\%}_{-3.3\%} $	&	\texttt{2.976e+00} &	$^{+2.1\%}_{-2.1\%} $ & $^{+3.7\%}_{-3.3\%} $	&	$1.158$\\
$425$	&	\texttt{1.490e+00} &	$^{+6\%}_{-6\%} $ &	$^{+3.8\%}_{-3.4\%} $	&	\texttt{1.725e+00} &	$^{+2.2\%}_{-2.3\%} $ & $^{+3.8\%}_{-3.5\%} $	&	$1.158$\\
$450$	&	\texttt{1.151e+00} &	$^{+7\%}_{-6\%} $ &	$^{+3.9\%}_{-3.5\%} $	&	\texttt{1.337e+00} &	$^{+2.2\%}_{-2.3\%} $ & $^{+3.9\%}_{-3.5\%} $	&	$1.162$\\
$500$	&	\texttt{7.068e-01} &	$^{+7\%}_{-7\%} $ &	$^{+4.0\%}_{-3.6\%} $	&	\texttt{8.230e-01} &	$^{+2.3\%}_{-2.4\%} $ & $^{+4.1\%}_{-3.7\%} $	&	$1.164$\\
$525$	&	\texttt{5.616e-01} &	$^{+7\%}_{-7\%} $ &	$^{+4.1\%}_{-3.7\%} $	&	\texttt{6.528e-01} &	$^{+2.3\%}_{-2.5\%} $ & $^{+4.2\%}_{-3.7\%} $	&	$1.162$\\
$575$	&	\texttt{3.596e-01} &	$^{+8\%}_{-7\%} $ &	$^{+4.3\%}_{-3.8\%} $	&	\texttt{4.193e-01} &	$^{+2.4\%}_{-2.7\%} $ & $^{+4.3\%}_{-3.9\%} $	&	$1.166$\\
$600$	&	\texttt{2.907e-01} &	$^{+8\%}_{-7\%} $ &	$^{+4.4\%}_{-3.9\%} $	&	\texttt{3.398e-01} &	$^{+2.5\%}_{-2.7\%} $ & $^{+4.4\%}_{-3.9\%} $	&	$1.169$\\
$650$	&	\texttt{1.926e-01} &	$^{+9\%}_{-8\%} $ &	$^{+4.5\%}_{-4.0\%} $	&	\texttt{2.256e-01} &	$^{+2.6\%}_{-2.9\%} $ & $^{+4.6\%}_{-4.1\%} $	&	$1.171$\\
$675$	&	\texttt{1.575e-01} &	$^{+9\%}_{-8\%} $ &	$^{+4.6\%}_{-4.1\%} $	&	\texttt{1.851e-01} &	$^{+2.5\%}_{-2.9\%} $ & $^{+4.6\%}_{-4.1\%} $	&	$1.175$\\
$725$	&	\texttt{1.072e-01} &	$^{+9\%}_{-8\%} $ &	$^{+4.8\%}_{-4.2\%} $	&	\texttt{1.259e-01} &	$^{+2.7\%}_{-3.1\%} $ & $^{+4.8\%}_{-4.2\%} $	&	$1.174$\\
$750$	&	\texttt{8.865e-02} &	$^{+10\%}_{-8\%} $ &	$^{+4.9\%}_{-4.3\%} $	&	\texttt{1.045e-01} &	$^{+2.7\%}_{-3.1\%} $ & $^{+4.9\%}_{-4.3\%} $	&	$1.179$\\
$800$	&	\texttt{6.132e-02} &	$^{+10\%}_{-9\%} $ &	$^{+5.1\%}_{-4.4\%} $	&	\texttt{7.260e-02} &	$^{+2.8\%}_{-3.2\%} $ & $^{+5.1\%}_{-4.4\%} $	&	$1.184$\\
$825$	&	\texttt{5.141e-02} &	$^{+10\%}_{-9\%} $ &	$^{+5.2\%}_{-4.5\%} $	&	\texttt{6.081e-02} &	$^{+2.8\%}_{-3.3\%} $ & $^{+5.2\%}_{-4.5\%} $	&	$1.183$\\
$875$	&	\texttt{3.617e-02} &	$^{+11\%}_{-9\%} $ &	$^{+5.4\%}_{-4.6\%} $	&	\texttt{4.289e-02} &	$^{+2.9\%}_{-3.4\%} $ & $^{+5.4\%}_{-4.6\%} $	&	$1.186$\\
$900$	&	\texttt{3.041e-02} &	$^{+11\%}_{-9\%} $ &	$^{+5.5\%}_{-4.7\%} $	&	\texttt{3.614e-02} &	$^{+2.9\%}_{-3.5\%} $ & $^{+5.5\%}_{-4.7\%} $	&	$1.188$\\
$950$	&	\texttt{2.171e-02} &	$^{+11\%}_{-10\%} $ &	$^{+5.8\%}_{-4.8\%} $	&	\texttt{2.584e-02} &	$^{+3.0\%}_{-3.6\%} $ & $^{+5.8\%}_{-4.9\%} $	&	$1.190$\\
$975$	&	\texttt{1.837e-02} &	$^{+12\%}_{-10\%} $ &	$^{+5.9\%}_{-4.9\%} $	&	\texttt{2.194e-02} &	$^{+3.1\%}_{-3.7\%} $ & $^{+5.9\%}_{-4.9\%} $	&	$1.194$\\
$1025$	&	\texttt{1.326e-02} &	$^{+12\%}_{-10\%} $ &	$^{+6.2\%}_{-5.1\%} $	&	\texttt{1.584e-02} &	$^{+3.2\%}_{-3.8\%} $ & $^{+6.2\%}_{-5.1\%} $	&	$1.195$\\
$1050$	&	\texttt{1.129e-02} &	$^{+12\%}_{-10\%} $ &	$^{+6.3\%}_{-5.1\%} $	&	\texttt{1.352e-02} &	$^{+3.2\%}_{-3.8\%} $ & $^{+6.3\%}_{-5.2\%} $	&	$1.198$\\
$1100$	&	\texttt{8.190e-03} &	$^{+12\%}_{-10\%} $ &	$^{+6.6\%}_{-5.3\%} $	&	\texttt{9.856e-03} &	$^{+3.3\%}_{-3.9\%} $ & $^{+6.6\%}_{-5.3\%} $	&	$1.203$\\
$1125$	&	\texttt{7.008e-03} &	$^{+13\%}_{-10\%} $ &	$^{+6.7\%}_{-5.4\%} $	&	\texttt{8.446e-03} &	$^{+3.3\%}_{-4.0\%} $ & $^{+6.7\%}_{-5.4\%} $	&	$1.205$\\
$1175$	&	\texttt{5.132e-03} &	$^{+13\%}_{-11\%} $ &	$^{+7.0\%}_{-5.5\%} $	&	\texttt{6.212e-03} &	$^{+3.4\%}_{-4.1\%} $ & $^{+7.0\%}_{-5.6\%} $	&	$1.210$\\
$1200$	&	\texttt{4.396e-03} &	$^{+13\%}_{-11\%} $ &	$^{+7.2\%}_{-5.6\%} $	&	\texttt{5.333e-03} &	$^{+3.4\%}_{-4.2\%} $ & $^{+7.2\%}_{-5.6\%} $	&	$1.213$\\
$1250$	&	\texttt{3.243e-03} &	$^{+13\%}_{-11\%} $ &	$^{+7.5\%}_{-5.8\%} $	&	\texttt{3.943e-03} &	$^{+3.5\%}_{-4.3\%} $ & $^{+7.5\%}_{-5.8\%} $	&	$1.216$\\
$1275$	&	\texttt{2.789e-03} &	$^{+14\%}_{-11\%} $ &	$^{+7.7\%}_{-5.9\%} $	&	\texttt{3.396e-03} &	$^{+3.6\%}_{-4.3\%} $ & $^{+7.6\%}_{-5.9\%} $	&	$1.218$\\
$1325$	&	\texttt{2.070e-03} &	$^{+14\%}_{-12\%} $ &	$^{+8.0\%}_{-6.1\%} $	&	\texttt{2.524e-03} &	$^{+3.7\%}_{-4.5\%} $ & $^{+7.9\%}_{-6.1\%} $	&	$1.219$\\
$1350$	&	\texttt{1.783e-03} &	$^{+14\%}_{-12\%} $ &	$^{+8.2\%}_{-6.2\%} $	&	\texttt{2.179e-03} &	$^{+3.7\%}_{-4.5\%} $ & $^{+8.1\%}_{-6.2\%} $	&	$1.222$\\
$1400$	&	\texttt{1.327e-03} &	$^{+14\%}_{-12\%} $ &	$^{+8.5\%}_{-6.4\%} $	&	\texttt{1.629e-03} &	$^{+3.8\%}_{-4.6\%} $ & $^{+8.5\%}_{-6.4\%} $	&	$1.228$\\

\hline\hline
\end{tabular}
} 
\caption{For representative masses $m_k$ [GeV] (first column), the predicted cross sections [fb] for $\sqrt{s}=13\TeV$ at LO (second column) and NLO (third column) in QCD for inclusive $pp\to k^{--}k^{++}+X$ via the Drell-Yan process (DY). Also shown are scale uncertainties [\%], PDF uncertainties [\%], and the QCD $K$-factor. See Sec.~\ref{sec:setup_sm} for SM inputs.}
\label{tab:app_xsec_LHCX13_kk}
\end{center}
\end{table*}

\begin{table*}[t!]
\begin{center}
\resizebox{.95\textwidth}{!}{
\begin{tabular}{c | c c c | c c c | c}
\hline\hline
\multicolumn{8}{c}{$pp\to h^{-}h^{+}+X$}\\
\hline
\multicolumn{8}{c}{$\sqrt{s}=13\TeV$ LHC}
\\
mass [GeV] & $\sigma^{\rm LO}_{13\TeV}$ [fb] & $\delta_{\rm RG~scale}$ [\%] &
	 $\delta_{\rm PDF}$ [\%] & $\sigma^{\rm NLO}_{13\TeV}$ [fb] & $\delta_{\rm RG~scale}$ [\%] &
	 $\delta_{\rm PDF}$ [\%] & $K^{\rm NLO}$ \\
\hline
$50$	&	\texttt{1.239e+03} &	$^{+9\%}_{-10\%} $ &	$^{+1.7\%}_{-2.0\%} $	&	\texttt{1.525e+03} &	$^{+3.3\%}_{-5.4\%} $ & $^{+1.6\%}_{-1.9\%} $	&	$1.231$\\
$75$	&	\texttt{2.392e+02} &	$^{+5\%}_{-6\%} $ &	$^{+2.0\%}_{-2.3\%} $	&	\texttt{2.869e+02} &	$^{+2.5\%}_{-3.2\%} $ & $^{+2.0\%}_{-2.2\%} $	&	$1.199$\\
$125$	&	\texttt{4.153e+01} &	$^{+1\%}_{-2\%} $ &	$^{+2.4\%}_{-2.5\%} $	&	\texttt{4.882e+01} &	$^{+1.8\%}_{-1.6\%} $ & $^{+2.4\%}_{-2.5\%} $	&	$1.176$\\
$150$	&	\texttt{2.195e+01} &	$^{+0\%}_{-0\%} $ &	$^{+2.6\%}_{-2.6\%} $	&	\texttt{2.564e+01} &	$^{+1.5\%}_{-1.1\%} $ & $^{+2.6\%}_{-2.6\%} $	&	$1.168$\\
$200$	&	\texttt{7.743e+00} &	$^{+2\%}_{-2\%} $ &	$^{+2.9\%}_{-2.7\%} $	&	\texttt{9.007e+00} &	$^{+1.6\%}_{-1.3\%} $ & $^{+2.9\%}_{-2.8\%} $	&	$1.163$\\
$225$	&	\texttt{4.987e+00} &	$^{+2\%}_{-2\%} $ &	$^{+3.0\%}_{-2.8\%} $	&	\texttt{5.781e+00} &	$^{+1.8\%}_{-1.5\%} $ & $^{+3.0\%}_{-2.8\%} $	&	$1.159$\\
$275$	&	\texttt{2.295e+00} &	$^{+4\%}_{-4\%} $ &	$^{+3.3\%}_{-3.0\%} $	&	\texttt{2.661e+00} &	$^{+1.8\%}_{-1.6\%} $ & $^{+3.3\%}_{-3.0\%} $	&	$1.159$\\
$300$	&	\texttt{1.623e+00} &	$^{+4\%}_{-4\%} $ &	$^{+3.4\%}_{-3.1\%} $	&	\texttt{1.879e+00} &	$^{+1.9\%}_{-1.8\%} $ & $^{+3.4\%}_{-3.1\%} $	&	$1.158$\\
$350$	&	\texttt{8.611e-01} &	$^{+5\%}_{-5\%} $ &	$^{+3.5\%}_{-3.2\%} $	&	\texttt{9.982e-01} &	$^{+2.0\%}_{-2.0\%} $ & $^{+3.6\%}_{-3.3\%} $	&	$1.159$\\
$375$	&	\texttt{6.429e-01} &	$^{+5\%}_{-5\%} $ &	$^{+3.6\%}_{-3.3\%} $	&	\texttt{7.440e-01} &	$^{+2.1\%}_{-2.1\%} $ & $^{+3.7\%}_{-3.3\%} $	&	$1.157$\\
$425$	&	\texttt{3.721e-01} &	$^{+6\%}_{-6\%} $ &	$^{+3.8\%}_{-3.4\%} $	&	\texttt{4.314e-01} &	$^{+2.2\%}_{-2.3\%} $ & $^{+3.8\%}_{-3.5\%} $	&	$1.159$\\
$450$	&	\texttt{2.877e-01} &	$^{+7\%}_{-6\%} $ &	$^{+3.9\%}_{-3.5\%} $	&	\texttt{3.343e-01} &	$^{+2.2\%}_{-2.3\%} $ & $^{+3.9\%}_{-3.5\%} $	&	$1.162$\\
$500$	&	\texttt{1.767e-01} &	$^{+7\%}_{-7\%} $ &	$^{+4.0\%}_{-3.6\%} $	&	\texttt{2.058e-01} &	$^{+2.3\%}_{-2.4\%} $ & $^{+4.1\%}_{-3.7\%} $	&	$1.165$\\
$525$	&	\texttt{1.404e-01} &	$^{+7\%}_{-7\%} $ &	$^{+4.1\%}_{-3.7\%} $	&	\texttt{1.632e-01} &	$^{+2.3\%}_{-2.5\%} $ & $^{+4.2\%}_{-3.7\%} $	&	$1.162$\\
$575$	&	\texttt{8.995e-02} &	$^{+8\%}_{-7\%} $ &	$^{+4.3\%}_{-3.8\%} $	&	\texttt{1.048e-01} &	$^{+2.4\%}_{-2.7\%} $ & $^{+4.3\%}_{-3.9\%} $	&	$1.165$\\
$600$	&	\texttt{7.266e-02} &	$^{+8\%}_{-7\%} $ &	$^{+4.4\%}_{-3.9\%} $	&	\texttt{8.497e-02} &	$^{+2.5\%}_{-2.7\%} $ & $^{+4.4\%}_{-3.9\%} $	&	$1.169$\\
$650$	&	\texttt{4.816e-02} &	$^{+9\%}_{-8\%} $ &	$^{+4.5\%}_{-4.0\%} $	&	\texttt{5.640e-02} &	$^{+2.6\%}_{-2.9\%} $ & $^{+4.6\%}_{-4.1\%} $	&	$1.171$\\
$675$	&	\texttt{3.939e-02} &	$^{+9\%}_{-8\%} $ &	$^{+4.6\%}_{-4.1\%} $	&	\texttt{4.627e-02} &	$^{+2.5\%}_{-2.9\%} $ & $^{+4.6\%}_{-4.1\%} $	&	$1.175$\\
$725$	&	\texttt{2.680e-02} &	$^{+9\%}_{-8\%} $ &	$^{+4.8\%}_{-4.2\%} $	&	\texttt{3.148e-02} &	$^{+2.7\%}_{-3.1\%} $ & $^{+4.8\%}_{-4.2\%} $	&	$1.175$\\
$750$	&	\texttt{2.218e-02} &	$^{+10\%}_{-8\%} $ &	$^{+4.9\%}_{-4.3\%} $	&	\texttt{2.613e-02} &	$^{+2.7\%}_{-3.1\%} $ & $^{+4.9\%}_{-4.3\%} $	&	$1.178$\\
$800$	&	\texttt{1.533e-02} &	$^{+10\%}_{-9\%} $ &	$^{+5.1\%}_{-4.4\%} $	&	\texttt{1.815e-02} &	$^{+2.8\%}_{-3.2\%} $ & $^{+5.1\%}_{-4.4\%} $	&	$1.184$\\
$825$	&	\texttt{1.286e-02} &	$^{+10\%}_{-9\%} $ &	$^{+5.2\%}_{-4.5\%} $	&	\texttt{1.520e-02} &	$^{+2.8\%}_{-3.3\%} $ & $^{+5.2\%}_{-4.5\%} $	&	$1.182$\\
$875$	&	\texttt{9.037e-03} &	$^{+11\%}_{-9\%} $ &	$^{+5.4\%}_{-4.6\%} $	&	\texttt{1.072e-02} &	$^{+2.9\%}_{-3.4\%} $ & $^{+5.4\%}_{-4.6\%} $	&	$1.186$\\
$900$	&	\texttt{7.602e-03} &	$^{+11\%}_{-9\%} $ &	$^{+5.5\%}_{-4.7\%} $	&	\texttt{9.035e-03} &	$^{+2.9\%}_{-3.5\%} $ & $^{+5.5\%}_{-4.7\%} $	&	$1.189$\\
$950$	&	\texttt{5.429e-03} &	$^{+11\%}_{-10\%} $ &	$^{+5.8\%}_{-4.8\%} $	&	\texttt{6.460e-03} &	$^{+3.0\%}_{-3.6\%} $ & $^{+5.8\%}_{-4.9\%} $	&	$1.190$\\
$975$	&	\texttt{4.595e-03} &	$^{+11\%}_{-10\%} $ &	$^{+5.9\%}_{-4.9\%} $	&	\texttt{5.484e-03} &	$^{+3.1\%}_{-3.7\%} $ & $^{+5.9\%}_{-4.9\%} $	&	$1.193$\\
$1025$	&	\texttt{3.316e-03} &	$^{+12\%}_{-10\%} $ &	$^{+6.2\%}_{-5.1\%} $	&	\texttt{3.959e-03} &	$^{+3.2\%}_{-3.8\%} $ & $^{+6.2\%}_{-5.1\%} $	&	$1.194$\\
$1050$	&	\texttt{2.821e-03} &	$^{+12\%}_{-10\%} $ &	$^{+6.3\%}_{-5.1\%} $	&	\texttt{3.381e-03} &	$^{+3.2\%}_{-3.8\%} $ & $^{+6.3\%}_{-5.2\%} $	&	$1.199$\\
$1100$	&	\texttt{2.048e-03} &	$^{+12\%}_{-10\%} $ &	$^{+6.6\%}_{-5.3\%} $	&	\texttt{2.464e-03} &	$^{+3.3\%}_{-3.9\%} $ & $^{+6.6\%}_{-5.3\%} $	&	$1.203$\\
$1125$	&	\texttt{1.751e-03} &	$^{+13\%}_{-10\%} $ &	$^{+6.7\%}_{-5.4\%} $	&	\texttt{2.111e-03} &	$^{+3.3\%}_{-4.0\%} $ & $^{+6.7\%}_{-5.4\%} $	&	$1.206$\\
$1175$	&	\texttt{1.283e-03} &	$^{+13\%}_{-11\%} $ &	$^{+7.0\%}_{-5.5\%} $	&	\texttt{1.553e-03} &	$^{+3.4\%}_{-4.1\%} $ & $^{+7.0\%}_{-5.6\%} $	&	$1.210$\\
$1200$	&	\texttt{1.099e-03} &	$^{+13\%}_{-11\%} $ &	$^{+7.2\%}_{-5.6\%} $	&	\texttt{1.333e-03} &	$^{+3.4\%}_{-4.2\%} $ & $^{+7.2\%}_{-5.6\%} $	&	$1.213$\\
$1250$	&	\texttt{8.111e-04} &	$^{+13\%}_{-11\%} $ &	$^{+7.5\%}_{-5.8\%} $	&	\texttt{9.856e-04} &	$^{+3.5\%}_{-4.3\%} $ & $^{+7.5\%}_{-5.8\%} $	&	$1.215$\\
$1275$	&	\texttt{6.973e-04} &	$^{+14\%}_{-11\%} $ &	$^{+7.7\%}_{-5.9\%} $	&	\texttt{8.491e-04} &	$^{+3.6\%}_{-4.3\%} $ & $^{+7.6\%}_{-5.9\%} $	&	$1.218$\\
$1325$	&	\texttt{5.180e-04} &	$^{+14\%}_{-11\%} $ &	$^{+8.0\%}_{-6.1\%} $	&	\texttt{6.311e-04} &	$^{+3.7\%}_{-4.5\%} $ & $^{+7.9\%}_{-6.1\%} $	&	$1.218$\\
$1350$	&	\texttt{4.459e-04} &	$^{+14\%}_{-12\%} $ &	$^{+8.2\%}_{-6.2\%} $	&	\texttt{5.446e-04} &	$^{+3.7\%}_{-4.5\%} $ & $^{+8.1\%}_{-6.2\%} $	&	$1.221$\\
$1400$	&	\texttt{3.315e-04} &	$^{+14\%}_{-12\%} $ &	$^{+8.5\%}_{-6.4\%} $	&	\texttt{4.071e-04} &	$^{+3.8\%}_{-4.6\%} $ & $^{+8.5\%}_{-6.4\%} $	&	$1.228$\\

\hline\hline
\end{tabular}
} 
\caption{For representative masses $m_h$ [GeV] (first column), the predicted cross sections [fb] for $\sqrt{s}=13\TeV$ at LO (second column) and NLO (third column) in QCD for inclusive $pp\to h^{-}h^{+}+X$ via the Drell-Yan process (DY). Also shown are scale uncertainties [\%], PDF uncertainties [\%], and the QCD $K$-factor. See Sec.~\ref{sec:setup_sm} for SM inputs.}
\label{tab:app_xsec_LHCX13_hh}
\end{center}
\end{table*}

\begin{table*}[t!]
\begin{center}
\resizebox{.95\textwidth}{!}{
\begin{tabular}{c | c c c | c c c | c}
\hline\hline
\multicolumn{8}{c}{$pp\to k^{--}k^{++}+X$}\\
\hline
\multicolumn{8}{c}{$\sqrt{s}=14\TeV$ LHC}
\\
mass [GeV] & $\sigma^{\rm LO}_{14\TeV}$ [fb] & $\delta_{\rm RG~scale}$ [\%] &
	 $\delta_{\rm PDF}$ [\%] & $\sigma^{\rm NLO}_{14\TeV}$ [fb] & $\delta_{\rm RG~scale}$ [\%] &
	 $\delta_{\rm PDF}$ [\%] & $K^{\rm NLO}$ \\
\hline
$50$	&	\texttt{5.391e+03} &	$^{+9\%}_{-11\%} $ &	$^{+1.7\%}_{-1.9\%} $	&	\texttt{6.650e+03} &	$^{+3.4\%}_{-5.6\%} $ & $^{+1.6\%}_{-1.9\%} $	&	$1.234$\\
$75$	&	\texttt{1.049e+03} &	$^{+5\%}_{-6\%} $ &	$^{+2.0\%}_{-2.2\%} $	&	\texttt{1.260e+03} &	$^{+2.5\%}_{-3.4\%} $ & $^{+2.0\%}_{-2.2\%} $	&	$1.201$\\
$125$	&	\texttt{1.845e+02} &	$^{+1\%}_{-2\%} $ &	$^{+2.4\%}_{-2.4\%} $	&	\texttt{2.170e+02} &	$^{+1.8\%}_{-1.7\%} $ & $^{+2.4\%}_{-2.4\%} $	&	$1.176$\\
$150$	&	\texttt{9.798e+01} &	$^{+0\%}_{-1\%} $ &	$^{+2.5\%}_{-2.5\%} $	&	\texttt{1.146e+02} &	$^{+1.6\%}_{-1.2\%} $ & $^{+2.5\%}_{-2.5\%} $	&	$1.170$\\
$200$	&	\texttt{3.496e+01} &	$^{+1\%}_{-2\%} $ &	$^{+2.8\%}_{-2.7\%} $	&	\texttt{4.067e+01} &	$^{+1.6\%}_{-1.2\%} $ & $^{+2.8\%}_{-2.7\%} $	&	$1.163$\\
$225$	&	\texttt{2.264e+01} &	$^{+2\%}_{-2\%} $ &	$^{+2.9\%}_{-2.7\%} $	&	\texttt{2.625e+01} &	$^{+1.7\%}_{-1.4\%} $ & $^{+3.0\%}_{-2.8\%} $	&	$1.159$\\
$275$	&	\texttt{1.053e+01} &	$^{+3\%}_{-3\%} $ &	$^{+3.2\%}_{-2.9\%} $	&	\texttt{1.221e+01} &	$^{+1.8\%}_{-1.6\%} $ & $^{+3.2\%}_{-2.9\%} $	&	$1.160$\\
$300$	&	\texttt{7.494e+00} &	$^{+4\%}_{-4\%} $ &	$^{+3.3\%}_{-3.0\%} $	&	\texttt{8.673e+00} &	$^{+1.8\%}_{-1.7\%} $ & $^{+3.3\%}_{-3.0\%} $	&	$1.157$\\
$350$	&	\texttt{4.016e+00} &	$^{+5\%}_{-4\%} $ &	$^{+3.4\%}_{-3.1\%} $	&	\texttt{4.655e+00} &	$^{+1.9\%}_{-1.8\%} $ & $^{+3.5\%}_{-3.2\%} $	&	$1.159$\\
$375$	&	\texttt{3.017e+00} &	$^{+5\%}_{-5\%} $ &	$^{+3.5\%}_{-3.2\%} $	&	\texttt{3.486e+00} &	$^{+2.0\%}_{-2.0\%} $ & $^{+3.6\%}_{-3.2\%} $	&	$1.155$\\
$425$	&	\texttt{1.767e+00} &	$^{+6\%}_{-5\%} $ &	$^{+3.7\%}_{-3.3\%} $	&	\texttt{2.045e+00} &	$^{+2.1\%}_{-2.1\%} $ & $^{+3.7\%}_{-3.4\%} $	&	$1.157$\\
$450$	&	\texttt{1.375e+00} &	$^{+6\%}_{-6\%} $ &	$^{+3.8\%}_{-3.4\%} $	&	\texttt{1.594e+00} &	$^{+2.2\%}_{-2.2\%} $ & $^{+3.8\%}_{-3.4\%} $	&	$1.159$\\
$500$	&	\texttt{8.540e-01} &	$^{+7\%}_{-6\%} $ &	$^{+3.9\%}_{-3.5\%} $	&	\texttt{9.922e-01} &	$^{+2.2\%}_{-2.3\%} $ & $^{+4.0\%}_{-3.6\%} $	&	$1.162$\\
$525$	&	\texttt{6.828e-01} &	$^{+7\%}_{-6\%} $ &	$^{+4.0\%}_{-3.6\%} $	&	\texttt{7.912e-01} &	$^{+2.2\%}_{-2.4\%} $ & $^{+4.0\%}_{-3.6\%} $	&	$1.159$\\
$575$	&	\texttt{4.423e-01} &	$^{+8\%}_{-7\%} $ &	$^{+4.1\%}_{-3.7\%} $	&	\texttt{5.141e-01} &	$^{+2.4\%}_{-2.6\%} $ & $^{+4.2\%}_{-3.8\%} $	&	$1.162$\\
$600$	&	\texttt{3.598e-01} &	$^{+8\%}_{-7\%} $ &	$^{+4.2\%}_{-3.8\%} $	&	\texttt{4.193e-01} &	$^{+2.4\%}_{-2.6\%} $ & $^{+4.3\%}_{-3.8\%} $	&	$1.165$\\
$650$	&	\texttt{2.414e-01} &	$^{+8\%}_{-7\%} $ &	$^{+4.4\%}_{-3.9\%} $	&	\texttt{2.816e-01} &	$^{+2.5\%}_{-2.7\%} $ & $^{+4.4\%}_{-3.9\%} $	&	$1.167$\\
$675$	&	\texttt{1.986e-01} &	$^{+9\%}_{-7\%} $ &	$^{+4.5\%}_{-4.0\%} $	&	\texttt{2.325e-01} &	$^{+2.4\%}_{-2.8\%} $ & $^{+4.5\%}_{-4.0\%} $	&	$1.171$\\
$725$	&	\texttt{1.368e-01} &	$^{+9\%}_{-8\%} $ &	$^{+4.6\%}_{-4.1\%} $	&	\texttt{1.601e-01} &	$^{+2.5\%}_{-2.9\%} $ & $^{+4.6\%}_{-4.1\%} $	&	$1.170$\\
$750$	&	\texttt{1.139e-01} &	$^{+9\%}_{-8\%} $ &	$^{+4.7\%}_{-4.1\%} $	&	\texttt{1.337e-01} &	$^{+2.6\%}_{-3.0\%} $ & $^{+4.7\%}_{-4.2\%} $	&	$1.174$\\
$800$	&	\texttt{7.976e-02} &	$^{+10\%}_{-8\%} $ &	$^{+4.9\%}_{-4.3\%} $	&	\texttt{9.403e-02} &	$^{+2.6\%}_{-3.1\%} $ & $^{+4.9\%}_{-4.3\%} $	&	$1.179$\\
$825$	&	\texttt{6.730e-02} &	$^{+10\%}_{-9\%} $ &	$^{+5.0\%}_{-4.3\%} $	&	\texttt{7.920e-02} &	$^{+2.7\%}_{-3.1\%} $ & $^{+5.0\%}_{-4.3\%} $	&	$1.177$\\
$875$	&	\texttt{4.799e-02} &	$^{+10\%}_{-9\%} $ &	$^{+5.2\%}_{-4.4\%} $	&	\texttt{5.660e-02} &	$^{+2.8\%}_{-3.2\%} $ & $^{+5.2\%}_{-4.5\%} $	&	$1.179$\\
$900$	&	\texttt{4.059e-02} &	$^{+10\%}_{-9\%} $ &	$^{+5.3\%}_{-4.5\%} $	&	\texttt{4.801e-02} &	$^{+2.8\%}_{-3.3\%} $ & $^{+5.3\%}_{-4.5\%} $	&	$1.183$\\
$950$	&	\texttt{2.937e-02} &	$^{+11\%}_{-9\%} $ &	$^{+5.5\%}_{-4.6\%} $	&	\texttt{3.477e-02} &	$^{+2.9\%}_{-3.4\%} $ & $^{+5.5\%}_{-4.7\%} $	&	$1.184$\\
$975$	&	\texttt{2.503e-02} &	$^{+11\%}_{-9\%} $ &	$^{+5.6\%}_{-4.7\%} $	&	\texttt{2.972e-02} &	$^{+2.9\%}_{-3.5\%} $ & $^{+5.6\%}_{-4.7\%} $	&	$1.187$\\
$1025$	&	\texttt{1.831e-02} &	$^{+11\%}_{-10\%} $ &	$^{+5.8\%}_{-4.8\%} $	&	\texttt{2.174e-02} &	$^{+3.0\%}_{-3.6\%} $ & $^{+5.8\%}_{-4.9\%} $	&	$1.187$\\
$1050$	&	\texttt{1.568e-02} &	$^{+11\%}_{-10\%} $ &	$^{+5.9\%}_{-4.9\%} $	&	\texttt{1.869e-02} &	$^{+3.0\%}_{-3.6\%} $ & $^{+5.9\%}_{-4.9\%} $	&	$1.192$\\
$1100$	&	\texttt{1.155e-02} &	$^{+12\%}_{-10\%} $ &	$^{+6.2\%}_{-5.0\%} $	&	\texttt{1.381e-02} &	$^{+3.1\%}_{-3.7\%} $ & $^{+6.1\%}_{-5.1\%} $	&	$1.196$\\
$1125$	&	\texttt{9.939e-03} &	$^{+12\%}_{-10\%} $ &	$^{+6.3\%}_{-5.1\%} $	&	\texttt{1.191e-02} &	$^{+3.1\%}_{-3.8\%} $ & $^{+6.3\%}_{-5.1\%} $	&	$1.198$\\
$1175$	&	\texttt{7.387e-03} &	$^{+12\%}_{-10\%} $ &	$^{+6.5\%}_{-5.3\%} $	&	\texttt{8.884e-03} &	$^{+3.2\%}_{-3.9\%} $ & $^{+6.5\%}_{-5.3\%} $	&	$1.203$\\
$1200$	&	\texttt{6.376e-03} &	$^{+12\%}_{-10\%} $ &	$^{+6.7\%}_{-5.3\%} $	&	\texttt{7.685e-03} &	$^{+3.3\%}_{-3.9\%} $ & $^{+6.7\%}_{-5.3\%} $	&	$1.205$\\
$1250$	&	\texttt{4.776e-03} &	$^{+13\%}_{-11\%} $ &	$^{+7.0\%}_{-5.5\%} $	&	\texttt{5.762e-03} &	$^{+3.4\%}_{-4.0\%} $ & $^{+6.9\%}_{-5.5\%} $	&	$1.206$\\
$1275$	&	\texttt{4.136e-03} &	$^{+13\%}_{-11\%} $ &	$^{+7.1\%}_{-5.6\%} $	&	\texttt{5.001e-03} &	$^{+3.4\%}_{-4.1\%} $ & $^{+7.1\%}_{-5.6\%} $	&	$1.209$\\
$1325$	&	\texttt{3.118e-03} &	$^{+13\%}_{-11\%} $ &	$^{+7.4\%}_{-5.7\%} $	&	\texttt{3.772e-03} &	$^{+3.5\%}_{-4.2\%} $ & $^{+7.3\%}_{-5.7\%} $	&	$1.210$\\
$1350$	&	\texttt{2.704e-03} &	$^{+13\%}_{-11\%} $ &	$^{+7.5\%}_{-5.8\%} $	&	\texttt{3.281e-03} &	$^{+3.5\%}_{-4.2\%} $ & $^{+7.5\%}_{-5.8\%} $	&	$1.213$\\
$1400$	&	\texttt{2.044e-03} &	$^{+14\%}_{-11\%} $ &	$^{+7.9\%}_{-6.0\%} $	&	\texttt{2.489e-03} &	$^{+3.6\%}_{-4.3\%} $ & $^{+7.8\%}_{-6.0\%} $	&	$1.218$\\

\hline\hline
\end{tabular}
} 
\caption{For representative masses $m_k$ [GeV] (first column), the predicted cross sections [fb] for $\sqrt{s}=14\TeV$ at LO (second column) and NLO (third column) in QCD for inclusive $pp\to k^{--}k^{++}+X$ via the Drell-Yan process (DY). Also shown are scale uncertainties [\%], PDF uncertainties [\%], and the QCD $K$-factor. See Sec.~\ref{sec:setup_sm} for SM inputs.}
\label{tab:app_xsec_LHCX14_kk}
\end{center}
\end{table*}

\begin{table*}[t!]
\begin{center}
\resizebox{.95\textwidth}{!}{
\begin{tabular}{c | c c c | c c c | c}
\hline\hline
\multicolumn{8}{c}{$pp\to h^{-}h^{+}+X$}\\
\hline
\multicolumn{8}{c}{$\sqrt{s}=14\TeV$ LHC}
\\
mass [GeV] & $\sigma^{\rm LO}_{14\TeV}$ [fb] & $\delta_{\rm RG~scale}$ [\%] &
	 $\delta_{\rm PDF}$ [\%] & $\sigma^{\rm NLO}_{14\TeV}$ [fb] & $\delta_{\rm RG~scale}$ [\%] &
	 $\delta_{\rm PDF}$ [\%] & $K^{\rm NLO}$ \\
\hline
$50$	&	\texttt{1.348e+03} &	$^{+9\%}_{-11\%} $ &	$^{+1.7\%}_{-1.9\%} $	&	\texttt{1.663e+03} &	$^{+3.4\%}_{-5.6\%} $ & $^{+1.6\%}_{-1.9\%} $	&	$1.234$\\
$75$	&	\texttt{2.624e+02} &	$^{+5\%}_{-6\%} $ &	$^{+2.0\%}_{-2.2\%} $	&	\texttt{3.149e+02} &	$^{+2.5\%}_{-3.4\%} $ & $^{+2.0\%}_{-2.2\%} $	&	$1.200$\\
$125$	&	\texttt{4.611e+01} &	$^{+1\%}_{-2\%} $ &	$^{+2.4\%}_{-2.4\%} $	&	\texttt{5.424e+01} &	$^{+1.8\%}_{-1.7\%} $ & $^{+2.4\%}_{-2.4\%} $	&	$1.176$\\
$150$	&	\texttt{2.449e+01} &	$^{+0\%}_{-1\%} $ &	$^{+2.5\%}_{-2.5\%} $	&	\texttt{2.865e+01} &	$^{+1.6\%}_{-1.2\%} $ & $^{+2.5\%}_{-2.5\%} $	&	$1.170$\\
$200$	&	\texttt{8.743e+00} &	$^{+1\%}_{-2\%} $ &	$^{+2.8\%}_{-2.7\%} $	&	\texttt{1.017e+01} &	$^{+1.6\%}_{-1.2\%} $ & $^{+2.8\%}_{-2.7\%} $	&	$1.163$\\
$225$	&	\texttt{5.664e+00} &	$^{+2\%}_{-2\%} $ &	$^{+2.9\%}_{-2.7\%} $	&	\texttt{6.561e+00} &	$^{+1.7\%}_{-1.4\%} $ & $^{+3.0\%}_{-2.8\%} $	&	$1.158$\\
$275$	&	\texttt{2.633e+00} &	$^{+3\%}_{-3\%} $ &	$^{+3.2\%}_{-2.9\%} $	&	\texttt{3.052e+00} &	$^{+1.8\%}_{-1.6\%} $ & $^{+3.2\%}_{-2.9\%} $	&	$1.159$\\
$300$	&	\texttt{1.872e+00} &	$^{+4\%}_{-4\%} $ &	$^{+3.3\%}_{-3.0\%} $	&	\texttt{2.168e+00} &	$^{+1.8\%}_{-1.7\%} $ & $^{+3.3\%}_{-3.0\%} $	&	$1.158$\\
$350$	&	\texttt{1.004e+00} &	$^{+5\%}_{-4\%} $ &	$^{+3.4\%}_{-3.1\%} $	&	\texttt{1.163e+00} &	$^{+1.9\%}_{-1.8\%} $ & $^{+3.5\%}_{-3.2\%} $	&	$1.158$\\
$375$	&	\texttt{7.545e-01} &	$^{+5\%}_{-5\%} $ &	$^{+3.5\%}_{-3.2\%} $	&	\texttt{8.716e-01} &	$^{+2.0\%}_{-2.0\%} $ & $^{+3.6\%}_{-3.2\%} $	&	$1.155$\\
$425$	&	\texttt{4.421e-01} &	$^{+6\%}_{-5\%} $ &	$^{+3.7\%}_{-3.3\%} $	&	\texttt{5.111e-01} &	$^{+2.1\%}_{-2.1\%} $ & $^{+3.7\%}_{-3.4\%} $	&	$1.156$\\
$450$	&	\texttt{3.438e-01} &	$^{+6\%}_{-6\%} $ &	$^{+3.8\%}_{-3.4\%} $	&	\texttt{3.984e-01} &	$^{+2.2\%}_{-2.2\%} $ & $^{+3.8\%}_{-3.4\%} $	&	$1.159$\\
$500$	&	\texttt{2.137e-01} &	$^{+7\%}_{-6\%} $ &	$^{+3.9\%}_{-3.5\%} $	&	\texttt{2.480e-01} &	$^{+2.2\%}_{-2.3\%} $ & $^{+4.0\%}_{-3.6\%} $	&	$1.161$\\
$525$	&	\texttt{1.706e-01} &	$^{+7\%}_{-6\%} $ &	$^{+4.0\%}_{-3.6\%} $	&	\texttt{1.978e-01} &	$^{+2.2\%}_{-2.4\%} $ & $^{+4.0\%}_{-3.6\%} $	&	$1.159$\\
$575$	&	\texttt{1.106e-01} &	$^{+8\%}_{-7\%} $ &	$^{+4.1\%}_{-3.7\%} $	&	\texttt{1.285e-01} &	$^{+2.4\%}_{-2.6\%} $ & $^{+4.2\%}_{-3.8\%} $	&	$1.162$\\
$600$	&	\texttt{8.991e-02} &	$^{+8\%}_{-7\%} $ &	$^{+4.2\%}_{-3.8\%} $	&	\texttt{1.048e-01} &	$^{+2.4\%}_{-2.6\%} $ & $^{+4.3\%}_{-3.8\%} $	&	$1.166$\\
$650$	&	\texttt{6.032e-02} &	$^{+8\%}_{-7\%} $ &	$^{+4.4\%}_{-3.9\%} $	&	\texttt{7.040e-02} &	$^{+2.5\%}_{-2.7\%} $ & $^{+4.4\%}_{-3.9\%} $	&	$1.167$\\
$675$	&	\texttt{4.965e-02} &	$^{+9\%}_{-7\%} $ &	$^{+4.5\%}_{-4.0\%} $	&	\texttt{5.811e-02} &	$^{+2.4\%}_{-2.8\%} $ & $^{+4.5\%}_{-4.0\%} $	&	$1.170$\\
$725$	&	\texttt{3.418e-02} &	$^{+9\%}_{-8\%} $ &	$^{+4.6\%}_{-4.1\%} $	&	\texttt{4.003e-02} &	$^{+2.5\%}_{-2.9\%} $ & $^{+4.6\%}_{-4.1\%} $	&	$1.171$\\
$750$	&	\texttt{2.846e-02} &	$^{+9\%}_{-8\%} $ &	$^{+4.7\%}_{-4.1\%} $	&	\texttt{3.343e-02} &	$^{+2.6\%}_{-3.0\%} $ & $^{+4.7\%}_{-4.2\%} $	&	$1.175$\\
$800$	&	\texttt{1.993e-02} &	$^{+10\%}_{-8\%} $ &	$^{+4.9\%}_{-4.3\%} $	&	\texttt{2.351e-02} &	$^{+2.6\%}_{-3.1\%} $ & $^{+4.9\%}_{-4.3\%} $	&	$1.180$\\
$825$	&	\texttt{1.682e-02} &	$^{+10\%}_{-9\%} $ &	$^{+5.0\%}_{-4.3\%} $	&	\texttt{1.980e-02} &	$^{+2.7\%}_{-3.1\%} $ & $^{+5.0\%}_{-4.3\%} $	&	$1.177$\\
$875$	&	\texttt{1.199e-02} &	$^{+10\%}_{-9\%} $ &	$^{+5.2\%}_{-4.4\%} $	&	\texttt{1.415e-02} &	$^{+2.8\%}_{-3.2\%} $ & $^{+5.2\%}_{-4.5\%} $	&	$1.180$\\
$900$	&	\texttt{1.015e-02} &	$^{+10\%}_{-9\%} $ &	$^{+5.3\%}_{-4.5\%} $	&	\texttt{1.200e-02} &	$^{+2.8\%}_{-3.3\%} $ & $^{+5.3\%}_{-4.5\%} $	&	$1.182$\\
$950$	&	\texttt{7.342e-03} &	$^{+11\%}_{-9\%} $ &	$^{+5.5\%}_{-4.6\%} $	&	\texttt{8.693e-03} &	$^{+2.9\%}_{-3.4\%} $ & $^{+5.5\%}_{-4.7\%} $	&	$1.184$\\
$975$	&	\texttt{6.256e-03} &	$^{+11\%}_{-9\%} $ &	$^{+5.6\%}_{-4.7\%} $	&	\texttt{7.431e-03} &	$^{+2.9\%}_{-3.5\%} $ & $^{+5.6\%}_{-4.7\%} $	&	$1.188$\\
$1025$	&	\texttt{4.579e-03} &	$^{+11\%}_{-10\%} $ &	$^{+5.8\%}_{-4.8\%} $	&	\texttt{5.434e-03} &	$^{+3.0\%}_{-3.6\%} $ & $^{+5.8\%}_{-4.9\%} $	&	$1.187$\\
$1050$	&	\texttt{3.920e-03} &	$^{+11\%}_{-10\%} $ &	$^{+5.9\%}_{-4.9\%} $	&	\texttt{4.672e-03} &	$^{+3.0\%}_{-3.6\%} $ & $^{+5.9\%}_{-4.9\%} $	&	$1.192$\\
$1100$	&	\texttt{2.887e-03} &	$^{+12\%}_{-10\%} $ &	$^{+6.2\%}_{-5.0\%} $	&	\texttt{3.452e-03} &	$^{+3.1\%}_{-3.7\%} $ & $^{+6.1\%}_{-5.1\%} $	&	$1.196$\\
$1125$	&	\texttt{2.485e-03} &	$^{+12\%}_{-10\%} $ &	$^{+6.3\%}_{-5.1\%} $	&	\texttt{2.978e-03} &	$^{+3.1\%}_{-3.8\%} $ & $^{+6.3\%}_{-5.1\%} $	&	$1.198$\\
$1175$	&	\texttt{1.847e-03} &	$^{+12\%}_{-10\%} $ &	$^{+6.5\%}_{-5.3\%} $	&	\texttt{2.221e-03} &	$^{+3.2\%}_{-3.9\%} $ & $^{+6.5\%}_{-5.3\%} $	&	$1.202$\\
$1200$	&	\texttt{1.594e-03} &	$^{+12\%}_{-10\%} $ &	$^{+6.7\%}_{-5.3\%} $	&	\texttt{1.921e-03} &	$^{+3.3\%}_{-3.9\%} $ & $^{+6.7\%}_{-5.3\%} $	&	$1.205$\\
$1250$	&	\texttt{1.194e-03} &	$^{+13\%}_{-11\%} $ &	$^{+7.0\%}_{-5.5\%} $	&	\texttt{1.441e-03} &	$^{+3.4\%}_{-4.0\%} $ & $^{+6.9\%}_{-5.5\%} $	&	$1.207$\\
$1275$	&	\texttt{1.034e-03} &	$^{+13\%}_{-11\%} $ &	$^{+7.1\%}_{-5.6\%} $	&	\texttt{1.250e-03} &	$^{+3.4\%}_{-4.1\%} $ & $^{+7.1\%}_{-5.6\%} $	&	$1.209$\\
$1325$	&	\texttt{7.791e-04} &	$^{+13\%}_{-11\%} $ &	$^{+7.4\%}_{-5.7\%} $	&	\texttt{9.428e-04} &	$^{+3.5\%}_{-4.2\%} $ & $^{+7.3\%}_{-5.7\%} $	&	$1.210$\\
$1350$	&	\texttt{6.764e-04} &	$^{+13\%}_{-11\%} $ &	$^{+7.5\%}_{-5.8\%} $	&	\texttt{8.202e-04} &	$^{+3.5\%}_{-4.2\%} $ & $^{+7.5\%}_{-5.8\%} $	&	$1.213$\\
$1400$	&	\texttt{5.111e-04} &	$^{+14\%}_{-11\%} $ &	$^{+7.9\%}_{-6.0\%} $	&	\texttt{6.222e-04} &	$^{+3.6\%}_{-4.3\%} $ & $^{+7.8\%}_{-6.0\%} $	&	$1.217$\\

\hline\hline
\end{tabular}
} 
\caption{For representative masses $m_h$ [GeV] (first column), the predicted cross sections [fb] for $\sqrt{s}=14\TeV$ at LO (second column) and NLO (third column) in QCD for inclusive $pp\to h^{-}h^{+}+X$ via the Drell-Yan process (DY). Also shown are scale uncertainties [\%], PDF uncertainties [\%], and the QCD $K$-factor. See Sec.~\ref{sec:setup_sm} for SM inputs.}
\label{tab:app_xsec_LHCX14_hh}
\end{center}
\end{table*}

\begin{table*}[t!]
\begin{center}
\resizebox{.95\textwidth}{!}{
\begin{tabular}{c | c c c | c c c | c}
\hline\hline
\multicolumn{8}{c}{$pp\to k^{--}k^{++}+X$}\\
\hline
\multicolumn{8}{c}{$\sqrt{s}=100\TeV$ LHC}
\\
mass [GeV] & $\sigma^{\rm LO}_{100\TeV}$ [fb] & $\delta_{\rm RG~scale}$ [\%] &
	 $\delta_{\rm PDF}$ [\%] & $\sigma^{\rm NLO}_{100\TeV}$ [fb] & $\delta_{\rm RG~scale}$ [\%] &
	 $\delta_{\rm PDF}$ [\%] & $K^{\rm NLO}$ \\
\hline
$250$	&	\texttt{2.440e+02} &	$^{+6\%}_{-7\%} $ &	$^{+1.6\%}_{-1.7\%} $	&	\texttt{2.884e+02} &	$^{+2.2\%}_{-3.4\%} $ & $^{+1.6\%}_{-1.7\%} $	&	$1.182$\\
$450$	&	\texttt{3.711e+01} &	$^{+3\%}_{-4\%} $ &	$^{+1.7\%}_{-1.9\%} $	&	\texttt{4.300e+01} &	$^{+1.6\%}_{-2.0\%} $ & $^{+1.7\%}_{-1.9\%} $	&	$1.159$\\
$850$	&	\texttt{4.293e+00} &	$^{+0\%}_{-1\%} $ &	$^{+2.1\%}_{-2.1\%} $	&	\texttt{4.890e+00} &	$^{+1.1\%}_{-0.8\%} $ & $^{+2.1\%}_{-2.1\%} $	&	$1.139$\\
$1050$	&	\texttt{2.018e+00} &	$^{+0\%}_{-1\%} $ &	$^{+2.3\%}_{-2.2\%} $	&	\texttt{2.294e+00} &	$^{+1.1\%}_{-0.8\%} $ & $^{+2.3\%}_{-2.2\%} $	&	$1.137$\\
$1450$	&	\texttt{6.126e-01} &	$^{+2\%}_{-2\%} $ &	$^{+2.6\%}_{-2.4\%} $	&	\texttt{6.934e-01} &	$^{+1.3\%}_{-1.1\%} $ & $^{+2.6\%}_{-2.4\%} $	&	$1.132$\\
$1650$	&	\texttt{3.730e-01} &	$^{+2\%}_{-3\%} $ &	$^{+2.8\%}_{-2.5\%} $	&	\texttt{4.223e-01} &	$^{+1.3\%}_{-1.2\%} $ & $^{+2.8\%}_{-2.6\%} $	&	$1.132$\\
$2050$	&	\texttt{1.579e-01} &	$^{+3\%}_{-3\%} $ &	$^{+3.0\%}_{-2.7\%} $	&	\texttt{1.789e-01} &	$^{+1.4\%}_{-1.3\%} $ & $^{+3.0\%}_{-2.8\%} $	&	$1.133$\\
$2250$	&	\texttt{1.078e-01} &	$^{+4\%}_{-4\%} $ &	$^{+3.1\%}_{-2.8\%} $	&	\texttt{1.221e-01} &	$^{+1.5\%}_{-1.4\%} $ & $^{+3.2\%}_{-2.9\%} $	&	$1.133$\\
$2650$	&	\texttt{5.419e-02} &	$^{+5\%}_{-4\%} $ &	$^{+3.3\%}_{-3.0\%} $	&	\texttt{6.120e-02} &	$^{+1.5\%}_{-1.6\%} $ & $^{+3.4\%}_{-3.0\%} $	&	$1.129$\\
$2850$	&	\texttt{3.938e-02} &	$^{+5\%}_{-5\%} $ &	$^{+3.5\%}_{-3.1\%} $	&	\texttt{4.465e-02} &	$^{+1.5\%}_{-1.6\%} $ & $^{+3.5\%}_{-3.1\%} $	&	$1.134$\\
$3250$	&	\texttt{2.185e-02} &	$^{+6\%}_{-5\%} $ &	$^{+3.6\%}_{-3.3\%} $	&	\texttt{2.481e-02} &	$^{+1.6\%}_{-1.8\%} $ & $^{+3.7\%}_{-3.3\%} $	&	$1.135$\\
$3450$	&	\texttt{1.659e-02} &	$^{+6\%}_{-5\%} $ &	$^{+3.7\%}_{-3.3\%} $	&	\texttt{1.883e-02} &	$^{+1.6\%}_{-1.8\%} $ & $^{+3.8\%}_{-3.4\%} $	&	$1.135$\\
$3850$	&	\texttt{9.824e-03} &	$^{+6\%}_{-6\%} $ &	$^{+3.9\%}_{-3.5\%} $	&	\texttt{1.119e-02} &	$^{+1.7\%}_{-1.9\%} $ & $^{+3.9\%}_{-3.5\%} $	&	$1.139$\\
$4050$	&	\texttt{7.681e-03} &	$^{+7\%}_{-6\%} $ &	$^{+4.0\%}_{-3.6\%} $	&	\texttt{8.755e-03} &	$^{+1.7\%}_{-2.0\%} $ & $^{+4.0\%}_{-3.6\%} $	&	$1.140$\\
$4450$	&	\texttt{4.779e-03} &	$^{+7\%}_{-6\%} $ &	$^{+4.2\%}_{-3.7\%} $	&	\texttt{5.465e-03} &	$^{+1.8\%}_{-2.1\%} $ & $^{+4.2\%}_{-3.7\%} $	&	$1.144$\\
$4650$	&	\texttt{3.808e-03} &	$^{+7\%}_{-6\%} $ &	$^{+4.3\%}_{-3.8\%} $	&	\texttt{4.361e-03} &	$^{+1.8\%}_{-2.1\%} $ & $^{+4.3\%}_{-3.8\%} $	&	$1.145$\\
$5050$	&	\texttt{2.461e-03} &	$^{+8\%}_{-7\%} $ &	$^{+4.5\%}_{-3.9\%} $	&	\texttt{2.821e-03} &	$^{+1.9\%}_{-2.2\%} $ & $^{+4.5\%}_{-3.9\%} $	&	$1.146$\\
$5250$	&	\texttt{1.991e-03} &	$^{+8\%}_{-7\%} $ &	$^{+4.6\%}_{-4.0\%} $	&	\texttt{2.283e-03} &	$^{+1.9\%}_{-2.3\%} $ & $^{+4.6\%}_{-4.0\%} $	&	$1.147$\\
$5650$	&	\texttt{1.321e-03} &	$^{+8\%}_{-7\%} $ &	$^{+4.8\%}_{-4.1\%} $	&	\texttt{1.519e-03} &	$^{+2.0\%}_{-2.4\%} $ & $^{+4.8\%}_{-4.1\%} $	&	$1.150$\\
$5850$	&	\texttt{1.082e-03} &	$^{+8\%}_{-7\%} $ &	$^{+4.9\%}_{-4.2\%} $	&	\texttt{1.245e-03} &	$^{+2.0\%}_{-2.4\%} $ & $^{+4.9\%}_{-4.2\%} $	&	$1.151$\\
$6250$	&	\texttt{7.321e-04} &	$^{+9\%}_{-8\%} $ &	$^{+5.2\%}_{-4.3\%} $	&	\texttt{8.471e-04} &	$^{+2.1\%}_{-2.5\%} $ & $^{+5.2\%}_{-4.4\%} $	&	$1.157$\\
$6450$	&	\texttt{6.045e-04} &	$^{+9\%}_{-8\%} $ &	$^{+5.3\%}_{-4.4\%} $	&	\texttt{7.014e-04} &	$^{+2.1\%}_{-2.6\%} $ & $^{+5.3\%}_{-4.4\%} $	&	$1.160$\\
$6850$	&	\texttt{4.180e-04} &	$^{+9\%}_{-8\%} $ &	$^{+5.5\%}_{-4.6\%} $	&	\texttt{4.844e-04} &	$^{+2.2\%}_{-2.7\%} $ & $^{+5.5\%}_{-4.6\%} $	&	$1.159$\\
$7050$	&	\texttt{3.486e-04} &	$^{+9\%}_{-8\%} $ &	$^{+5.7\%}_{-4.6\%} $	&	\texttt{4.043e-04} &	$^{+2.2\%}_{-2.7\%} $ & $^{+5.6\%}_{-4.7\%} $	&	$1.160$\\
$7450$	&	\texttt{2.434e-04} &	$^{+10\%}_{-8\%} $ &	$^{+5.9\%}_{-4.8\%} $	&	\texttt{2.837e-04} &	$^{+2.3\%}_{-2.8\%} $ & $^{+5.9\%}_{-4.8\%} $	&	$1.166$\\
$7650$	&	\texttt{2.044e-04} &	$^{+10\%}_{-8\%} $ &	$^{+6.1\%}_{-4.9\%} $	&	\texttt{2.384e-04} &	$^{+2.3\%}_{-2.8\%} $ & $^{+6.1\%}_{-4.9\%} $	&	$1.166$\\
$8050$	&	\texttt{1.446e-04} &	$^{+10\%}_{-9\%} $ &	$^{+6.4\%}_{-5.0\%} $	&	\texttt{1.691e-04} &	$^{+2.4\%}_{-2.9\%} $ & $^{+6.3\%}_{-5.0\%} $	&	$1.169$\\
$8250$	&	\texttt{1.221e-04} &	$^{+10\%}_{-9\%} $ &	$^{+6.5\%}_{-5.1\%} $	&	\texttt{1.428e-04} &	$^{+2.4\%}_{-3.0\%} $ & $^{+6.5\%}_{-5.1\%} $	&	$1.170$\\
$8650$	&	\texttt{8.718e-05} &	$^{+10\%}_{-9\%} $ &	$^{+6.8\%}_{-5.3\%} $	&	\texttt{1.023e-04} &	$^{+2.5\%}_{-3.1\%} $ & $^{+6.8\%}_{-5.3\%} $	&	$1.173$\\
$8850$	&	\texttt{7.365e-05} &	$^{+11\%}_{-9\%} $ &	$^{+7.0\%}_{-5.4\%} $	&	\texttt{8.680e-05} &	$^{+2.5\%}_{-3.1\%} $ & $^{+6.9\%}_{-5.4\%} $	&	$1.179$\\
$9250$	&	\texttt{5.297e-05} &	$^{+11\%}_{-9\%} $ &	$^{+7.3\%}_{-5.6\%} $	&	\texttt{6.270e-05} &	$^{+2.6\%}_{-3.2\%} $ & $^{+7.3\%}_{-5.6\%} $	&	$1.184$\\
$9450$	&	\texttt{4.508e-05} &	$^{+11\%}_{-9\%} $ &	$^{+7.5\%}_{-5.7\%} $	&	\texttt{5.336e-05} &	$^{+2.6\%}_{-3.2\%} $ & $^{+7.4\%}_{-5.7\%} $	&	$1.184$\\
$9850$	&	\texttt{3.268e-05} &	$^{+11\%}_{-10\%} $ &	$^{+7.8\%}_{-5.9\%} $	&	\texttt{3.878e-05} &	$^{+2.7\%}_{-3.3\%} $ & $^{+7.8\%}_{-5.8\%} $	&	$1.187$\\
$10050$	&	\texttt{2.785e-05} &	$^{+11\%}_{-10\%} $ &	$^{+8.0\%}_{-6.0\%} $	&	\texttt{3.308e-05} &	$^{+2.7\%}_{-3.4\%} $ & $^{+7.9\%}_{-5.9\%} $	&	$1.188$\\
$10450$	&	\texttt{2.027e-05} &	$^{+12\%}_{-10\%} $ &	$^{+8.3\%}_{-6.2\%} $	&	\texttt{2.418e-05} &	$^{+2.8\%}_{-3.5\%} $ & $^{+8.3\%}_{-6.1\%} $	&	$1.193$\\

\hline\hline
\end{tabular}
} 
\caption{For representative masses $m_k$ [GeV] (first column), the predicted cross sections [fb] for $\sqrt{s}=100\TeV$ at LO (second column) and NLO (third column) in QCD for inclusive $pp\to k^{--}k^{++}+X$ via the Drell-Yan process (DY). Also shown are scale uncertainties [\%], PDF uncertainties [\%], and the QCD $K$-factor. See Sec.~\ref{sec:setup_sm} for SM inputs.}
\label{tab:app_xsec_LHC100_kk}
\end{center}
\end{table*}

\begin{table*}[t!]
\begin{center}
\resizebox{.95\textwidth}{!}{
\begin{tabular}{c | c c c | c c c | c}
\hline\hline
\multicolumn{8}{c}{$pp\to h^{-}h^{+}+X$}\\
\hline
\multicolumn{8}{c}{$\sqrt{s}=100\TeV$ LHC}
\\
mass [GeV] & $\sigma^{\rm LO}_{100\TeV}$ [fb] & $\delta_{\rm RG~scale}$ [\%] &
	 $\delta_{\rm PDF}$ [\%] & $\sigma^{\rm NLO}_{100\TeV}$ [fb] & $\delta_{\rm RG~scale}$ [\%] &
	 $\delta_{\rm PDF}$ [\%] & $K^{\rm NLO}$ \\
\hline
$250$	&	\texttt{6.102e+01} &	$^{+6\%}_{-7\%} $ &	$^{+1.6\%}_{-1.7\%} $	&	\texttt{7.210e+01} &	$^{+2.2\%}_{-3.4\%} $ & $^{+1.6\%}_{-1.7\%} $	&	$1.182$\\
$450$	&	\texttt{9.278e+00} &	$^{+3\%}_{-4\%} $ &	$^{+1.7\%}_{-1.9\%} $	&	\texttt{1.075e+01} &	$^{+1.6\%}_{-2.0\%} $ & $^{+1.7\%}_{-1.9\%} $	&	$1.159$\\
$850$	&	\texttt{1.072e+00} &	$^{+0\%}_{-1\%} $ &	$^{+2.1\%}_{-2.1\%} $	&	\texttt{1.222e+00} &	$^{+1.1\%}_{-0.8\%} $ & $^{+2.1\%}_{-2.1\%} $	&	$1.140$\\
$1050$	&	\texttt{5.044e-01} &	$^{+0\%}_{-1\%} $ &	$^{+2.3\%}_{-2.2\%} $	&	\texttt{5.735e-01} &	$^{+1.1\%}_{-0.8\%} $ & $^{+2.3\%}_{-2.2\%} $	&	$1.137$\\
$1450$	&	\texttt{1.529e-01} &	$^{+2\%}_{-2\%} $ &	$^{+2.6\%}_{-2.4\%} $	&	\texttt{1.734e-01} &	$^{+1.3\%}_{-1.1\%} $ & $^{+2.6\%}_{-2.4\%} $	&	$1.134$\\
$1650$	&	\texttt{9.323e-02} &	$^{+3\%}_{-2\%} $ &	$^{+2.8\%}_{-2.5\%} $	&	\texttt{1.056e-01} &	$^{+1.3\%}_{-1.2\%} $ & $^{+2.8\%}_{-2.6\%} $	&	$1.133$\\
$2050$	&	\texttt{3.952e-02} &	$^{+4\%}_{-3\%} $ &	$^{+3.0\%}_{-2.7\%} $	&	\texttt{4.472e-02} &	$^{+1.4\%}_{-1.3\%} $ & $^{+3.0\%}_{-2.8\%} $	&	$1.132$\\
$2250$	&	\texttt{2.695e-02} &	$^{+4\%}_{-4\%} $ &	$^{+3.1\%}_{-2.8\%} $	&	\texttt{3.053e-02} &	$^{+1.5\%}_{-1.4\%} $ & $^{+3.2\%}_{-2.9\%} $	&	$1.133$\\
$2650$	&	\texttt{1.354e-02} &	$^{+5\%}_{-4\%} $ &	$^{+3.4\%}_{-3.0\%} $	&	\texttt{1.530e-02} &	$^{+1.5\%}_{-1.6\%} $ & $^{+3.4\%}_{-3.0\%} $	&	$1.130$\\
$2850$	&	\texttt{9.843e-03} &	$^{+5\%}_{-5\%} $ &	$^{+3.5\%}_{-3.1\%} $	&	\texttt{1.116e-02} &	$^{+1.5\%}_{-1.6\%} $ & $^{+3.5\%}_{-3.1\%} $	&	$1.134$\\
$3250$	&	\texttt{5.456e-03} &	$^{+5\%}_{-5\%} $ &	$^{+3.6\%}_{-3.3\%} $	&	\texttt{6.203e-03} &	$^{+1.6\%}_{-1.8\%} $ & $^{+3.7\%}_{-3.3\%} $	&	$1.137$\\
$3450$	&	\texttt{4.149e-03} &	$^{+6\%}_{-5\%} $ &	$^{+3.7\%}_{-3.3\%} $	&	\texttt{4.708e-03} &	$^{+1.6\%}_{-1.8\%} $ & $^{+3.8\%}_{-3.4\%} $	&	$1.135$\\
$3850$	&	\texttt{2.456e-03} &	$^{+6\%}_{-6\%} $ &	$^{+3.9\%}_{-3.5\%} $	&	\texttt{2.798e-03} &	$^{+1.7\%}_{-1.9\%} $ & $^{+3.9\%}_{-3.5\%} $	&	$1.139$\\
$4050$	&	\texttt{1.919e-03} &	$^{+7\%}_{-6\%} $ &	$^{+4.0\%}_{-3.6\%} $	&	\texttt{2.189e-03} &	$^{+1.7\%}_{-2.0\%} $ & $^{+4.0\%}_{-3.6\%} $	&	$1.141$\\
$4450$	&	\texttt{1.194e-03} &	$^{+7\%}_{-6\%} $ &	$^{+4.2\%}_{-3.7\%} $	&	\texttt{1.366e-03} &	$^{+1.8\%}_{-2.1\%} $ & $^{+4.2\%}_{-3.7\%} $	&	$1.144$\\
$4650$	&	\texttt{9.523e-04} &	$^{+7\%}_{-6\%} $ &	$^{+4.3\%}_{-3.8\%} $	&	\texttt{1.090e-03} &	$^{+1.8\%}_{-2.1\%} $ & $^{+4.3\%}_{-3.8\%} $	&	$1.145$\\
$5050$	&	\texttt{6.152e-04} &	$^{+8\%}_{-7\%} $ &	$^{+4.5\%}_{-3.9\%} $	&	\texttt{7.052e-04} &	$^{+1.9\%}_{-2.2\%} $ & $^{+4.5\%}_{-3.9\%} $	&	$1.146$\\
$5250$	&	\texttt{4.978e-04} &	$^{+8\%}_{-7\%} $ &	$^{+4.6\%}_{-4.0\%} $	&	\texttt{5.707e-04} &	$^{+1.9\%}_{-2.3\%} $ & $^{+4.6\%}_{-4.0\%} $	&	$1.146$\\
$5650$	&	\texttt{3.300e-04} &	$^{+8\%}_{-7\%} $ &	$^{+4.8\%}_{-4.1\%} $	&	\texttt{3.796e-04} &	$^{+2.0\%}_{-2.4\%} $ & $^{+4.8\%}_{-4.1\%} $	&	$1.150$\\
$5850$	&	\texttt{2.706e-04} &	$^{+8\%}_{-7\%} $ &	$^{+4.9\%}_{-4.2\%} $	&	\texttt{3.113e-04} &	$^{+2.0\%}_{-2.4\%} $ & $^{+4.9\%}_{-4.2\%} $	&	$1.150$\\
$6250$	&	\texttt{1.831e-04} &	$^{+9\%}_{-8\%} $ &	$^{+5.2\%}_{-4.3\%} $	&	\texttt{2.118e-04} &	$^{+2.1\%}_{-2.5\%} $ & $^{+5.2\%}_{-4.4\%} $	&	$1.157$\\
$6450$	&	\texttt{1.511e-04} &	$^{+9\%}_{-8\%} $ &	$^{+5.3\%}_{-4.4\%} $	&	\texttt{1.754e-04} &	$^{+2.1\%}_{-2.6\%} $ & $^{+5.3\%}_{-4.4\%} $	&	$1.161$\\
$6850$	&	\texttt{1.045e-04} &	$^{+9\%}_{-8\%} $ &	$^{+5.5\%}_{-4.6\%} $	&	\texttt{1.211e-04} &	$^{+2.2\%}_{-2.7\%} $ & $^{+5.5\%}_{-4.6\%} $	&	$1.159$\\
$7050$	&	\texttt{8.712e-05} &	$^{+9\%}_{-8\%} $ &	$^{+5.7\%}_{-4.6\%} $	&	\texttt{1.011e-04} &	$^{+2.2\%}_{-2.7\%} $ & $^{+5.6\%}_{-4.7\%} $	&	$1.160$\\
$7450$	&	\texttt{6.080e-05} &	$^{+10\%}_{-8\%} $ &	$^{+5.9\%}_{-4.8\%} $	&	\texttt{7.094e-05} &	$^{+2.3\%}_{-2.8\%} $ & $^{+5.9\%}_{-4.8\%} $	&	$1.167$\\
$7650$	&	\texttt{5.109e-05} &	$^{+10\%}_{-8\%} $ &	$^{+6.1\%}_{-4.9\%} $	&	\texttt{5.961e-05} &	$^{+2.3\%}_{-2.8\%} $ & $^{+6.1\%}_{-4.9\%} $	&	$1.167$\\
$8050$	&	\texttt{3.615e-05} &	$^{+10\%}_{-9\%} $ &	$^{+6.4\%}_{-5.0\%} $	&	\texttt{4.228e-05} &	$^{+2.4\%}_{-2.9\%} $ & $^{+6.3\%}_{-5.0\%} $	&	$1.170$\\
$8250$	&	\texttt{3.054e-05} &	$^{+10\%}_{-9\%} $ &	$^{+6.5\%}_{-5.1\%} $	&	\texttt{3.570e-05} &	$^{+2.4\%}_{-3.0\%} $ & $^{+6.5\%}_{-5.1\%} $	&	$1.169$\\
$8650$	&	\texttt{2.179e-05} &	$^{+10\%}_{-9\%} $ &	$^{+6.8\%}_{-5.3\%} $	&	\texttt{2.558e-05} &	$^{+2.5\%}_{-3.1\%} $ & $^{+6.8\%}_{-5.3\%} $	&	$1.174$\\
$8850$	&	\texttt{1.842e-05} &	$^{+11\%}_{-9\%} $ &	$^{+7.0\%}_{-5.4\%} $	&	\texttt{2.170e-05} &	$^{+2.5\%}_{-3.1\%} $ & $^{+6.9\%}_{-5.4\%} $	&	$1.178$\\
$9250$	&	\texttt{1.324e-05} &	$^{+11\%}_{-9\%} $ &	$^{+7.3\%}_{-5.6\%} $	&	\texttt{1.567e-05} &	$^{+2.6\%}_{-3.2\%} $ & $^{+7.3\%}_{-5.6\%} $	&	$1.184$\\
$9450$	&	\texttt{1.128e-05} &	$^{+11\%}_{-9\%} $ &	$^{+7.5\%}_{-5.7\%} $	&	\texttt{1.334e-05} &	$^{+2.6\%}_{-3.2\%} $ & $^{+7.4\%}_{-5.7\%} $	&	$1.183$\\
$9850$	&	\texttt{8.171e-06} &	$^{+11\%}_{-10\%} $ &	$^{+7.8\%}_{-5.9\%} $	&	\texttt{9.695e-06} &	$^{+2.7\%}_{-3.3\%} $ & $^{+7.8\%}_{-5.8\%} $	&	$1.187$\\
$10050$	&	\texttt{6.967e-06} &	$^{+11\%}_{-10\%} $ &	$^{+8.0\%}_{-6.0\%} $	&	\texttt{8.270e-06} &	$^{+2.7\%}_{-3.4\%} $ & $^{+7.9\%}_{-5.9\%} $	&	$1.187$\\
$10450$	&	\texttt{5.064e-06} &	$^{+12\%}_{-10\%} $ &	$^{+8.3\%}_{-6.2\%} $	&	\texttt{6.045e-06} &	$^{+2.8\%}_{-3.5\%} $ & $^{+8.3\%}_{-6.1\%} $	&	$1.194$\\

\hline\hline
\end{tabular}
} 
\caption{For representative masses $m_h$ [GeV] (first column), the predicted cross sections [fb] for $\sqrt{s}=100\TeV$ at LO (second column) and NLO (third column) in QCD for inclusive $pp\to h^{-}h^{+}+X$ via the Drell-Yan process (DY). Also shown are scale uncertainties [\%], PDF uncertainties [\%], and the QCD $K$-factor. See Sec.~\ref{sec:setup_sm} for SM inputs.}
\label{tab:app_xsec_LHC100_hh}
\end{center}
\end{table*}

\
\newpage
\
\newpage

{\tiny
\bibliography{zeeBabu_LHCupdate_refs}
}

\end{document}